\documentclass[fleqn,usenatbib]{mnras}

\usepackage[T1]{fontenc}
\usepackage{ae,aecompl}

\usepackage{graphicx}	
\usepackage{amsmath}	
\usepackage{amssymb}	

\usepackage[dvipsnames]{xcolor}

\usepackage{hyperref}

\def \mum {$\rm \mu m$}
\def \lir {$\rm L_{IR}$}
\def \mstar {$\rm M_{*}$}
\def \mdust {$\rm M_{dust}$}

\def \msun{$\rm M_{\odot}$}

\def \maslim {$\rm M_{lim}$}
\def \tcool {$\rm t_{cool}$}
\def \mh{$\rm M_{200}$}
\def \fsf{$f_{\rm SF}$}
\def \fq{$f_{\rm Q}$}
\def \z{$z_{\rm spec}$}

\usepackage{newtxtext,newtxmath}

\title[BCG evolution in the Nearby Universe II]{The Evolution of Brightest Cluster Galaxies in the Nearby Universe II: The star-formation activity and the Stellar Mass from Spectral Energy Distribution.}

\author[G. Orellana-Gonz\'alez et al.]{Orellana-Gonz\'alez, G.$^{1}$\thanks{E-mail: \textcolor{blue}{gustavo.orellana@ucsc.cl}},
Cerulo, P.$^{2}$\thanks{E-mail: \textcolor{blue}{pcerulo@inf.udec.cl}},
Covone, G.$^{3,4,5}$,
Cheng, C.$^{6,7}$,
Leiton, R.$^{8}$,
\and
Demarco, R.$^{8}$ \&\ 
Marie-Lou Gendron-Marsolais $^{9}$
\\
$^{1}$Departamento de Matem\'atica y F\'isica Aplicadas, Universidad Cat\'olica de la Sant\'isima Concepci\'on, Alonso de Ribera 2850, Concepci\'on, Chile.\\
$^{2}$Departamento de Ingenier\'ia Inform\'atica y Ciencias de la Computaci\'on, Universidad de Concepci\'on, Chile.\\
$^{3}$Dipartimento di Fisca ``E. Pancini'', University of Naples ``Federico II'', Naples, Italy. \\
$^{4}$INFN, Sezione di Napoli,  Naples, Italy.\\
$^{5}$INAF, Osservatorio Astronomico di Capodimonte,  Naples, Italy.\\
$^{6}$Chinese Academy of Sciences South America Center for Astronomy, National Astronomical Observatories, CAS, Beijing 100101, China. \\
$^{7}$CAS Key Laboratory of Optical Astronomy, National Astronomical Observatories, Chinese Academy of Sciences, Beijing 100101, China. \\
$^{8}$Departamento de Astronom\'ia, Facultad de Ciencias F\'isicas y Matem\'aticas,
Universidad de Concepci\'on, Concepci\'on, Chile. \\
$^{9}$European Southern Observatory, Alonso de C\'ordova 3107, Vitacura, Casilla 19001, Santiago de Chile.
}

\date{Accepted 2021 December 22. Received 2021 December 15; in original form 2021 May 26}

\pubyear{2015}

\hypersetup{draft}

\begin{document}
\label{firstpage}
\pagerange{\pageref{firstpage}--\pageref{lastpage}}
\maketitle

\begin{abstract}
We study the star-formation activity in a sample of $\sim$ 56,000
brightest cluster galaxies (BCGs) at $0.05 < z < 0.42$ using optical and infra-red data from SDSS and WISE. We estimate stellar masses and star-formation rates (SFR) through SED fitting and study the evolution of the SFR  with redshift as well as the effects of BCG stellar mass, cluster halo mass  and cooling time on star formation. Our BCGs have $SFR = 1.4\times10^{-3}-275.2$ [\msun/yr] and $sSFR=5 \times 10^{-15}-6 \times 10^{-10}$ [yr$^{-1}$] . We find that star-forming BCGs are more abundant at higher redshifts and have higher $SFR$ than at lower redshifts. The fraction of star-forming BCGs (\fsf) varies from 30\% to 80\% at $0.05 < z < 0.42$. Despite the large values of \fsf, we show that only 13\% of the BCGs lie on the star-forming main sequence for field galaxies at the same redshifts. 
We also find that \fsf\ depends only weakly on \mh, while it sharply decreases with \mstar. 
We finally find that the $SFR$ in BCGs decreases with increasing \tcool, suggesting that star formation is related to the cooling of the intra-cluster medium. 
However, we also find a weak correlation of \mstar\ and \mh\ with \tcool\, suggesting that AGN are heating the intra-cluster gas around the BCGs. 
We compare our estimates of $SFR$ with the predictions from empirical models for the evolution of the $SFR$ with redshift, finding that the transition from a merger dominated to a cooling-dominated star formation may happen at $z < 0.6$.
\end{abstract}




\section{Introduction}\label{sec:intro}

Brightest Cluster Galaxies (BCGs) are the most massive and luminous galaxies in the Universe. The peculiar properties of these objects have so far attracted the interest of observational and theoretical astrophysicists as they are key to understand the formation and evolution of galaxies.
BCGs are usually located near the centers of clusters of galaxies, close to the minimum of their potential well. Their interaction with the surrounding intra-cluster medium (ICM; \citealt{Hu_1985}, \citealt{Cavagnolo_2008}) is highlighted, among other things, by the so-called X-ray cavities (see e.g.\ \citealt{Hlavacek_2015}), which represent one of the most striking evidence of feedback from an active galactic nucleus (AGN).

Most BCGs have elliptical morphologies, and the most massive of them also show extended stellar haloes (cD galaxies; e.g. \citealt{Tonry_1987}). 
The light profiles of BCGs are characterised by high S\'ersic indices, $n$, which can reach up to $n=10$ (\citealt{Graham_1996, Donzelli_2011}), while their stellar population is old and consistent with being formed at early cosmic epochs (i.e., z$>$2; see \citealt{Brough_2007,von_der_Linden_2007,Whiley_2008,Loubser_2009}), and their optical colours are consistent with passive evolution (\citealt{Cerulo_2019}, hereafter Paper I).

The stellar masses of BCGs range between $10^{10.5}$ and $10^{12.5}$ \msun\ and correlate with the masses of their hosting clusters (the halo mass), supporting the 
hypothesis that the stellar mass assembly of these galaxies is related to the assembly of the clusters in which they reside (see e.g., \citealt{Brough_2008}, \citealt{Lidman_2012}, \citealt{Oliva_Altamirano_2014}, \citealt{Bellstedt_2016}, \citealt{Lavoie_2016}, \citealt{Gozaliasl_2016}, \citealt{Zhao_2015b}).  
The stellar mass assembly of BCGs currently represents one of the open questions in galaxy evolution. There are two possible physical processes which can build up stellar mass in these galaxies, namely the formation of stars and mergers with satellite galaxies (e.g., \citealt{Cooke_2019}). 

\citet{De_Lucia_and_Blaizot_2007} show that BCGs grow in stellar mass through major mergers during the early stages of their evolution ($z > 1$), while afterwards minor mergers become the main driver for stellar mass build-up. This model is consistent with the results of \citet{Lidman_2012, Lidman_2013} and \citet{Ascaso_2014}, who show that BCGs nearly doubled their stellar mass in the last 9 Gyr. \citet{Webb_2015} has shown that star formation would be too low to produce the observed increase in stellar mass with redshift.

According to \citet{De_Lucia_and_Blaizot_2007}, BCGs should have nearly doubled their stellar mass since $z=0.5$. 
This conclusion is in disagreement with observations of BCGs in this redshift range, which report no significant growth in stellar mass (\citealt{Lin_2013}, \citealt{Oliva_Altamirano_2014}, Paper I). 
This slow growth at low redshifts can be explained if one considers that satellite galaxies that are accreted by the BCG lose part of their stars during the merging process. These stars end up constituting the intracluster light (ICL; e.g.\ \citealt{Contini_2018, Contini_2019}).

In a recent study \citet{Cooke_2019}, based on multi-wavelength photometry and spectral energy distribution (SED) fitting in a sample of BCGs from the Cosmic Evolution Survey (COSMOS, \citealt{Scoville_2007_COSMOS}), proposed a model with a three-stage sequence for the build-up of stellar mass: 
a) in the earliest epochs ($ z>2.5$) the BCG grows its stellar mass through star formation; 
b) at intermediate epochs ($ z \sim 1.25$) mergers start to become relevant, and the stellar mass growth occurs through both star formation and mergers; 
c) finally, at late epochs ($z\lesssim 1$) dry mergers are the dominant mechanism with no significant star formation contributing to the mass growth. 
This picture is in agreement with other observations and simulations where BCGs show a significant star formation at $ z > 0.5$ ($SFR \sim$ 10 - 100 $\rm M_\odot$/yr$^{-1}$), similar to spiral and starburst galaxies in the field at the same redshifts (e.g. \citealt{Webb_2015,Bonaventura_2017}).

BCGs in the nearby Universe are mostly quiescent, and only a small fraction of them are star-forming. \citet{Oliva_Altamirano_2014} find that at $z<0.4$ the fraction of star-forming BCGs is 27$\%$, whereas \citet{Fraser_McKelvie_2014} find a smaller fraction (1\%) of star-forming BCGs in a sample of 245 clusters at $z < 0.1$. In Paper I we showed that at $\rm 0.05 < z < 0.35$ the fraction of BCGs with star formation is $\sim 9\%$, and we found, in agreement with \cite{Oliva_Altamirano_2014}, that it decreases with cluster halo mass and BCG stellar mass. 
We also found that it  increases with redshift. The fractions of star-forming BCGs reported in \citet{Oliva_Altamirano_2014} and Paper I are higher than those expected in a scenario such as that proposed in \citet{Cooke_2019}, according to which most BCGs in the nearby Universe should be quenched.

It is not well understood what is triggering the formation of new stars in nearby BCGs and, in order to understand the onset of the star formation in BCGs, it is necessary to take into consideration the interaction between them and the surrounding ICM. 
The temperature of the ICM is close to the virial temperature for a system of the mass of a group or cluster of galaxies ($\sim 10^7$ K) and it loses energy by the emission of X-ray photons. 
Near the cluster centre the ICM is densest and a flow generates in which the gas from more external layers is accreted on to the central overdense region, losing energy through radiation. 
This region thus grows in mass and size accreting more gas that cools down. The weight of the upper layers of gas in this region causes a slow inflow towards the cluster centre. 
This process is called {\itshape{cooling flow}} \citep{Fabian_1994_ARA}.  
According to the cooling flow scenario, BCGs should be fed with large amounts of cold gas ($\sim 100-1000$ \msun/yr, \citealt{Mcdonald_2018}), resulting in extreme star formation. 
However, despite several works have observed the existence of cooling flows (e.g. \citealt{McNamara_1989, Edge_2001, Olivares_2019}), the star formation rates of BCGs are lower than what would result if all the accreted cool gas ended up fuelling star formation.

This "cooling flow problem" may be overcome if one takes into account the AGN activity of BCGs, since the feedback from an AGN could halt the formation of new stars \citep{De_Lucia_2006, McCarthy_2008}. 
Furthermore, \citet{best_2007} show that at $0.02 < z < 0.16$ radio-loud AGN are more abundant in BCGs than in field galaxies with similar stellar masses. 
The theoretical models of \citet{Voit_2017} and \citet{Gaspari_2017} support a scenario in which the local thermal instabilities in the ICM cause the cold 
gas to condense and fall on to the BCG, feeding its central supermassive black hole. 
This process may result in the activation of the BCG nucleus with the result that the outflows generated by the AGN will heat the ICM. On the other hand, several works (e.g., \citealt{Bildfell_2008,Pipino_2009}) show evidence that although the AGN feedback drives the heating of the ICM, it is not strong enough to completely stop the cooling flow.
They argue that when the cooling flow starts, the central galaxy will not  
stop forming stars. 

\citet{Mcdonald_2016} propose an evolutionary picture in which the principal drivers of star formation in BCGs change with redshift. They show that they can reproduce the increasing trend of star-formation rate ($SFR$) with redshfit with a model in which the star formation is mainly triggered through gas-rich mergers at $z > 0.5$ and by ICM cooling at $z < 0.5$. 
Interestingly, \cite{Hlavacek_2020} report the detection of a cooling flow in a cluster at $z=1.5$ in which the BCG is forming stars, indicating that more than one physical driver should be taken into account to explain what triggers star formation in these massive galaxies. 

In the present paper we study the star formation in a large sample of BCGs drawn from the Sloan Digital Sky Survey (SDSS, \citealt{York_2000}) in the redshift range $0.05 < z < 0.42$. 
This sample upgrades the one used in Paper I, extending it up to $z=0.42$ and including only BCGs with spectroscopic redsfhifts and high signal-to-noise ($S/N$) infra-red (IR) photometry from the Wide-field Infrared Survey Explorer (WISE; \citealt{Wright_2010}). This allows us to perform SED fitting and obtain accurate estimates of star-formation rate ($SFR$) and stellar mass (\mstar). 
This is the second in a series of papers devoted to the study of the physics of BCGs in the nearby Universe. 
In this work we update the sample and focus on the evolution of the SFR and its relation with BCG and cluster properties.

The paper is organized as follows: we present the data and the SED fitting procedure in Sect. 2. 
Section 3 shows the analysis and the results, which are discussed in Sect. 4. 
Section 5 finally summarises the main conclusions of the paper. 
Throughout the paper we use a cosmological model with $\rm \Omega_m = 0.27$, $\rm \Omega_\Lambda = 0.73$, and $\rm H_0 = 70.5~ km~ s^{-1}  Mpc^{-1} $ (\citealt{Hinshaw_2009}), unless otherwise stated. 
SDSS magnitudes are reported in the AB system, while WISE magnitudes and fluxes are reported in the Vega system. We will interchangeably use the symbols $z$ and \z\ to refer to spectroscopic redshifts. We will use the notation $\rm R_{200}$ to indicate the radius within which the local density is 200 times the critical density of the Universe at the redshift of each BCG and \mh\ to indicate the total mass enclosed within this radius.

\section{Data}\label{sec:data_description}

\subsection{The Brightest Galaxy Clusters Sample}
\label{subsec:cluster_catalog}

We use the catalogue of brightest galaxy clusters presented in \citet{Wen_2015b} (hereafter WHL15), which is an updated version of the catalogue of \citet{Wen_2012} (hereafter WHL12).

The WHL12 catalogue comprises 132,684 clusters at $0.05\leq z_{\rm photo} <0.75$ detected in the Data Release 8 of the Sloan Digital Sky Survey (SDSS DR8; \citealt{Aihara_2011}) with an algorithm based on galaxy photometric redshifts and projected positions. The WHL12 sample is 75\% complete at $ 0.05 \leq z_{\rm photo} < 0.42$ for clusters with halo masses $\rm M_{200} > 6 \times 10^{13}$ [\msun]. The centroids of the clusters are defined as the positions of the BCGs. The WHL15 catalogue adds the spectroscopic redshifts from the SDSS DR12 (\citealt{Alam_2015}), which increases the fraction of spectroscopically confirmed clusters from 64\% to 85\% in the entire sample. 
We refer the reader to WHL12 and WHL15 for more details on the detection algorithm, and to Paper I for a summary of the detection procedures. Since the centroids of the clusters coincide with the positions of the BCGs in WHL12 and WHL15, the catalogue used in this paper can be considered both as a BCG and a cluster catalogue. Therefore, hereafter we will interchangeably refer to it as the cluster or BCG sample.

We estimated the dark matter halo masses \mh\ of the clusters using the scaling relation between cluster richness and \mh\ presented in Equation 2 of WHL12. This scaling relation was confirmed by \cite{Covone_2014} through a weak lensing analysis conducted on a sub-sample of clusters of the WHL12 sample.

\subsection{Optical data and Spectra} 
\label{subsec:optical_data}

We use the optical photometry in the $u$, $g$, $r$, $i$ and $z$ bands, from the SDSS DR12 database. We employed the scalar function {\ttfamily{fGetNearestObjEq}} with a search radius of $\sim0.02^{\prime}$ (i.e.\ $\sim$1$^{\prime\prime}$)\footnote{In Section 2.2 of Paper I, we erroneously quoted $\sim0.02^{\prime\prime}$ instead of $\sim0.02^{\prime}$ as the size of the search radius.} to find the closest matches with the WHL15 catalogue. In this work we use model magnitudes ({\ttfamily{modelMag}} which we correct for Galactic extinction with the coefficients provided in the DR12 database and obtained adopting the dust maps of \citet{Schlegel_1998}).

Spectroscopic redshifts (\z) were downloaded from the table {\ttfamily{SpecObjAll}} in the DR12 repository. These redshifts were measured on spectra observed with optical fibers with 3$^{\prime\prime}$ wide aperture diameters, centred in the target source. The estimates of the redshifts were performed by cross-correlating the spectra with templates of several source types including different kinds of galaxies, stars, cataclysmic variables and quasars.

\subsection{Infrared data}\label{subsec:IR_data}

We use near-to-mid-infrared (IR) photometric data from WISE in the four $W1$, $W2$, $W3$ and $W4$ bands, centred at 3.4 \mum, 4.6 \mum, 12.0 \mum\ and 22.8 \mum\ (see \citealt{Brown_2014}), respectively. While in Paper I we queried the ALLWISE repository to find matches with the WHL15 catalogue, in this work we decided to use the unblurred coadds of the WISE imaging (unWISE\footnote{\url{http://unwise.me/}}; \citealt{Lang_2014_unWISE, Lang_2016_unWISE}). The unWISE catalogue was constructed with the forced photometry technique, which uses the positions of the sources in the SDSS optical catalogue, the classification of a source as a star or a galaxy, and its optical light profile. Flux measurements take into account the point-spread function (PSF) and the noise mode from WISE (see \citealt{Lang_2014_unWISE} for more details). 

Since there is no PSF matching and the object detection is made on the SDSS r-band images, the forced photometry allows one to measure accurate fluxes from bright extended sources and to recover some weak sources detected in SDSS and not detectable in the PSF-matched images used in ALLWISE. Therefore, with unWISE we are able to build a high $S/N$ optical and IR multi-wavelength catalogue, significantly increasing the number of WISE-detected BCGs with respect to Paper I.

\subsection{Spectral energy distribution fitting}
\label{subsec:SED}

We use the Code Investigating GAlaxy Emission ({\ttfamily{CIGALE}}; \citealt{Boquien_2019}) to fit SEDs to the optical and IR data of our BCGs. This software package allows one to perform reliable multi-wavelength photometric analyses from far ultraviolet (FUV; 1,500 \AA) to radio (20 cm) wavelengths. {\ttfamily{CIGALE}} builds different SED templates, considering modules for star formation history (SFH), simple stellar populations (SSP), nebular emission, dust attenuation law, dust emission, AGN emission, and radio synchrotron emission.
Each module allows the selection of different models, and each model requires the selection of certain values to constrain the final result. 
The template SEDs that the software builds have each a given value of SFR, \mstar, dust mass (\mdust) and IR luminosity (\lir). 
 
For our sample, we selected the following four modules: \begin{itemize}
    \item SFH: We select the option \emph{sfhdelayed}, based on the so-called "delayed" SFH. In this SFH model the star formation increases exponentially with time until the time $\tau$ (the peak of the star-formation) and then decreases until the time t$_0$:
    \begin{equation}
    \nonumber
      \rm SFH(t)\propto\frac{t}{\tau^2}\times\exp\left(\frac{-t}{\tau}\right) ~\left[\frac{M_{\odot}}{yr}\right],~~ 0\leq t\leq t_0. 
    \end{equation}
    We selected 6 values for the $\tau$ of the main stellar population model ($\tau_{\rm main}$), in the range 0.3 -- 12 Gyr and seven values for the age of the oldest stars ($\rm Age_{old}$) in the range 2 -- 10 Gyr.
    \item SSP: We select the option $m05$ based on the SSP synthesis model developed by \citet{Maraston_2005}\footnote{Following \cite{Tonini_2012}, we adopted the \cite{Maraston_2005} models, as demonstrated by those authors, they are more reliable in tracing the stellar mass of galaxies.}, using a \citep{Chabrier_2003} initial mass function (IMF).
   We selected 4 values for metallicity (Z), in the range from 0.004 to 0.02, consistent with the metallicities inferred from optical colours in Paper I.  We also selected 7 equal age intervals between the young and the old stellar populations ($\rm Age_{sep}$) in the range 0.05 to 5 Gyr.
    \item Attenuation law: We selected the option $dustatt\_modified\_ $ $starburst$ based on a modified \citet{Calzetti_2000} dust attenuation law (see \citealt{Boquien_2019}), choosing 3 values of the colour excess for the stellar continuum light of the young population in the range $E_{B-V}$ = 0.1 -- 0.3 mag. 
    \item Dust emission: We selected the option $dl2007$, based on the dust emission model developed by \citet{Draine_2007}.
   We used 4 values for the mass fraction of polycyclic aromatic hydrocarbons (PAH) ($\rm q_{PAH}$) ranging from 0.47 to 4.58, 6 values for the minimum radiation field ($\rm u_{min}$) in the range 0.2-25.0, 2 values for the maximum radiation field ($\rm u_{max}$) in the range 10$^4$-10$^5$, and 2 values for the fraction of the dust illuminated by the maximum and minimum radiation field ($\gamma$) in the range 0.00-0.02.
\end{itemize}
In total, we use a library of 7,451,136 SED templates to characterize our sample.

\subsection{Final sample and sub-samples}
\label{subsec: final samp}

\begin{figure*}
\centering
\includegraphics[width=0.32\textwidth]{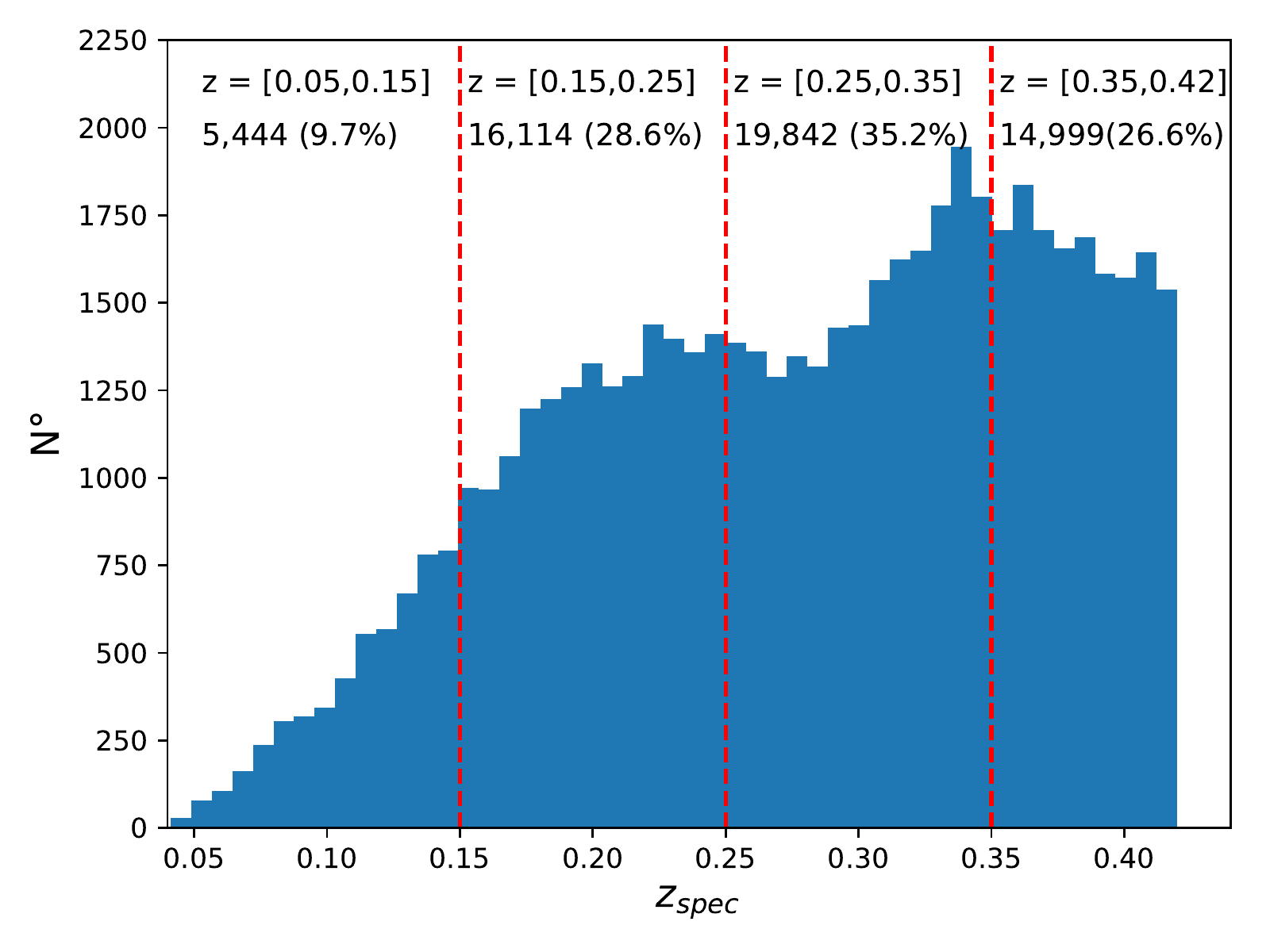}
\includegraphics[width=0.32\textwidth]{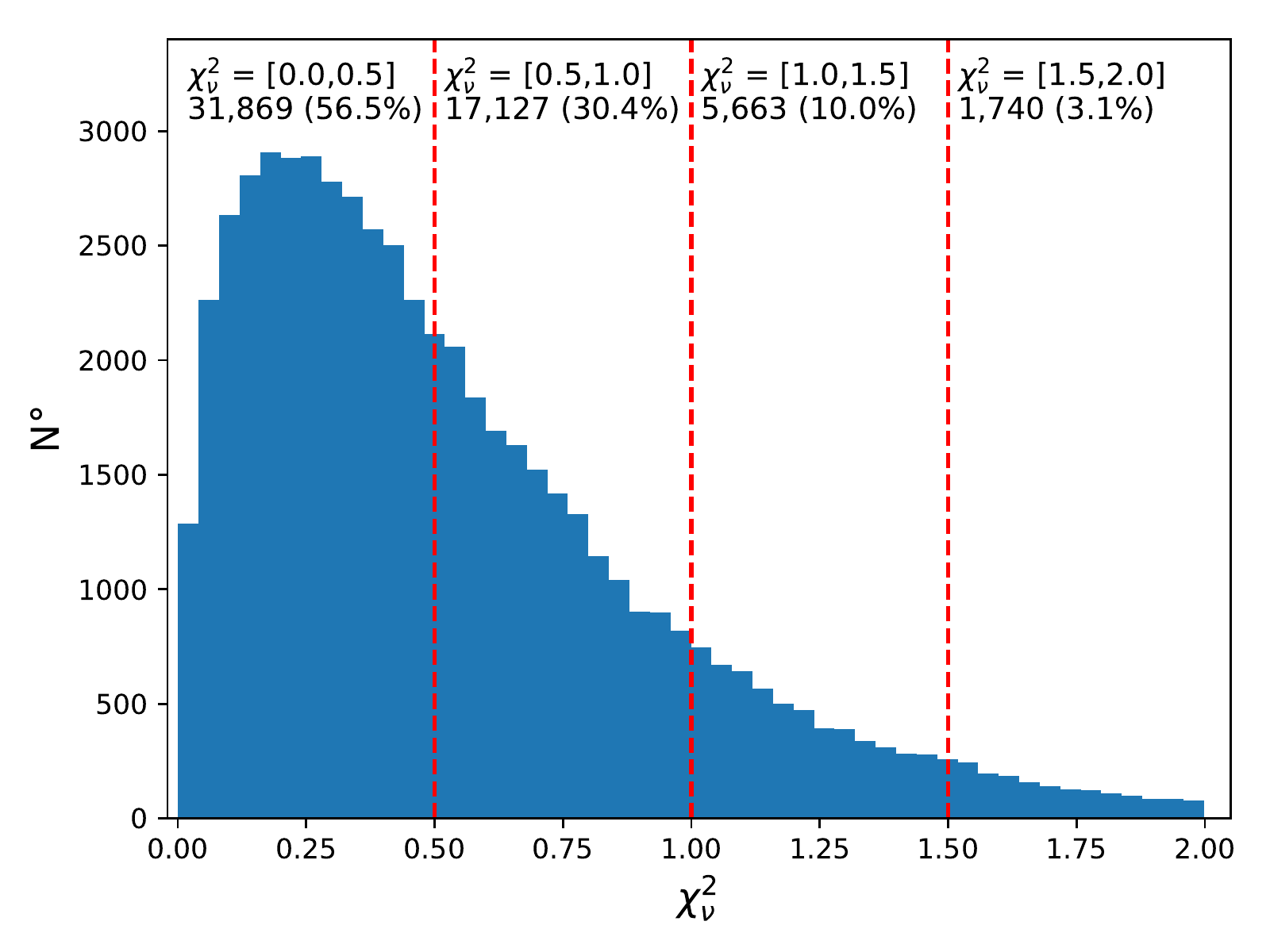}
\includegraphics[width=0.35\textwidth]{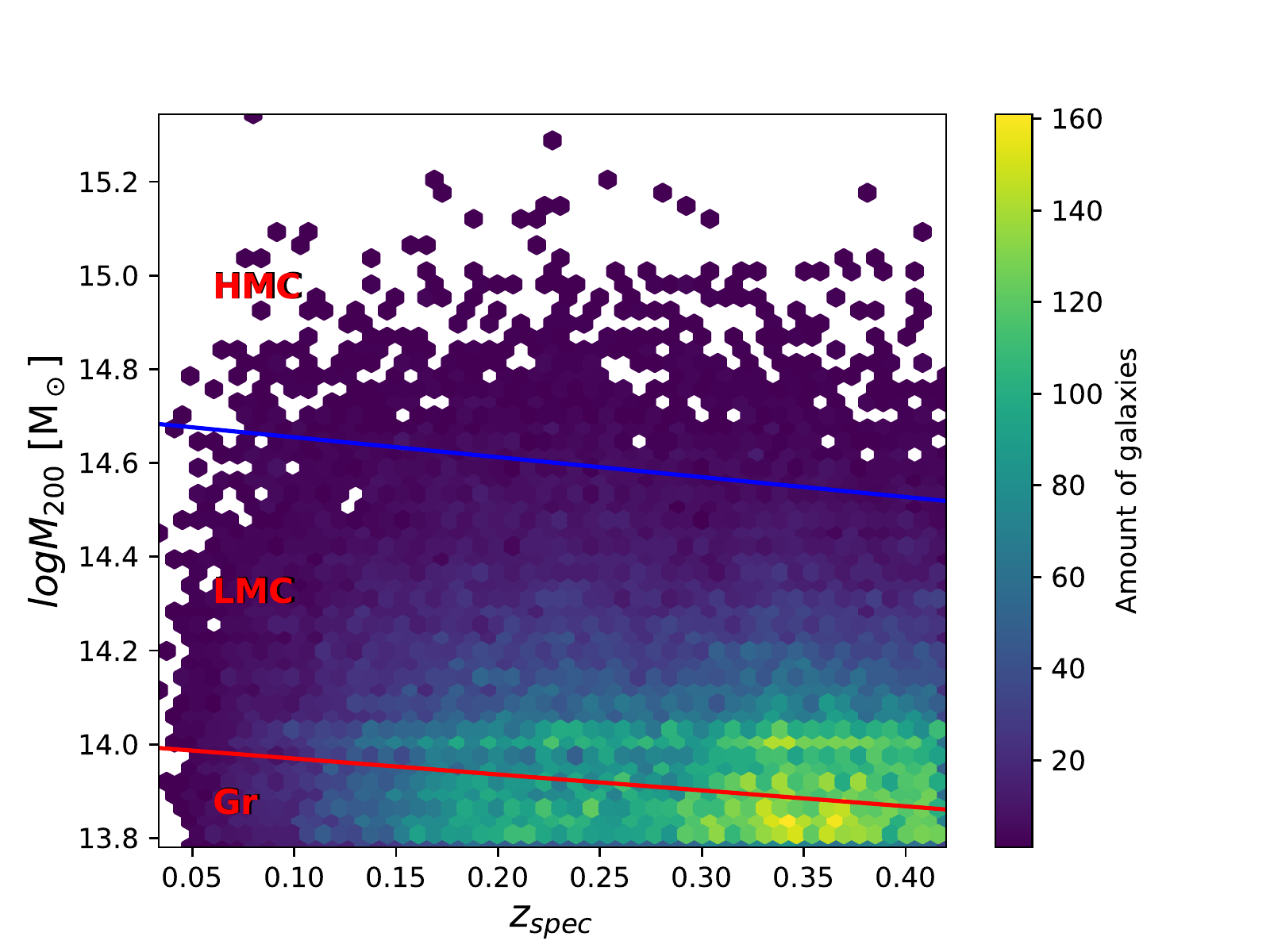}
\caption{Distributions of spectroscopic redshift (left-hand panel) and reduced $\chi^2$ $\left(\chi^2_{\nu}\right)$ (middle panel). The ranges covered by each quantity are divided into four bins delimited by the red vertical dashed lines. At the top of each plot we show the boundaries of each bin, the number of galaxies that they contain and the corresponding percentage with respect to the number of BCGs in the final sample.The right-hand panel shows a density plot representing the cluster halo mass (M$_{200}$) as a  function of the redshift. The plot is colour-coded according to the amount of galaxies in each cell. Red and blue lines separate the BCG in groups ($Gr$), low mass clusters (LMC) and high mass clusters (HMC). 
} 
\label{fig:histo gal with w3}	
\end{figure*}

\begin{table*}
    \centering
    \begin{tabular}{l@{\hspace{0.05cm}}cc@{\hspace{0.13cm}}c@{\hspace{0.13cm}}c@{\hspace{0.13cm}}c@{\hspace{0.13cm}}cc@{\hspace{0.13cm}}c@{\hspace{0.13cm}}c@{\hspace{0.13cm}}c@{\hspace{0.13cm}}c@{\hspace{0.13cm}}c@{\hspace{0.13cm}}}
\hline    
Redshift & Redshift                &  Frac.  &  Frac.  &  Frac. & Total & Total    & Median             & Median              & Median           & Median            &  Median            & Median\\
bin      &range                    & Gr      & LMC     &  HMC   & Frac. & number   & $SFR$              & $SFR^{\rm a}$       & log M$_*$        & log M$_*^{\rm a}$ &  log M$_{\rm 200}$ & log M$_{\rm 200}^{\rm a}$\\
         &                         &         &         &        &       &          &[$\rm M_{\odot}$/yr]& [$\rm M_{\odot}$/yr]& [$\rm M_{\odot}$]& [$\rm M_{\odot}$] & [$\rm M_{\odot}$] & [$\rm M_{\odot}$]\\
\hline
\hline
\vspace{0.1cm}
B1 & 0.05 $\leq$ z $<$ 0.15 &  43.6 \% & 57.0 \% & 2.4 \% & 9.6 \% & 5,402 & 0.7 $^{+ 1.6 } _{- 0.4}$ & 0.7 $^{+ 1.0 } _{- 0.4}$ & 11.41 $^{+ 0.50} _{- 0.21}$ & 11.52 $^{+ 0.31} _{- 0.12 }$ & 14.01 $^{+ 0.86} _{- 0.15}$ & 14.1 $^{+ 2.3 } _{- 0.2}$  \\ 
\vspace{0.1cm}
B2 & 0.15 $\leq$ z $<$ 0.25 &  41.1 \% & 59.9 \% & 2.2 \% & 28.6 \% & 16,114 & 1.2 $^{+ 4.0 } _{- 0.8}$ & 1.1 $^{+ 2.4} _{- 0.7}$ & 11.47 $^{+ 0.38} _{- 0.19}$ & 11.54 $^{+ 0.27} _{- 0.13}$ & 13.99 $^{+ 0.64} _{- 0.15}$ & 14.01 $^{+ 0.92} _{- 0.16}$  \\ 
\vspace{0.1cm}
B3 & 0.25 $\leq$ z $<$ 0.35 &  34.4 \% & 66.8 \% & 2.1 \% & 35.2 \% & 19,842 & 2.8 $^{+ 6.2} _{-1.9}$ & 2.6$^{+ 5.3} _{- 1.6}$ & 11.50 $^{+ 0.34} _{-0.18}$ & 11.55 $^{+ 0.28} _{- 0.13}$ & 13.98 $^{+ 0.49} _{-0.14}$ & 14.0 $^{+ 0.57} _{- 0.15}$  \\ 
\vspace{0.1cm}
B4 & 0.35 $\leq$ z $<$ 0.42 &  29.8 \% & 75.6 \% & 1.7 \% & 26.6 \% & 14,999 & 3.3 $^{+ 6.6 } _{- 2.1 }$ & 3.0 $^{+ 5.7 } _{- 1.9 }$ & 11.53 $^{+ 0.33} _{- 0.18}$ & 11.57 $^{+ 0.28} _{- 0.13}$ & 13.97 $^{+ 0.40} _{- 0.13}$ & 13.98 $^{+ 0.43} _{- 0.14}$  \\ 
\hline
\hline
BT & 0.05 $\leq$ z $<$ 0.42 &  33.1 \% & 64.9 \% & 2.0 \% & 100.0 \% &  & 2.1 $^{+ 5.6}_{- 1.5 }$ & 2.0 $^{+ 4.7 }_{- 1.5 }$ & 11.49 $^{+ 0.36 }_{- 0.19 }$ & 11.55 $^{+ 0.28 }_{- 0.13 }$ & 13.98 $^{+ 0.52 }_{- 0.14 }$ & 14.00 $^{+ 0.62 }_{- 0.15 }$  \\ 
Total &  &                     18,676 & 36,580 & 1,143 & 56,399 &  &  &  &  &  & \\ 

\hline
    \end{tabular}
    \caption{Fraction of galaxies per redshift and halo mass (log M$_{200}$) bins. 
    Percentages for the redshift bins B1 to B4, are derived with respect to the total amount of galaxies in the bin, indicated in the Total number column.
    The Total Frac. column indicates the fraction of BCGs in each redshift bin.
    BT corresponds to the entire redshift range of the sample, and the fractions reported for this bin are derived with respect to the total amount of BCGs in our sample. Total reports the number of BCGs in each of the $Gr$, LMC and HMC sub-samples across the entire redshift range.
    The last six columns show the median values of $\log SFR$, $\rm \log{(M_*)}$ and $\rm log{(M_{200})}$ with the $1\sigma$ widths of the distributions of these quantities in each redshift bin and in the  entire sample (BT). See section \ref{subsec:star formation and stellar mass dist} for more details. \newline
     $^{\rm a}$ Median value derived using galaxies with \mstar$>$\maslim\ (see section \ref{subsec:star formation and stellar mass dist} for more details).
     }
    \label{tab:bins_def}
\end{table*}

The BCG sample used for the analysis in the present work was obtained applying the following criteria to the WHL15 sample:

\begin{itemize}
    \item Spectroscopic redshift selection: we restrict our analysis to spectroscopically confirmed BCGs. 
    This selection removes 36,996 galaxies, which are the 27.9\% of the galaxies contained in the original WHL15 sample.
    \item Redshift boundaries: we selected BCGs with $\rm 0.05 \leq z_{\rm spec} < 0.42$, the upper boundary corresponding to the redshift at which the WHL15 sample is 75\% complete. 
    In this selection we discard 36,860 galaxies, corresponding to 27.8\% of the galaxies in the original sample.
    \item Cluster halo mass limit: we selected clusters with masses \mh$\geq 6\times10^{13}$ \msun, as done in Paper I, and corresponding to the mass at which the WHL15 sample is 75\% complete at $ z_{\rm spec} < 0.42$. 
    With this cut, we discard 1,153 galaxies, the 0.9\% of the entire sample.
    \item $S/N$: in order to obtain a reliable SED fitting, we only considered galaxies with $S/N >10$ in the SDSS optical bands and with $S/N > 5$ in the WISE $W1$ and $W2$ bands. For the WISE $W3$ and $W4$ bands we imposed the cut $S/N > 3$. The minimum requirement for a galaxy to be considered in the SED fitting was that it were above the $S/N$ cuts in the $g$, $r$, $i$, $z$, $W1$ and $W2$ filters, which are the deepest bands in the sample. There were 588 galaxies, corresponding to 0.4\% of the sample, which did not meet this requirement and were discarded from the analysis. 
    \item Reduced $\chi^2$: We selected the galaxies that had SED fits with a reduced $\chi^2$ ($\chi^2_{\nu} = \chi^2/\nu $, where $\nu$ are the degrees of freedom) less than 2.0. 
    This removes 3,405, the 2.6\% of the entire sample.
\end{itemize}

At the end of this selection process, the sample used for the analysis of this work contained 56,399 BCGs. 
These objects all have spectroscopic redshifts and SEDs fitted on at least six of the $u$, $g$, $r$, $i$, $z$, $W1$, $W2$, $W3$ and $W4$ bands (in 808 cases we have fits on all the nine photometric bands), covering a portion of the electromagnetic spectrum that goes from near UV ($\lambda=0.35$\mum) to mid IR ($\lambda=22.8$ \mum).
The redshift distribution of the sample (Fig. \ref{fig:histo gal with w3}, left-hand panel), shows that the number of galaxies grows with redshift, with a median $z_{\rm spec} = 0.29$.  
The distribution of the $\chi^2_{\nu}$ values of the SED fitting (Fig. \ref{fig:histo gal with w3}, middle panel), has a median $\chi^2_{\nu}$ = 0.53  in the range 0.001 < $\chi^2_{\nu}$ < 2.0. We note that, because of our strict selection in $S/N$, the values $\chi_\nu^2 < 1.0$ are not a result of large flux uncertainties.

To characterize the sample, we divide it into four major bins of redshift and $\chi_{\nu}^2$, and three bins of cluster halo mass (\mh).
We found that 56.5\% of the sample has excellent fits ($\chi^2_{\nu}<0.5$), 30.4 \% has very good fits ($0.5<\chi^2_{\nu}<1.0$), 10.0\% has good fits ($1.0<\chi^2_{\nu}<1.5$) and just 3.1\% has satisfactory fits ($1.5<\chi^2_{\nu}<2.0$).

The four bins defined for the redshift distribution are: B1 $\equiv$ $0.05 \leq z < 0.15$; B2 $\equiv$ $0.15 \leq z < 0.25$; B3 $\equiv$ $0.25 \leq z < 0.35$, and B4 $\equiv$ $0.35 \leq z < 0.42$. We find that 9.6\% of the galaxies are in B1; 28.6\% in B2; 35.2\% in B3; and 26.6\% in B4 (see details in Table \ref{tab:bins_def}).

We divided the cluster sample in \mh\ bins following the criterion used in Paper I and splitting it into Groups (Gr), Low-Mass Clusters (LMC), and High-Mass Clusters (HMC). We find that 33.1\% of our BCGs are in Gr, 64.9\% in LMC, and just 2.0\% in HMC (see details in Table \ref{tab:bins_def}). We remind the reader that the division in \mh\ bins takes into account the accretion history of dark matter haloes so that the clusters in the nearest redshift bin fall in the same \mh\ bin of their progenitors. This allows us to control the progenitor bias in the sample (see \citealt{Correa_2015a, Correa_2015b, Correa_2015c}).\footnote{We used the {\ttfamily{commah PYTHON}} package to derive the mass accretion histories of the clusters.}

\begin{figure}
\centering
\includegraphics[width=0.4\textwidth]{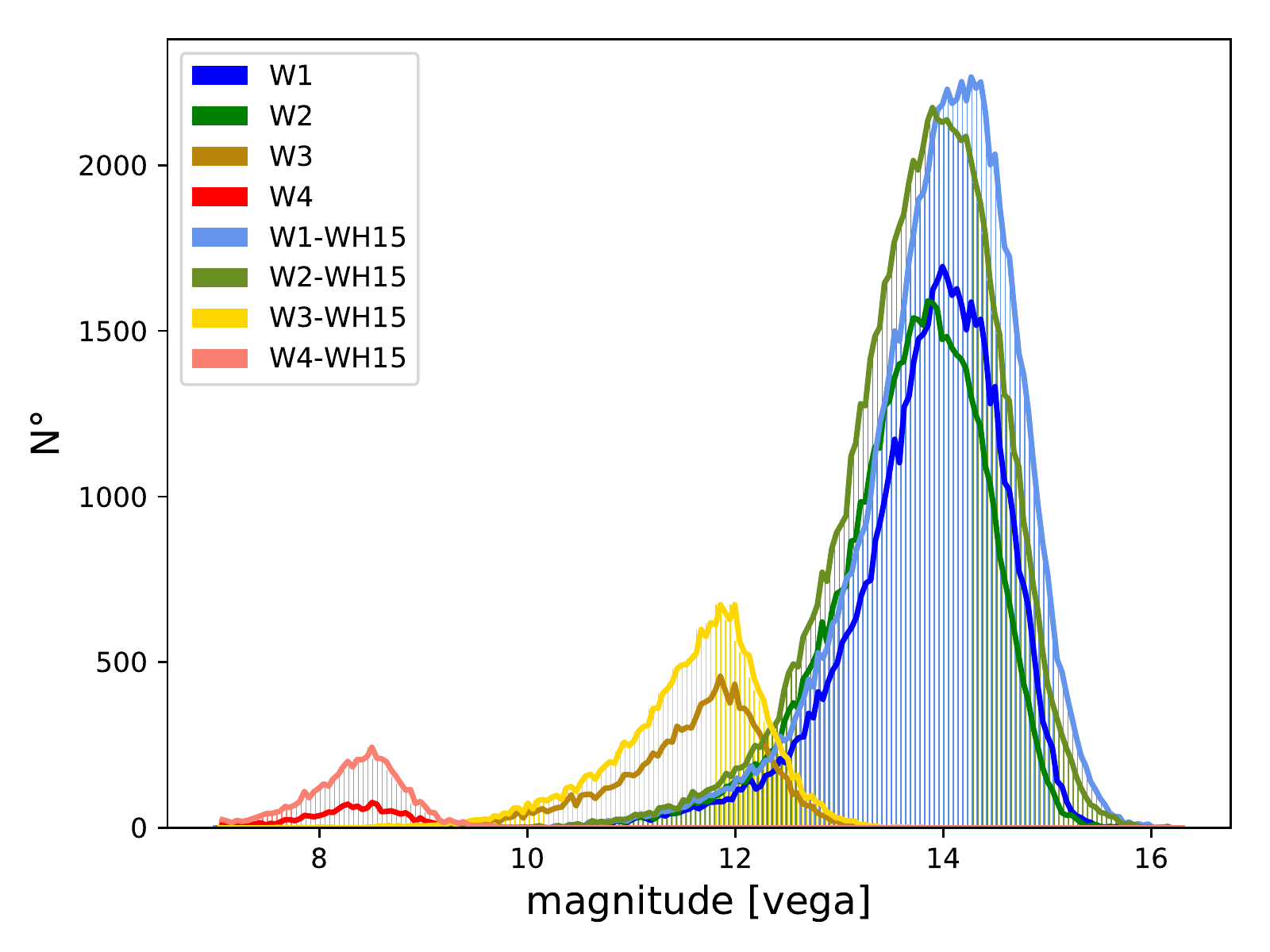}
\caption{Magnitude distributions of the WISE $W1$, $W2$, $W3$ and $W4$ bands for both the final sample and the original WH15 sample restricted at $ z_{\rm phot}<0.42$.}
\label{fig:histo_wise_filt}	
\end{figure}

\begin{table}
    \centering
    \begin{tabular}{ccccc}
\hline    
Filter  &   Min &   Max &   Median  & Amount\\
        & [mag] & [mag] &  [mag]    & of galaxies\\
\hline

W1 & 9.54 & 15.95 & 13.96 & 56,399  (100.0 \% ) \\ 
W2 & 9.52 & 16.13 & 13.73 & 56,399  (100.0 \% ) \\ 
W3 & 5.73 & 13.26 & 11.65 & 12,625  (22.4 \% ) \\ 
W4 & 2.56 & 9.84  & 8.27  & 1,670   (3.0 \% ) \\ 

\hline
    \end{tabular}
    \caption{ Characterization of the magnitude distribution in each WISE filter. 
    Percentages are obtained with respect to the final sample.}
    \label{tab:wisedata}
\end{table}

Figure \ref{fig:histo_wise_filt} shows the magnitude distribution in each WISE filter for the final sample and for the original sample (WHL15 with the photometric redshift restriction $z_{\rm phot}<0.42$). The main difference between these two samples appears to be in the numbers of galaxies that each of them contains. The distributions of the $W1$ and $W2$ magnitudes have median values $W1 = 13.96$ mag and $W2 = 13.73$ mag, and the entire final BCG sample has measurements in both filters (see Table \ref{tab:wisedata}). The $W3$ and $W4$ magnitude distributions have median values $W3 = 11.65$ mag  and $W4 = 8.27$ mag. 22.4\% of the final BCG sample have measurements in $W3$ and 3\% in $W4$.

\subsection{The Error Estimation on Stellar Mass and Star Formation Rate}
\label{secubsec:error estimation}

\begin{table}
    \centering
    \begin{tabular}{ccccccc}
\hline    
bins            &   Median  & Min  &  Max & Median & Min & Max  \\
$\chi^2_{\nu}$  &   error   & error &  error &error   & error & error  \\
                &   $SFR$     & $SFR$   &  $SFR$ & M$_*$   & M$_*$ & M$_*$  \\
\hline
\hline
0.0 - 0.5       & 16.0\%    &  6.9\% &  34.7\% &1.8\%  & 0.4\%  &  5.3\% \\
0.5 - 1.0       & 19.8\%    &  5.7\% &  43.5\% &1.4\%  &  0.6\% &  6.3\% \\
1.0 - 1.5       & 21.2\%    &  5.8\% &  40.1\% &1.6\%  &  0.2\% &  5.2\% \\
1.5 - 2.0       & 18.2\%    &  5.9\% &  34.7\% &1.5\%  &  0.4\% &  5.2\% \\
\hline
    \end{tabular}
    \caption{Percentage of error in the $SFR$ and the M$_*$, obtained from a Monte Carlo Simulation.}
    \label{tab:err_sym}
\end{table}

The errors on the measurements of $SFR$ and \mstar\ were obtained performing Monte Carlo simulations and running {\ttfamily{CIGALE}} on the simulated input catalogues. The $\chi^2_{\nu}$ was used as an indicator of the goodness of the SED fitting and therefore of the uncertainty on the fitted parameters (larger values of $\chi^2_{\nu}$, indicate larger errors).

The procedure that we adopted consists of the following steps: 
\begin{enumerate}
    \item We subdivided the sample using four $\chi^2_{\nu}$ bins between 0.0 and 2.0 as shown in Table \ref{tab:err_sym};
    \item we randomly selected 50 galaxies in each $\chi^2_{\nu}$ bin, creating a sub-sample of 200 galaxies; 
    \item we created 100 versions of each galaxy, randomly perturbing their fluxes within the photometric errors; 
    \item we ran {\ttfamily{CIGALE}} for each simulated galaxy, obtaining a distribution of $SFR$ and \mstar values; 
    \item for each distribution we derived the mean and the standard deviation and calculated the fractional deviation defined as the ratio between the standard deviation and the mean;
    \item in each bin we derived the median value of the fractional deviations, which we used as an estimate of the fractional error. We also derived the boundaries of the $1\sigma$ width of the distributions of the fractional deviation. 
\end{enumerate}

Table \ref{tab:err_sym} reports the values of the fractional errors (median and boundaries of the $1\sigma$ width of the distributions) for $SFR$ and \mstar\ as a function of $\chi^2_{\nu}$ bin. It can be seen that the fractional error in $SFR$ increases with the central values of the $\chi^2_{\nu}$ bins, indicating that we can characterise the uncertainty in this parameter with $\chi^2_{\nu}$. The last bin (1.5 $< \chi^2_{\nu} < 2.0$), exhibits a 3\% decrease in the fractional error with respect to the previous bin, and we attribute this effect to low-number statistics. We notice, however, that this value of the fractional error is consistent with the errors obtained in the other bins.

While the median values of the percentage of error for $SFR$ lie in the range 16-21\%, for \mstar\ the percentage of error shows values in the range between 1.4\% and 1.8\% with no correlation with $\chi^2_{\nu}$ (see Table \ref{tab:err_sym}). Since there is no apparent dependence of the \mstar\ error on the goodness of the SED fit, we can affirm that the uncertainty in this parameter reflects the flux uncertainty in the filters used in the fit. We note here that \mstar\ is better constrained than $SFR$.

We therefore use as a definition of the uncertainty on $SFR$ the fractional error, obtained from the SED fitting, according to the median value showed in Table \ref{tab:err_sym}. For the M$_*$, because the median percentage of error is less than 2\%, for each $\chi^2_{\nu}$ bin, we consider an upper limit of 2\% on the fractional error for each galaxy.

\section{Results}
\label{sec:results}

\subsection{Stellar mass and star formation rate in BCGs}
\label{subsec:star formation and stellar mass dist}

\begin{figure*}
\centering
\includegraphics[width=0.32\textwidth]{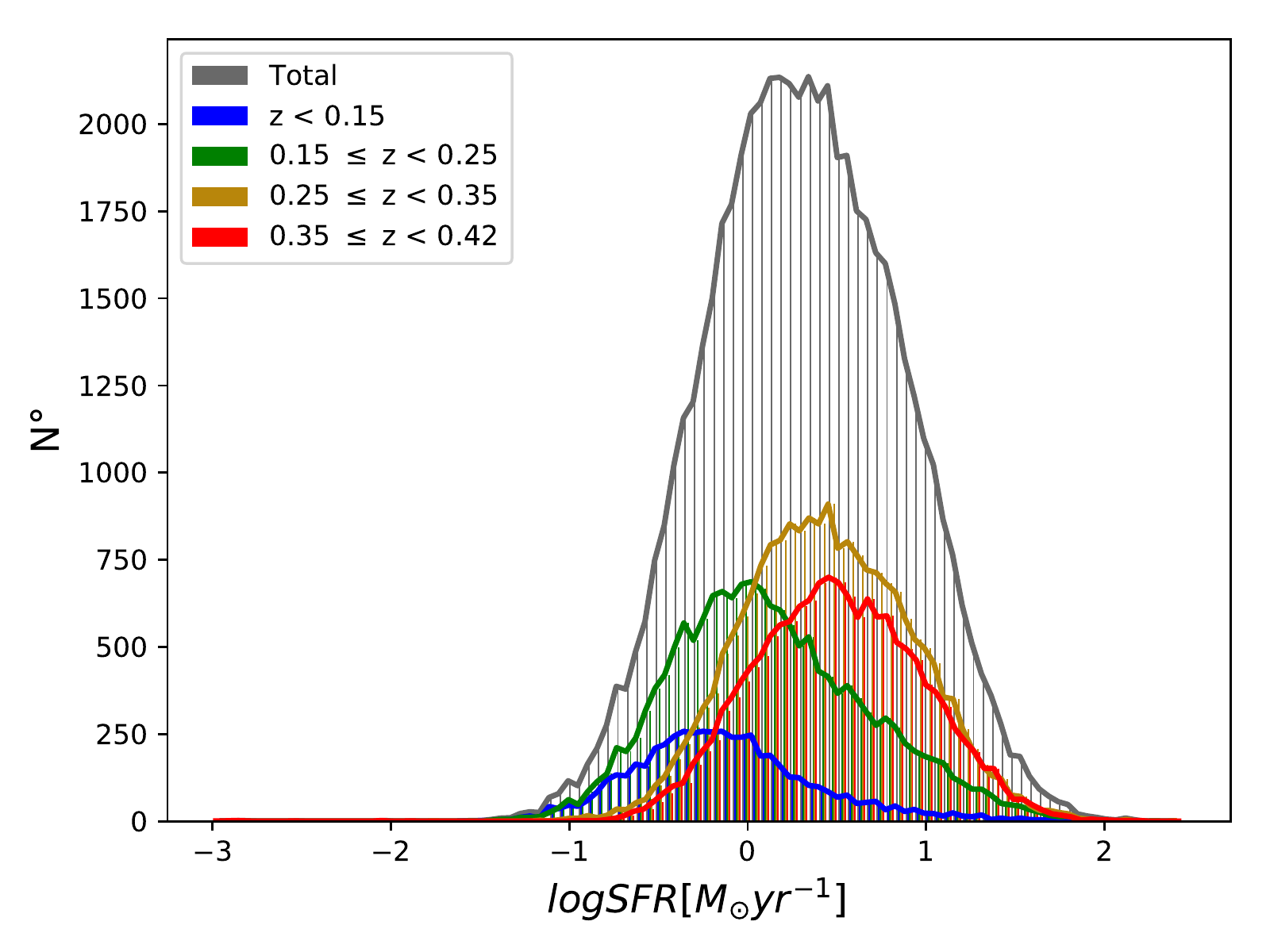}
\includegraphics[width=0.32\textwidth]{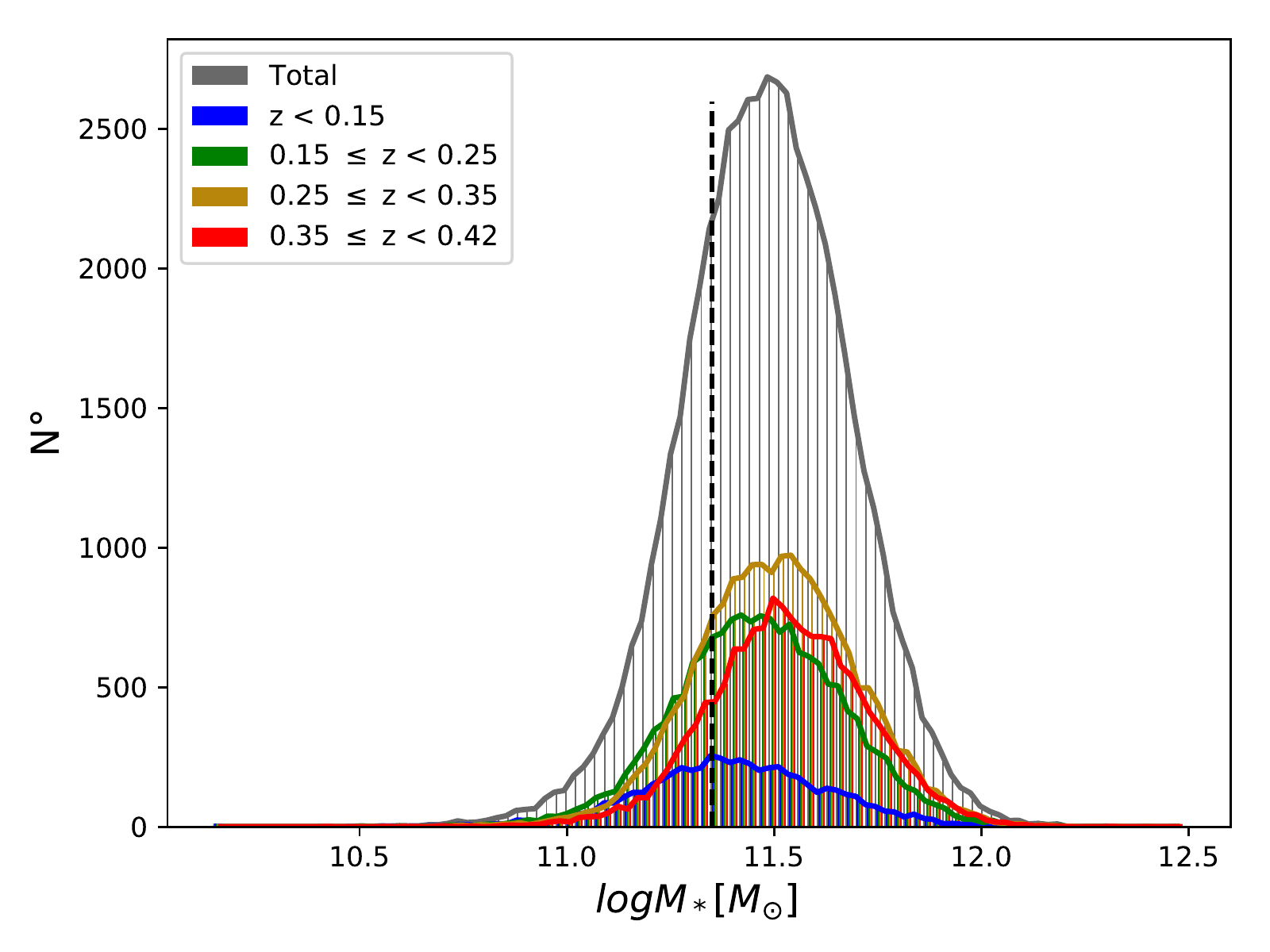}
\includegraphics[width=0.35\textwidth]{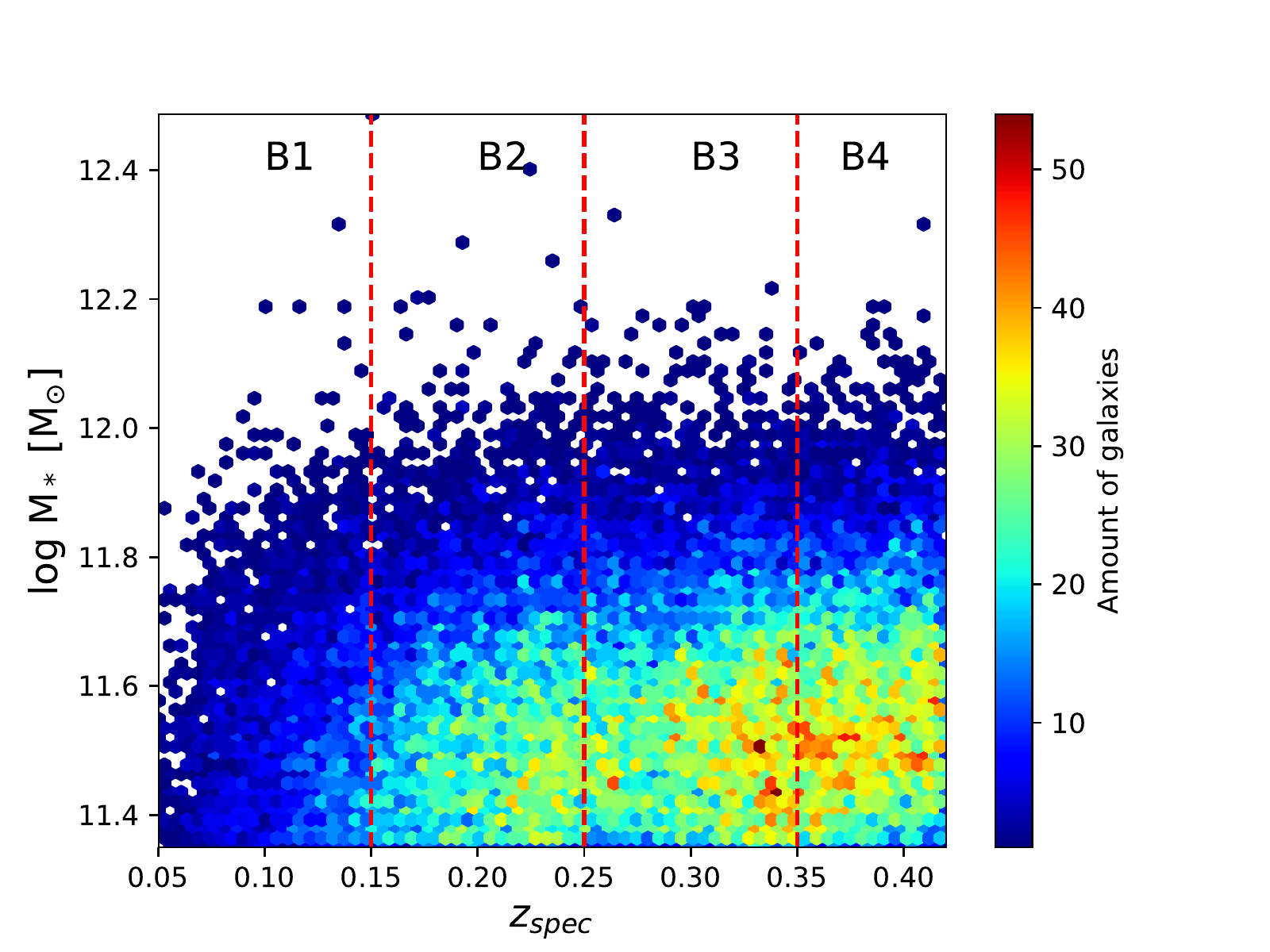}
\caption{ Distributions of $SFR$ (left-hand panel) and \mstar\ (middle panel). The grey histograms correspond to the distribution for the entire sample (not \mstar\-limited), and the coloured histograms correspond the distributions for the sub-samples in each of the B1, B2, B3 and B4 redshift bins.
The vertical black line in the \mstar\ distribution represents the stellar mass completeness limit $\rm M_{lim}$ of the sample.
 The right-hand panel shows a density plot of \mstar\ as a function of cluster redshift, colour coded according to the amount of galaxies in each cell.
}
\label{fig:histo sfr and mstar}	
\end{figure*}

\begin{figure}
\centering
\includegraphics[width=0.4\textwidth]{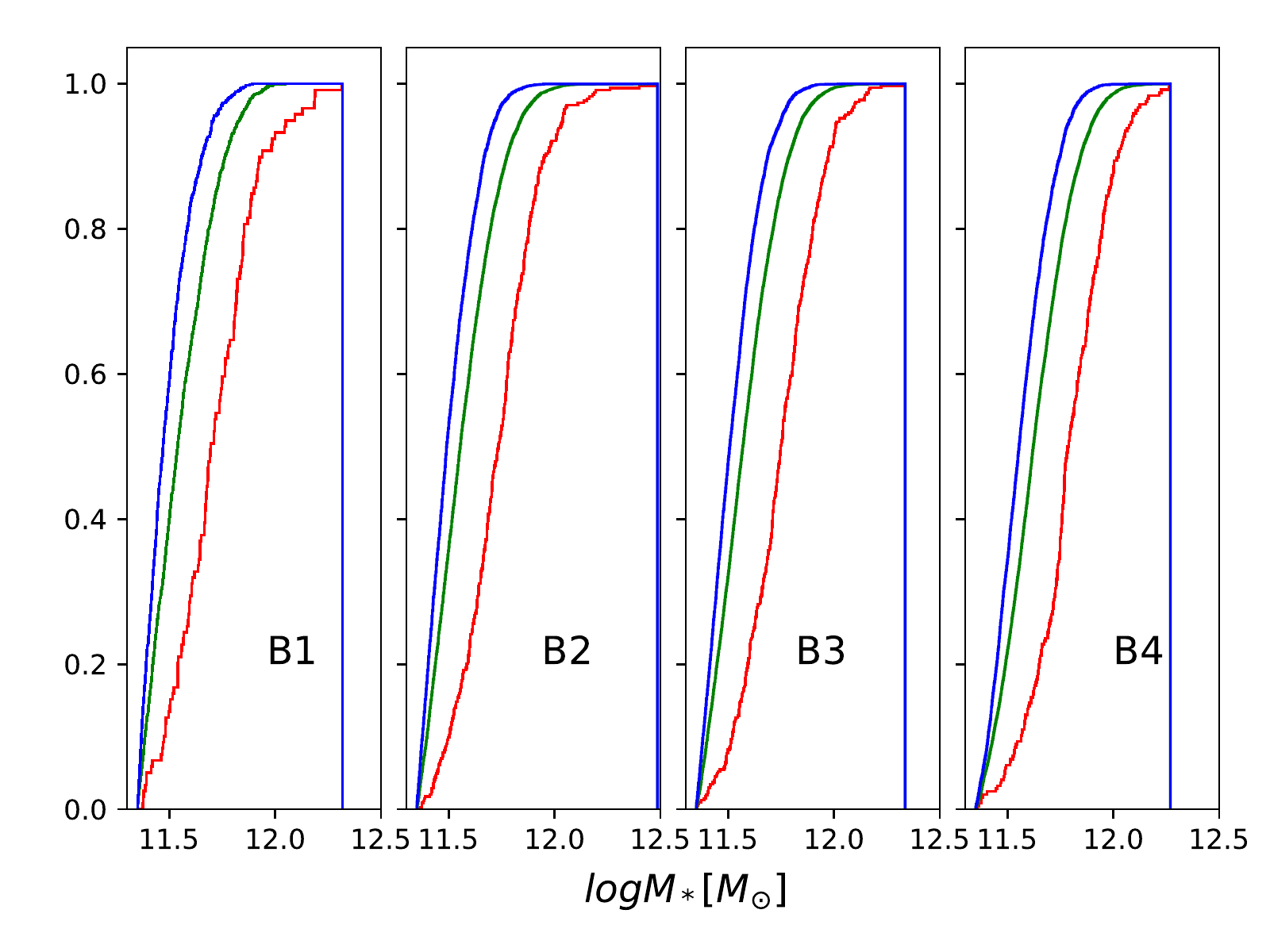}
\includegraphics[width=0.4\textwidth]{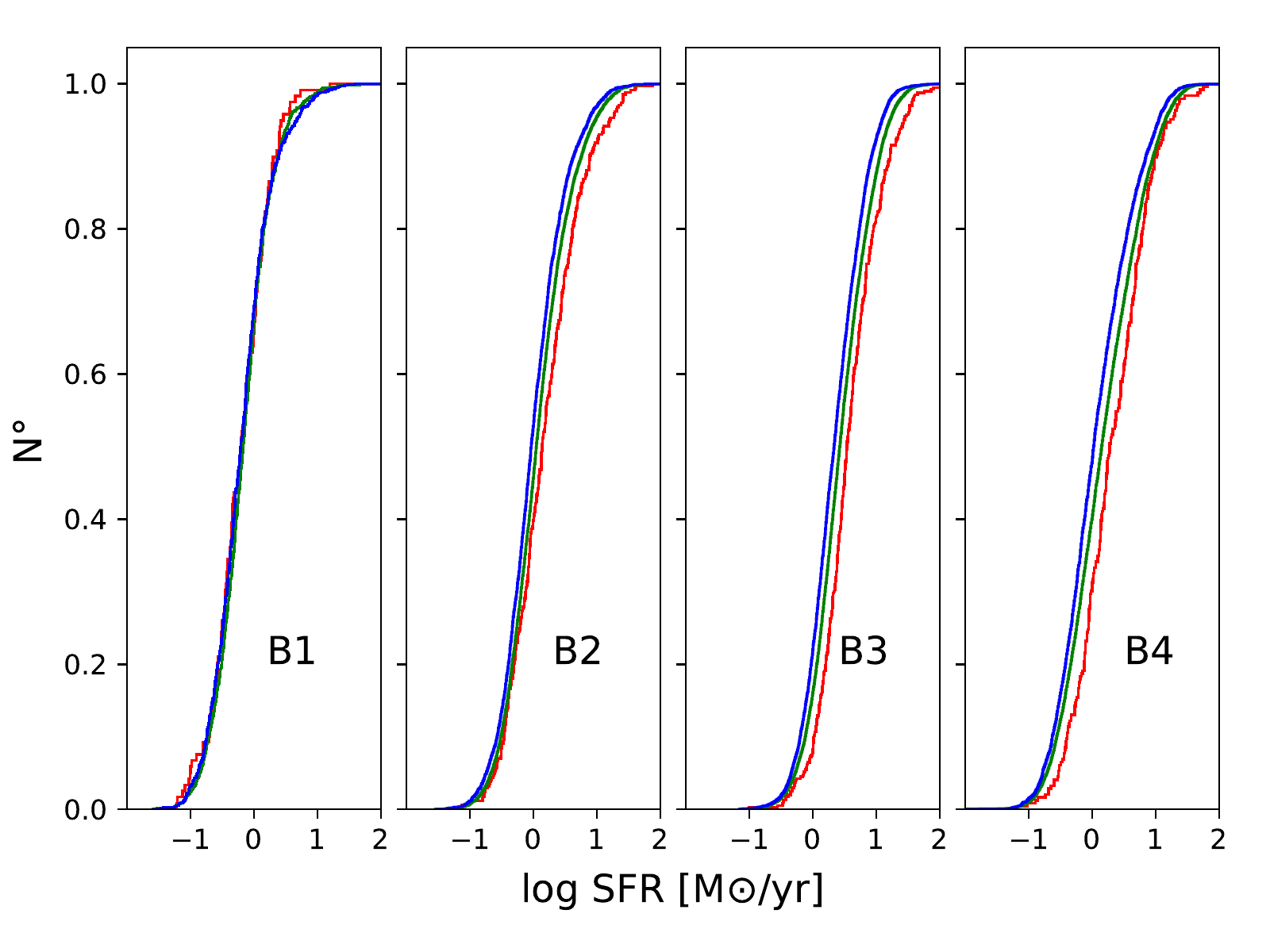}
\caption{ Plots of the cumulative distribution of \mstar\ (top panel) and $SFR$ (bottom panel) in each redshift bin. The red, green and blue histograms are respectively the distributions for the Group, LMC and HMC stellar-mass limited sub-samples.
}
\label{fig:cumulative sfr and mstar}	
\end{figure}

Figures \ref{fig:histo sfr and mstar} and \ref{fig:cumulative sfr and mstar} show the histograms of the distributions and cumulative distributions of \mstar\ and $SFR$. We plot in each figure the distributions for the entire sample and for the sub-samples of clusters with different \mh\ and at different redshifts.
We do not further investigate the stellar mass evolution in our sample of BCGs, since this work is focused on the study of the evolution of the star-formation activity. 
The stellar mass build-up will be the subject of a work that is already ongoing. In order to study the evolution of the star-formation properties of the BCGs, we built a stellar mass complete sample that allowed us to minimise the effects of the Malmquist Bias. 
We determined the \mstar\ completeness limits following two different approaches as detailed in the subsequent paragraphs. 

The first approach follows the method described in \cite{Pozzetti_2010} and used in \cite{Cerulo_2017}. It consists in supposing that at all redshifts the BCGs are observed at the limiting magnitude of the sample in the SDSS $r$ band. With the assumption that the stellar mass-to-light ratio of galaxies is constant, one can derive a limiting stellar mass, \maslim, from the equation 
\begin{equation}
    \rm log (M_{lim}/M_{\odot}) = log(M_*/M_{\odot}) + 0.4 \times (r - r_{lim}),
    \label{eq:masslimit}
\end{equation}
where $\rm r_{lim}$ is the Petrosian  magnitude ($\rm r_{lim} = 20.2$ mag) of the faintest BCG of the sample in the $r$-band, $r$ is the $r$-band magnitude and \mstar, the stellar mass of the galaxies.

We take the \maslim\ distributions of the 20\% faintest galaxies in each redshift bin and for them we derive the 95\% highest \maslim. The latter defines the 95\% stellar mass limit in a given redshift bin of the BCG sample. 
Because the stellar mass limits monotonically increase with redshift, the stellar mass limit for the highest redshift bin defines the stellar mass limit for the entire sample. 
Using this criterion, we find \maslim\ $ = 10^{11.3}$\msun.

The second method was adopted in \citet{Cerulo_2017} and Paper I and is based on the stellar mass of a model SSP with formation redshift $z=5$, \citet{Salpeter_1955} IMF and solar metallicity drawn from the \citet{Maraston_2005} stellar population library. At $z=0.42$ the stellar mass for such a model observed at the $r$-band magnitude limit of the sample is \maslim $= 10^{11.4}$\msun.\footnote{We used the PYTHON {\ttfamily{ezgal}} package \citep{Mancone_2012} to derive the stellar mass evolution of the SSP.}

Since the stellar mass limits obtained with the two methods are similar, we chose to set the stellar mass limit for the BCG sample at \maslim\ $= 10^{11.35}$ M$_\odot$, which represents a compromise between the two estimates. In Figure \ref{fig:histo sfr and mstar} we plot the stellar mass limit on the \mstar\ distributions in each redshift bin. It can be seen that by limiting our sample to the completeness limit we lose galaxies at all redshifts, with the fraction of lost galaxies increasing towards low redshift. As it can be seen from Table\ref{tab:bins_def}, the median \mstar\ as a function of redshift has a slight increase, even after limiting the sample to \maslim. Although the four median values are consistent with each other within the $1\sigma$ widths of the \mstar\ distributions, this represents an evidence that there is a residual, non significant effect of the Malmquist Bias in the stellar mass limited sample. The results presented from this point refer to BCGs with \mstar$>$\maslim.

We see that the median $SFR$ of the entire sample is 2.0 \msun/yr, indicating that it is not correct to assume that the BCGs at low redshift are  passively evolving galaxies with no ongoing star formation. From the analysis of the $SFR$ distribution (Figure \ref{fig:histo sfr and mstar} bottom panel) we find that 63.52\% of the BCGs have $SFR > 1$ \msun/yr, 18.66\% have $SFR > 5$ \msun/yr, 7.78\% have $SFR > 10$ \msun/yr, and 0.17\% have $SFR > 50$ \msun/yr. Therefore, at least 63.52\% of the sample is composed of non-quiescent galaxies. The redshift bins from B2 to B4 have a median $SFR$ greater than 1 \msun/yr (see Table \ref{tab:bins_def}) and there is a slight increase in the median $SFR$ with redshift, although the median values remain all consistent within the $1 \sigma$ widths of the $SFR$ distributions. 

We performed a two-sample Kolmogorov-Smirnov (KS) test on the $SFR$ distributions and found that there is a probability $p < 0.003$ that the $SFR$ distributions in B2, B3, and B4 are drawn from the same parent distribution as in B1. So the $SFR$ distribution in the lowest-redshift bin is statistically different from any of the distributions at higher redshifts. 
We further find that the two-sample KS test for the comparison of the distributions in B2 and B3 returns a p-value of 0.34, and for the comparison of B2 and B4 a p-value of 0.44, indicating that the $SFR$ distributions in B3 and B4 are drawn from the same parent distribution of B2. 

In order to investigate the effects of cluster mass build-up, we separately studied the distributions of \mstar\ and $SFR$ as a function of redshift in each of the three sub-samples of groups, low-halo-mass clusters and high-halo-mass clusters. 
The top panel of Figure \ref{fig:cumulative sfr and mstar} and Table \ref{tab:bins_def} show that \mstar\ has three different distributions in the three cluster sub-samples at all redsfhits.
We find, in particular, that the Gr sub-sample contains the BCGs with the lowest \mstar, followed by the LMC and then the HMC, which contain the most massive BCGs. 

We performed a two-sample KS test to compare the \mstar\ distributions in all the cluster sub-samples in pair, and we found that the distributions are all different between them  (obtaining p values $<< 10^{-20}$). We repeated the same exercise for the $SFR$ distributions, finding that all the sub-samples have significantly different distributions with the exceptions of Gr and HMC in B1 ($p=0.97$), and LMC and HMC in B1 ($p=0.28$).

These results indicate that the BCGs in clusters with different halo masses differ in terms of their star-formation activity and \mstar\ distributions, and that these differences persist with redshift with few exceptions that correspond to the $SFR$ distributions in the lowest redshift bin.
These results suggest that at $0.05 \leq z < 0.42$ the mass of the hosting cluster affects the star-formation activity of BCGs at earlier epochs, with BCGs in lower-mass clusters having higher $SFR$ than BCGs in higher-mass clusters. At the lowest-redshift end of the sample, the star-formation properties of BCGs become similar in all cluster sub-samples.

\subsection{The star-forming Main Sequence of Brightest Cluster Galaxies}
\label{subsec: main_sequence}

\begin{figure}
\centering
\includegraphics[width=0.45\textwidth]{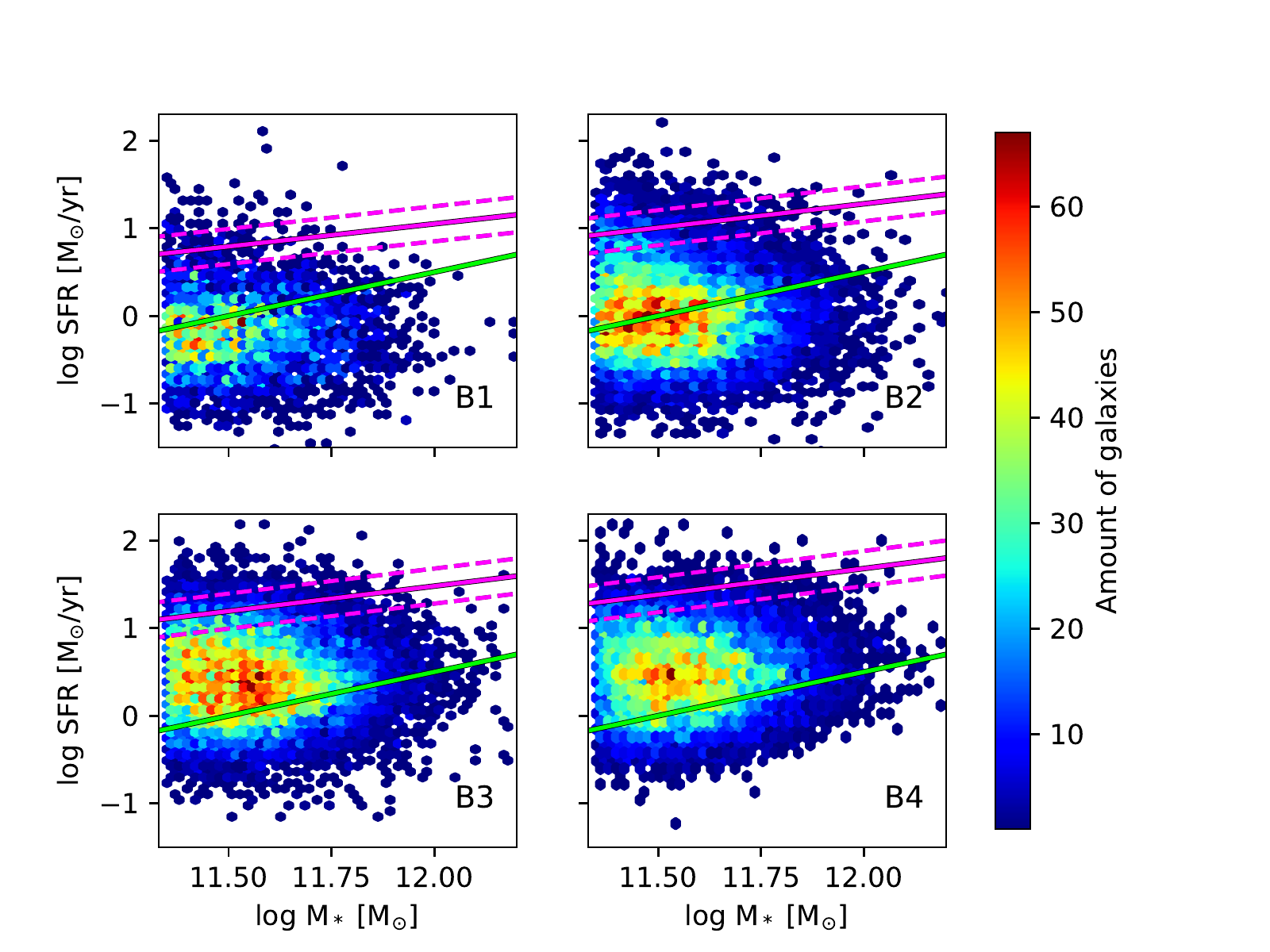}
\caption{ Density plot representing $SFR$ as a function of \mstar\  for the four redshift bins (B1, B2, B3 and B4).
The plot is colour-coded according to the amount of galaxies in each cell.
The solid magenta line corresponds to the main-sequence straight line derived from Equation \ref{subsec: main_sequence}, using the median redshift of each bin ($z = 0.1, 0.2, 0.3 ,0.415$). The dashed {magenta} lines correspond to the boundaries of the scatter of the main sequence ($\pm 0.2$), while the  green line divides the $SFR$ vs \mstar\ plane in two, to separate star-forming ($\log{(sSFR/\mbox{yr}^{-1})} > -11.5$), from quiescent ($\log{(sSFR/\mbox{yr}^{-1})} \leq -11.5$) BCGs.
}
\label{fig:sSFR}	
\end{figure}

\begin{table}
    \centering
    \renewcommand{\arraystretch}{1.5}
    \begin{tabular}{ccccc}
\hline    
Redshift &   $f_{Pa}$  & $f_{MS}$  & $f_{SB}$  &  $N_g$ \\
bin      &     (\%)     &   (\%)   &    (\%)   &         \\
\hline
\hline
B1 & $95.8_{-0.4}^{+0.3}$ & $3.1_{-0.3}^{+0.3}$ & $1.13_{-0.17}^{+0.19}$ & $3,319$ \\
B2 & $92.9_{-0.2}^{+0.2}$ & $5.6_{-0.2}^{+0.2}$ & $1.59_{-0.11}^{+0.12}$ & $11,646$ \\
B3 & $89.9_{-0.2}^{+0.2}$ & $8.6_{-0.2}^{+0.2}$ & $1.48_{-0.09}^{+0.10}$ & $15,738$ \\
B4 & $93.0_{-0.2}^{+0.2}$ & $6.3_{-0.2}^{+0.2}$ & $0.67_{-0.07}^{+0.08}$ & $12,534$ \\
\hline
\hline
BT  & $92.06_{-0.13}^{+0.13}$ & $6.70_{-0.12}^{+0.12}$ & $1.24_{-0.05}^{+0.05}$ & $43,257$ \\
Total  &   $39,826$                &  $2,895$                 &  $536$ \\
\hline
    \end{tabular}
    \caption{Fractions of galaxies (expressed as percentages) classified using the $SFR$ vs \mstar\ plane and the star-formation main sequence in each redsfhit bin. $Pa$ are passive galaxies, $MS$ are main-sequence galaxies and $SB$ are star-burst galaxies (see Section \ref{subsec: main_sequence} for details on the definition of each class). $Ng$ is the number of galaxies in each redshift bin. 
    The BT row shows the fractions derived across the entire redsfhit range $0.05 \leq z < 0.42$, while the Total row reports the number of BCGs in each class at $0.05 \leq z < 0.42$.
    \newline
    }
    \label{tab:bins MS}
\end{table}

Star-forming galaxies arrange themselves in the $SFR$ - \mstar\ plane along a sequence, called the main sequence ($MS$), in which the $SFR$ increases approximately linearly with \mstar\ (e.g. \citealt{Brinchmann_2004,Noeske_2007,Elbaz_2007}).
At a given \mstar, galaxies with $SFR$ ten times higher or lower than the linear fit to the MS are called star-burst ($SB$) or passive ($Pa$), respectively.

The MS is observed in several large samples of galaxies in the field and in clusters, at redshifts up to $z \sim$ 6 (e.g. \citealt{Elbaz_2007,Elbaz_2011,Daddi_2007,Dunne_2009,Santini_2009,Oliver_2010,Karim_2011,Rodighiero_2011,zahid_2012,Bouwens_2012,Sobral_2014,Steinhardt_2014, Old_2020, Nantais_2020}). 
However, the effect of the environment on the MS shape is a matter of debate (see e.g.: \citealt{Old_2020, Nantais_2020}).

To characterise the MS and its evolution in our sample, we use the parametrisation described in \citet{Speagle_2014} and expressed by the following equation: 
\begin{equation}
\begin{aligned}
 \rm \log SFR(M_*,t) = &\rm [0.84\pm0.02 - (0.026\pm0.03 \times t)]\log M_* \\
                    & \rm - [6.51\pm 0.24 - (0.11\pm0.03 \times t)]
\label{eq:MV}
\end{aligned} 
\end{equation}
where $t$ is the age of the Universe in Gyr.  
For the $MS$ scatter we use the range $\pm 0.2$ dex suggested in the same work. BCGs in our sample form a cloud in the $SFR$ - \mstar\ plane (see Figure \ref{fig:sSFR}), with no clear $MS$. 
However, using the redshift as a third parameter one can see that lower redshift galaxies tend to concentrate in regions with low $SFR$ ($-1.0 \lesssim \log{ (SFR/\mbox{\msun/yr})} \lesssim 0.4$) and low \mstar\ ($11.35 \lesssim \log{(\mbox{\mstar/\msun})} \lesssim 11.5$). 
On the other hand, galaxies at higher redshifts are distributed in regions with higher stellar masses and a larger range of $SFR$.
To see where the MS lies at the redshifts of our sample, we use Equation \ref{eq:MV} with the median redshift of the sample ($z_{\rm med} = 0.286$) used to derive the epoch $t$ in the equation.

By classifying the BCGs according to MS types we find that  the $92.06^{+0.02}_{-0.02}\%$ are $Pa$, $6.70^{+0.12}_{-0.12}\%$ are $MS$ and $1.24^{+0.05}_{-0.05}\%$ are $SB$ (see Table \ref{tab:bins MS}). We notice that the fraction $f_{Pa}$ of $Pa$ BCGs decreases from bin B1 to bin B3, with a corresponding increase in the fractions $f_{MS}$ and $f_{SB}$ of MS and SB galaxies. Bin B4 has a peculiar behaviour, since it exhibits an increase in $f_{Pa}$ and a decrease in $f_{SB}$ and $f_{MS}$ (see Table \ref{tab:bins MS}). 
We notice that this cannot be an effect of stellar mass incompleteness, since we are only considering BCGs that are above the stellar mass limit of the sample.

We argue that the increase in $f_{Pa}$ may arise from the fact that the MS modelling that we adopt was obtained for field galaxies with stellar masses lower ($< 10^{11}$\msun) than those of our BCGs. \citet{Lee_2015} propose the existence of a turn over of the main sequence at \mstar$\sim10^{10}$\msun. Hence we miss MS BCGs in bin B4 because the evolutionary MS model that we use does not take into account the turn-over. 

\begin{table*}
    \centering
    \renewcommand{\arraystretch}{1.5}  
    \begin{tabular}{l@{\hspace{0.15cm}}ccc@{\hspace{0.15cm}}cccc@{\hspace{0.15cm}}ccc@{\hspace{0.15cm}}ccc@{\hspace{0.15cm}}cc}
\hline 
        &      & B1   &       &  &     & B2    &        & &      & B3   &       &  &    & B4     &       \\ 
        & Gr   & LMC  &   HMC &  & Gr  & LMC   &   HMC  & & Gr   & LMC  &   HMC &  & Gr  & LMC   &   HMC  \\
\hline
\hline
$f_{SB}$ &  $1.5_{-0.4}^{+0.4}$  &  $1.0_{-0.2}^{+0.2}$  &  $1.4_{-0.8}^{+1.4}$  & &  $1.05_{-0.16}^{+0.18}$  &  $1.81_{-0.15}^{+0.16}$  &  $3.1_{-0.9}^{+1.0}$  & &  $0.83_{-0.13}^{+0.15}$  &  $1.65_{-0.12}^{+0.13}$  &  $4.5_{-1.0}^{+1.1}$  & &  $0.7_{-0.2}^{+0.2}$  &  $0.86_{-0.14}^{+0.15}$  &  $4.1_{-1.6}^{+2.0}$  \\ 
$f_{MS}$ & $4.4_{-0.6}^{+0.7}$  &  $2.7_{-0.3}^{+0.4}$  &  $2.3_{-1.1}^{+1.6}$  & &  $5.0_{-0.3}^{+0.4}$   &  $5.7_{-0.3}^{+0.3}$  &  $7.9_{-1.2}^{+1.7}$  & &  $7.2_{-0.4}^{+0.4}$   &  $9.2_{-0.3}^{+0.3}$  &  $10.5_{-1.3}^{+1.7}$  & &  $5.8_{-0.6}^{+0.7}$  &  $7.8_{-0.4}^{+0.4}$  &  $8.0_{-1.9}^{+3.0}$  \\ 
$f_{Pa}$   & $94.2_{-0.8}^{+0.7}$  &  $96.4_{-0.4}^{+0.4}$  &  $96.9_{-1.9}^{+1.3}$  & &  $94.0_{-0.4}^{+0.4}$  &  $92.5_{-0.3}^{+0.3}$  &  $89.1_{-1.9}^{+1.5}$  & &  $92.0_{-0.4}^{+0.4}$  &  $89.2_{-0.3}^{+0.3}$  &  $85.1_{-1.9}^{+1.6}$  & &  $93.6_{-0.7}^{+0.6}$  &  $91.4_{-0.5}^{+0.4}$  &  $88_{-4}^{+2}$  \\ 
\hline
$f_{SF}$ & $31.8_{-1.5}^{+1.5}$ &  $28.8_{-0.9}^{+1.0}$  &  $18_{-3}^{+4}$  & &  $45.2_{-0.8}^{+0.8}$  &  $48.1_{-0.6}^{+0.6}$  &  $42_{-3}^{+3}$  & &  $75.6_{-0.7}^{+0.7}$  &  $76.6_{-0.4}^{+0.4}$  &  $72_{-2}^{+2}$  & &  $82.2_{-1.1}^{+1.0}$  &  $82.8_{-0.6}^{+0.6}$  &  $87_{-4}^{+3}$  \\ 
$f_Q$    & $68.2_{-1.5}^{+1.5}$ &  $71.2_{-1.0}^{+0.9}$  &  $82_{-4}^{+3}$  & &  $54.8_{-0.8}^{+0.8}$  &  $51.9_{-0.6}^{+0.6}$  &  $58_{-3}^{+3}$  & &  $24.4_{-0.7}^{+0.7}$  &  $23.4_{-0.4}^{+0.4}$  &  $28_{-2}^{+2}$  & &  $17.8_{-1.0}^{+1.1}$  &  $17.2_{-0.6}^{+0.6}$  &  $13_{-3}^{+4}$  \\ 
\hline
$N_g$  &  $1,000$  &  $2,202$  &  $117$  & &  $3,862$ &  $7,440$  &  $344$  & &  $4,519$  &  $10,816$  &  $403$  & &  $2,998$  &  $9,294$  &  $246$  \\ 
\hline
    \end{tabular}
    \caption{ Fraction of starburst ($f_{SB}$), main sequence ($f_{MS}$), passive ($f_{Pa}$), star-forming (\fsf), and quiescent (\fq) BCGs as a function of spectroscopic redshift in each of the Gr, LMC and HMC sub-samples. The fractions are derived in the stellar mass limited sample and are expressed as percentages. $N_g$ is the number of BCGs in each sub-sample and in the redshift bins indicated.} 
    \label{tab:bins MS and M200}
\end{table*}

If we split the sample into Gr, LMC and HMC clusters (see Table \ref{tab:bins MS and M200}), we find that all the sub-samples exhibit a decrease in $f_{Pa}$ with redshift. We particularly find that the $f_{Pa}$ decrease is highest in the HMC sub-sample ($\sim12\%$). These results reflect the shift towards higher $SFR$ in the cumulative distributions of all the sub-samples plotted in Figure \ref{fig:cumulative sfr and mstar}. In the same figure one can appreciate that there is no similar shift in the cumulative distributions of \mstar.

The SB fraction keeps approximately constant across the redshift range of the sample, while the MS fraction increases up to B3 and then decreases (see Table \ref{tab:bins MS and M200}). Similar trends for $f_{MS}$ are seen in the Gr, LMC and HMC sub-samples. $f_{SB}$ in the Gr and LMC sub-samples shows trends with redshift that are similar to the trend observed in the entire sample, while it has a slight increase with redshift in the HMC sub-sample.

The \cite{Speagle_2014} modelling of the MS evolution does not take into account the existence of the turn-over for high-mass galaxies. Thus, subdividing the BCGs in our sample according to their position with respect to the MS would result in the loss of star-forming galaxies, especially at high stellar masses. For this reason we decided to select star-forming BCGs according to their specific SFR, $sSFR= SFR/$\mstar. 

Similarly to \citet{Wetzel_2012}, we define as star-forming galaxies those with $sSFR > 10^{-11.5}$ yr$^{-1}$, and as quenched or quiescent those with $sSFR \leq 10^{-11.5}$ yr$^{-1}$.
The blue line in Figure \ref{fig:sSFR} shows the separation between star-forming and quiescent BCGs based on this criterion. 
The visual inspection of the plot reveals that this classification is less restrictive with respect to the MS one in separating star-forming from quiescent galaxies.

The results on the trends with redshift of the $SFR$ and $sSFR$ obtained using these cuts are shown in Table \ref{tab:evol SFR}, in Figures \ref{fig:evolution of SFR} and \ref{fig:evolution of sSFR}, while the results on the fractions \fq\ and \fsf\ of quiescent and star-forming BCGs, respectively, are shown in Tables \ref{tab:bins MS and M200} and \ref{table_pcerulo_1}-\ref{table_pcerulo_3}. Overall, we find an increase with redshift of the fraction of star-forming BCGs together with an increase in their $SFR$ and $sSFR$. The next section discusses these results in detail, together with the study of the behaviour of \fq\ and \fsf\ with cluster mass and BCG stellar mass. We will also discuss the effects of the ICM cooling time \tcool\ on the star-formation activity in BCGs.

\section{Discussion}
\label{sec:discussion}

 BCGs are amongst the most massive galaxies in the Universe, and the study of their properties constitutes a crucial step towards a comprehensive understanding of the evolution of galaxies. In this paper we focus on the study of the star-formation activity in BCGs, starting from the sample used in Paper I and extending it to $z=0.42$. We selected spectroscopically confirmed BCGs in the WHL15 cluster catalogue and replaced the AllWISE photometry used in Paper I with the deeper unWISE data. All these improvements allow us to robustly fit SEDs to the galaxies in our sample and obtain reliable estimates of \mstar\ and $SFR$.

We notice that with a pure selection in spectroscopic redshift we are able to build a larger sample with respect to Paper I ($\sim 56,000$ vs $19,000$ galaxies), and we ascribe this to the conservative selection in photometric redshift that we applied in Paper I. On the other hand, selecting only spectroscopically confirmed BCGs results in a higher stellar mass completeness limit ( $10^{11.35}$ \msun) with respect to Paper I ($10^{11}$ \msun) 

The most significant change with respect to Paper I is the increase in the number of galaxies for which we can investigate star formation. While in Paper I we were restricted to a sub-sample of 1,857 BCGs with reliable WISE photometry, here we can extend the analysis of Paper I to a sample that is nearly 30 times larger. Furthermore, while in Paper I we had to use WISE IR colours as proxies for the presence of star formation, here we derive the $SFR$ of the BCGs from SED fitting, which allows us to distinguish between star-forming and quiescent galaxies on the basis of their $sSFR$. In Appendix \ref{subsec: wise colours} we present a calibration for the WISE colour-colour diagram that updates the boundaries presented in \citet{Jarrett_2011, Jarrett_2017} to distinguish between star-forming and quiescent BCGs. This can be used for studies in which $SFR$ and \mstar\ cannot be derived.

In addition to a more accurate distinction between star-forming and quiescent BCGs, we can also study the evolution of $SFR$ and $sSFR$, enabling us to directly compare our results with others in the literature (e.g.: \citealt{Webb_2015}, \citealt{Mcdonald_2016}, \citealt{Fogarty_2017}). Unlike Paper I, here we do not estimate the fraction of AGN hosts. We noticed that WISE colours are able to detect the galaxies that are dominated by AGN emission, while objects in which AGN coexist with star formation are below the $(W1-W2) = 0.8$ threshold for a galaxy that is considered an AGN host.
Active Galactic Nuclei will be the subject of a forthcoming analysis.

In this section we discuss the results presented in Section \ref{sec:results} focusing on three different aspects of star formation in BCGs, namely the evolution of the $SFR$ in nearby BCGs (Section \ref{subsec:evolution of the sfr}), the effect of the ICM cooling on BCG star-formation (Section \ref{subsec: cooling flows as driver of SF}), and the evolution of the star-formation activity as parametrized by the star-forming and quiescent fractions (Section \ref{subsec:fraction_evolution})

\subsection{The Evolution of the Star Formation Rate in nearby BCGs}
\label{subsec:evolution of the sfr}

\begin{table*}
    \centering
        \renewcommand{\arraystretch}{1.5}  
    \begin{tabular}{l c@{\hspace{0.15cm}}c@{\hspace{0.15cm}}c@{\hspace{0.15cm}}c c@{\hspace{0.15cm}}c@{\hspace{0.15cm}}c@{\hspace{0.15cm}}c c@{\hspace{0.15cm}}c@{\hspace{0.15cm}}c@{\hspace{0.15cm}}c}
\hline 
     &       & $SFR$     &       &                   &         &  log(\mstar)   &       &                   &       &  log($sSFR$)   &       &         \\                 
bin  & Gr    & LMC       &   HMC &   SF    &   Gr    & LMC       &   HMC &   SF    & Gr    & LMC       &   HMC &   SF \\ 
\hline
\hline
\vspace{0.1cm}
B1 &0.6$^{+1.0}_{-0.4}$ & 0.7$^{+1.0}_{-0.4}$ &0.6$^{+1.0}_{-0.4}$ & 1.8$^{+2.2}_{-0.7}$ &11.47$^{+0.19 }_{-0.09 }$ & 11.54$^{+0.29 }_{-0.13 }$ & 11.70$^{+0.35 }_{-0.18 }$ & 11.47 $^{+0.23 }_{-0.09 }$&-11.68 $^{+0.21 }_{-0.46 }$ & -11.72$^{+0.24 }_{-0.44 }$ & -11.90$^{+0.10 }_{-0.51 }$ & -11.28$^{+0.33 }_{-0.16 }$ \\ 
\vspace{0.1cm}
B2 &0.9$^{+2.0}_{-0.6}$ & 1.1$^{+2.7}_{-0.7}$ &1.4$^{+3.6}_{-1.0}$ & 2.4$^{+4.3}_{-1.1}$&11.49 $^{+0.22 }_{-0.10 }$ & 11.56$^{+0.29 }_{-0.13 }$ & 11.74$^{+0.28 }_{-0.18 }$ & 11.49$^{+0.24 }_{-0.10 }$&-11.55 $^{+0.15 }_{-0.43 }$ & -11.53 $^{+0.22 }_{-0.47 }$ & -11.61$^{+0.40 }_{-0.54 }$ & -11.16$^{+0.01 }_{-0.25 }$ \\  
\vspace{0.1cm}
B3 &2.2$^{+4.2}_{-1.3}$ & 2.7$^{+5.7}_{-1.7}$ &3.4$^{+8.2}_{-2.0}$ & 3.5$^{+6.0}_{-1.9}$ &11.51$^{+0.21 }_{-0.11 }$  & 11.57$^{+0.28}_{-0.14}$ & 11.75$^{+0.31}_{-0.18}$ & 11.54$^{+0.25}_{-0.12}$&-11.19$^{+0.08}_{-0.43}$ & -11.17$^{+0.17}_{-0.44}$ & -11.21$^{+ 0.27}_{-0.44}$ & -11.02$^{+0.05}_{-0.32}$ \\
\vspace{0.1cm}
B4 &2.6$^{+4.8}_{-1.6}$ & 3.2$^{+6.0}_{-2.0}$ &5$^{+10}_{-3}$ & 3.9$^{+6.3}_{-2.1}$ &11.51$^{+0.22}_{-0.10}$  & 11.58$^{+0.27}_{-0.14}$ & 11.74$^{+0.33}_{-0.17}$ & 11.55$^{+0.27}_{-0.13}$&-11.14$^{+0.02}_{-0.42}$ & -11.11$^{+0.08}_{-0.44}$ & -11.08$^{+0.24}_{-0.44}$ & -11.00$^{+0.09}_{-0.33}$ \\ 
\hline
\hline
    \end{tabular}
    \caption{Median and $1 \sigma$ width of the $SFR$ (in \msun/yr),the $\log($\mstar$/$\msun $)$ and the $\log{(sSFR/\mbox{yr}^{-1})}$ distribution as a function of redshift.
    Columns represent, from left to right, the redshift bin, the halo mass class (Gr, LMC and HMC) and the sub-sample of star-forming (SF) BCGs selected according to the criterion $ sSFR > 10^{-11.5}$ yr$^{-1}$. }
    \label{tab:evol SFR}
\end{table*}

\begin{figure}
\centering
\includegraphics[width=0.4\textwidth]{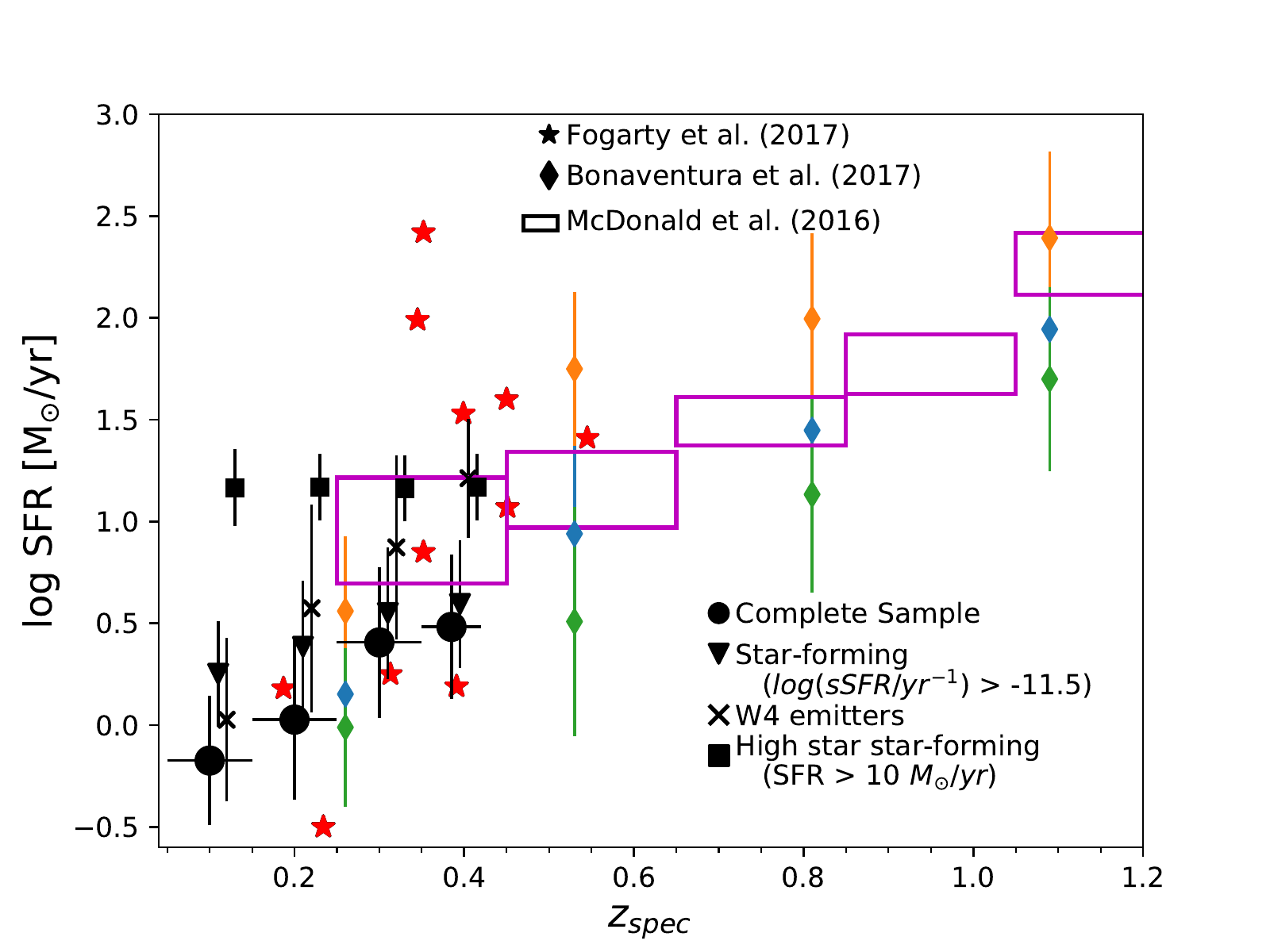}
\caption{ Evolution with redshift of the $SFR$ in nearby BCGs. 
Black circles show our sample divided in the four redshift bins (B1, B2, B3 and B4). 
The inverse triangles correspond to the sub-sample of BCGs catalogued as star-forming ($\log{(sSFR/\mbox{yr}^{-1})} > -11.5$).
The x's represent BCGs detected in $W4$ ($W4$ emitters). 
The squares represent BCGs with SFR $\ge$ 10 \mstar/yr.
We apply a small shift in redshift to aid visualising the symbols corresponding to each sub-sample. The horizontal error bars (only for black circles) represent the redshift bin size, while the vertical error bars show the $1 \sigma$ widths of the $SFR$ distributions in each bin and for each sub-sample. Purple boxes are the results from \citet{Mcdonald_2016}, green, blue and orange diamonds are results from \citet{Bonaventura_2017} respectively for 24\mum\ faint, detected and bright BCGs. Red stars are BCGs from the CLASH sample \citep{Fogarty_2017}.}
\label{fig:evolution of SFR}	
\end{figure}

\begin{figure}
\centering
\includegraphics[width=0.4\textwidth]{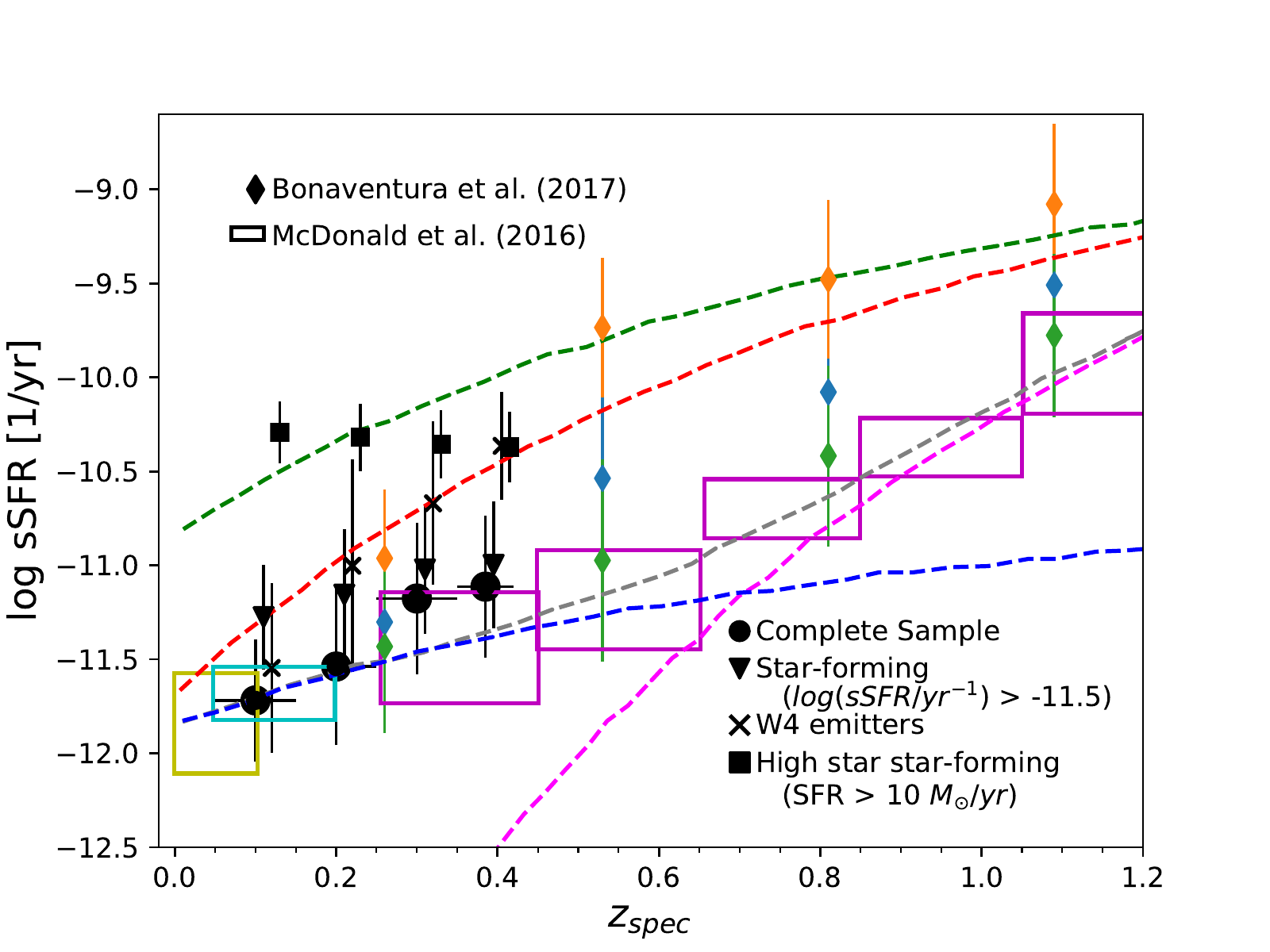}
\includegraphics[width=0.45\textwidth]{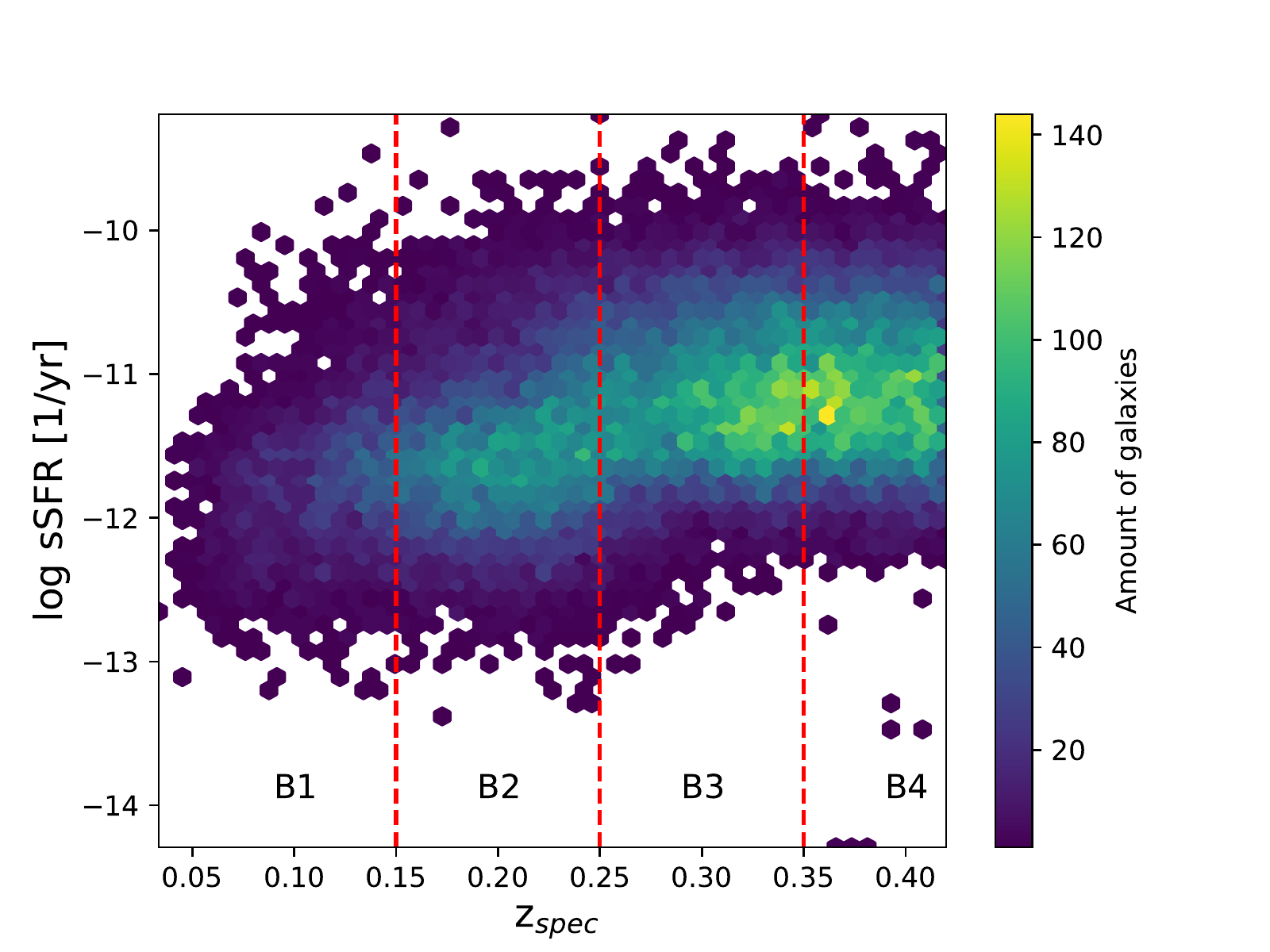}
\caption{ Top Panel: Specific SFR ($sSFR$) as a function of redshift for BCGs. Circles, inverse triangles, x's, diamonds  and squares, are described in Figure \ref{fig:evolution of SFR}. Purple, cyan and yellow rectangles show the results from \citet{Mcdonald_2016}, \citet{Haarsma_2010} and \citet{Fraser_McKelvie_2014}, respectively. The green and red dashed lines are the evolutionary paths of galaxies in the field and in clusters, respectively, according to \citet{Alberts_2014}. The magenta, blue and grey dashed lines respectively show the evolutionary tracks for the $sSFR$ of galaxies resulting from mergers, ICM cooling and the combination of the two processes, according to \citet{Mcdonald_2016}.
Bottom Panel: Density plot of the $sSFR$ as a function of redshift. The plot is colour-coded according to the amount of galaxies in each cell, and the red dashed lines represent the boundaries of the redshift bins used in this work.
}
\label{fig:evolution of sSFR}	
\end{figure}

In Section \ref{subsec: main_sequence} we showed that the use of the MS classification reveals that our entire sample of BCGs is dominated by Passive galaxies. 
However, this effect could be a consequence of the fact that the calibration that we adopt for the evolution of the MS does not take into account the presence of the break at high \mstar\ (see \citealt{Lee_2015}). When using the subdivision based on $sSFR$ we find that the fraction \fsf\ of star-forming galaxies is larger than the fraction \fq\ of quenched galaxies $\left(f_{\rm SF} = 0.660_{-0.002}^{+0.002} \right)$. 
This  result is significantly different from what we found in Paper I ($f_{\rm SF} \sim 0.09$), showing not only that star-formation is not rare among BCGs but also that star-forming systems dominate the population of BCGs at $z<0.42$. 
This is in contrast with the conclusions of  \citet{Fraser_McKelvie_2014}, who selected star-forming BCGs in the WISE colour-colour diagram and of \citet{Oliva_Altamirano_2014} who insted used the BPT diagram (\citealt{Baldwin_1981, Kewley_2001}) to divide BCGs with emission lines into star-forming and AGN hosts.

We note, however, that the values of $SFR$ for the BCGs in our sample are low and, as previously mentioned, most of them lie below the main sequence for field galaxies at $z < 0.42$. We particularly find that the median $SFR$ in our sample is $\sim 2$\msun/yr. Thus our results support the notion that BCGs have in their majority low levels of star formation, sufficient to classify them as star-forming according to our $sSFR = 10^{-11.5}$ yr$^{-1}$ threshold, but not high enough to place them on the MS.

It is important to stress at this point that our estimates of \mstar\ and $SFR$ were obtained using the stellar population library of \citet{Maraston_2005}. To assess the effects of this assumption we also performed the SED fitting using the \citet{Bruzual_2003} library and obtained similar stellar masses but lower values of $SFR$. The comparisons between the \fq\ and \fsf\ obtained with the two stellar population libraries and their implications for this paper are discussed in Section \ref{subsec:fraction_evolution}.

In agreement with \citet{Webb_2015} we find that the $SFR$ in BCGs increases with redsfhift for the entire, stellar-mass limited sample and for the three sub-samples of groups, low-halo-mass and high-halo-mass clusters (see Tables \ref{tab:bins MS and M200} and \ref{tab:evol SFR}). The median $SFR$ per redshift bin increases from $\sim 0.7$ \msun/yr  to $\sim 3.3$ \msun/yr  between the redshift bins B1 and B4. Figure \ref{fig:evolution of SFR} shows that the $SFR$ in our stellar mass limited sample  steadily increases in the range $z=0.05-0.42$ (black circles). 

In the same plot we also show the results obtained considering only star-forming BCGs ( $sSFR> 10^{-11.5}$ yr$^{-1}$; black inverted triangles), for BCGs with high star formation ($SFR > 10$ \msun/yr; squares), and BCGs with a detection in the $W4$ band (black x symbols). We will name the BCGs in the latter class  $W4$ emitters. We see that when adopting these two different criteria to define star-forming galaxies, the $SFR$ follows different trends with redshift. Indeed, while BCGs classified as star-forming according to their $sSFR$ have high $SFR$ in B1 and then undergo a small increase with redshift with a flattening between B3 and B4, the $W4$ emitters have low $SFR$ in B1 and then undergo a steep increase with redshift, reaching the highest $SFR$ values in the entire sample. The trend for the $W4$ emitters appears to be in agreement with the results obtained by \citet{Bonaventura_2017}, who classified BCGs, according to their emission at 24\mum, as faint, detected and bright (green, blue and yellow diamonds, respectively, in Figure \ref{fig:evolution of SFR}). We find that star-forming  BCGs in B2 have $SFR$ values comparable to BCGs classified as detected and bright, while BCGs in B3 and B4 have $SFR$ values comparable to BCGs classified as bright. 

Our results can be compared with the results of \citet{Mcdonald_2016} (magenta rectangles in Figure \ref{fig:evolution of SFR}) only in B3 and B4, because the sample studied by those authors contains BCGs at $z > 0.25$. The comparison shows that if we consider all the BCGs in the stellar mass limited sample or only the star-forming ones, in both cases the median $SFR$ resides below the lower bounds of $SFR$ in \citet{Mcdonald_2016}. 
However, if we restrict ourselves to the $W4$ emitters, which have higher $SFR$, we see a better agreement, and the median $SFR$ values fall inside the magenta rectangles. Finally the galaxies with higher star formation $SFR > 10$ \mstar/yr, reveals no evolution along the redshift bins, with flat values in concordance with the higher bounds of $SFR$ in \citet{Mcdonald_2016}

\citet{Fogarty_2017} studied star formation in BCGs in the Cluster Lensing and Supernova Survey with Hubble (CLASH, \citealt{Postman_2012}, red stars in Figure \ref{fig:evolution of SFR}). They found that the $SFR$ also increases with redsfhit, although with large scatter. We see that our results are consistent with some of the CLASH BCGs, and that there are two BCGs at $z \sim 0.35$ in the \cite{Fogarty_2017} sample that have $SFR > 100$ \msun\ yr$^{-1}$ (Clusters MACS 1931.8-2653 and RX 1532.9+3021). The median $SFR$ from \citet{Fogarty_2017} across the range $z=0.187-0.545$  is consistent with the median $SFR$ in the four redshift bins in which we split our sample. The  median $SFR$ from \citet{Fogarty_2017} in the same redshift range agrees with the median $SFR$ for the \citet{Mcdonald_2016} sample, with the median $SFR$ of the \citet{Bonaventura_2017} bright galaxies, and with the median $SFR$ of the $W4$ emitters in this work. 

When we split the sample according to \mh, we find that the median $SFR$ for Gr, LMC and HMC increases with redshift. In addition, the evolution of the median $SFR$ shows to be related with the evolution in \fsf: both increase with redshift (see Table \ref{tab:bins MS and M200}). Furthermore, the BCGs in the HMC sub-sample exhibit the strongest increase in the median $SFR$ and in \fsf, which respectively increase by 7.8 times and 69\% between the redshift bins B1 and B4. On the other hand, the median $SFR$ and the value of \fsf\ in the galaxies of the Gr and LMC sub-samples increase by 4.7 and 54\% (LMC), and 4.2 and 50\% (Gr).

The median \mstar\ in our BCG sample exhibits a flat trend with refshift (see Table \ref{tab:evol SFR}), implying that the median $sSFR$  increases with redshift. It can be seen, from Figure \ref{fig:evolution of sSFR}, that the median $sSFR$ in B1 and B2 is $\sim$0.5 dex lower than that in B3 and B4.
Similarly to what we observed for the $SFR$, star-forming galaxies and $W4$ emitters show different trends. 
In particular, star-forming galaxies have larger median values of $sSFR$ in B1 with a moderate increase towards higher redshifts, while the $W4$ emitters show a low median value of $sSFR$ in B1 and a steeper trend with redshift. 
The median $sSFR$ of $W4$ emitters in B4 is $\sim 0.6$ dex larger than that of star-forming BCGs. 

To explain the evolution of the sSFR in BCGs, \citet{Mcdonald_2016} proposed that the BCGs have two phases of star formation: at $z > 0.6$ star formation is driven by gas-rich mergers, while at $z < 0.6$ it is induced by cooling flows. 
The evolutionary tracks for the $sSFR$ that they derived from this assumption are different from those followed by field galaxies and non-BCG (satellite) cluster galaxies (see Figure \ref{fig:evolution of sSFR}). \citet{Mcdonald_2016} showed that the $sSFR$ evolutionary tracks obtained from their two-phase scenario can model the trend of $sSFR$ with redsfhit in their BCG sample at $0.25 < z < 1.25$. Those tracks are also able to reproduce the results of \citet{Haarsma_2010} and \citet{Fraser_McKelvie_2014} at lower redshifts.

The grey dashed line in Figure \ref{fig:evolution of sSFR} represents the model that best reproduces the results from \citet{Mcdonald_2016}. We find that while the median $sSFR$ in B1 and B2 fall on the evolutionary track, their counterparts in B3 and B4 are above the line. We note, however, that for the two highest-redsfhit bins, the evolutionary track falls within the lower $1 \sigma$ interval of the $sSFR$ distributions, suggesting that our results do not rule out the conclusions of \citet{Mcdonald_2016} at $0.25 \leq z < 0.42$.

If we only restrict ourselves to star-forming BCGs, we note that the median $sSFR$ lies above the grey line at any resdhift in the range $0.05 \leq z < 0.42$ and that the line is below the lower $1 \sigma$ bound of the $sSFR$ distribution. This suggests that the \citet{Mcdonald_2016} fiducial model underpredicts our measurements and that a model that describes the evolution of the $sSFR$ in BCGs as a function of redshift must account for higher $sSFR$. We note, however, that our results are less than $2 \sigma$ off the grey line, indicating that we cannot statistically rule out the evolutionary scenario for the $sSFR$ in BCGs proposed by \citet{Mcdonald_2016} when we only consider star-forming galaxies.

Interestingly, the median $sSFR$ for star-forming BCGs at $z < 0.25$ is closer to the evolutionary track for cluster satellite galaxies from \citet{Alberts_2014} (red dashed line) than to the \citet{Mcdonald_2016} fiducial model. At higher redshifts it is located between the red and the grey line (in B3 it is slightly closer to the red line).

$W4$ emitters show an even different behaviour. The median $sSFR$ for these galaxies is closest to the \citet{Mcdonald_2016} fiducial model at $0.05 \leq z < 0.15$, whereas it lies on or extremely close to the red line at higher redshifts. This would suggest that the $sSFR$ evolution for these galaxies is well described by the evolutionary tracks for satellite cluster galaxies. 
However, except for the result in B4 we cannot statistically rule out the \citet{Mcdonald_2016} evolutionary track.

We note that the results from \citet{Bonaventura_2017} are in agreement with the \citet{Mcdonald_2016} fiducial model only in the case of BCGs with faint emission at 24\mum, whereas galaxies with brighter emission at 24\mum\ are intermediate between the satellite track and the \citet{Mcdonald_2016} fiducial model. Interestingly, the $sSFR$ of bright 24\mum\ emitters agree with the track for the $sSFR$ evolution of field galaxies (green dashed line).

The results from our analysis of the evolution of the star formation activity in BCGs support a scenario in which the $SFR$ and $sSFR$ for these galaxies decrease as a function of cosmic time. \citet{Mcdonald_2016} propose a model in which the processes that drive the occurrence of star formation in BCGs change with redshift. In particular, they propose that the star formation in BCGs was merger-driven at $z > 0.6$ and cooling-flow driven at lower redshifts. 
The evolutionary tracks for the $sSFR$ resulting from this descriptive model lie below our median $sSFR$, suggesting that they underpredict the $sSFR$ in BCGs as observed in this work. We note, however, that the \citet{Mcdonald_2016} fiducial model cannot be statistically ruled out, since it is within the $2 \sigma$ width of the $sSFR$ distributions in all the bins in which we split the redshift range $0.05 \leq z < 0.42$. 

Our sample contains $\sim 56,000$ galaxies, providing large statistics for the analysis of star formation in BCGs. 
We argue, therefore, that the \citet{Mcdonald_2016} descriptive model can be improved taking our results into account and including AGN physics, since this affects the BCG star formation as discussed in the next section.

\subsection{Cooling flows as drivers of star-formation}
\label{subsec: cooling flows as driver of SF}

\begin{figure}
\centering
\includegraphics[width=0.4\textwidth]{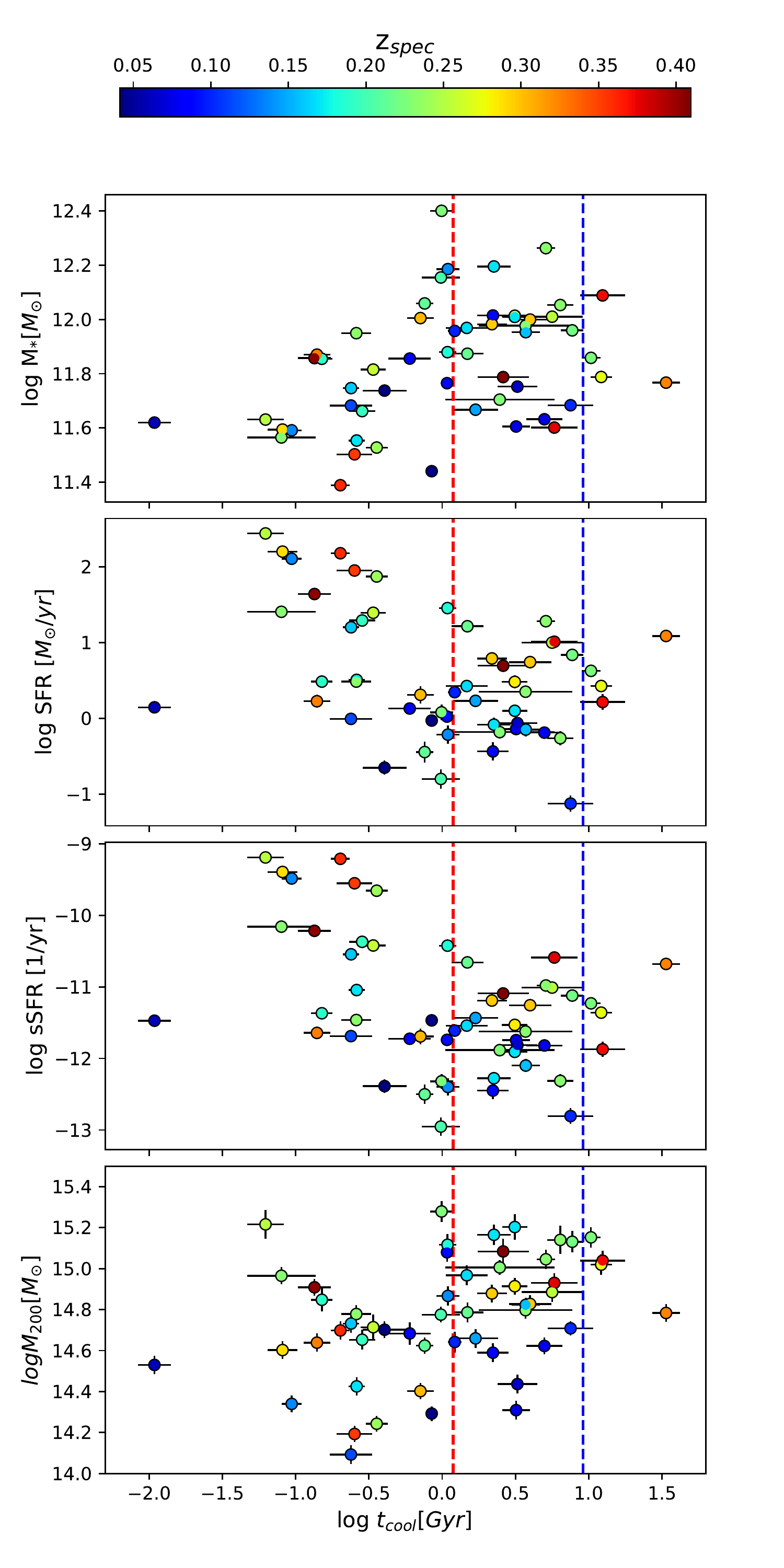}
\caption{ Stellar mass (top panel), $SFR$ (middle-top panel),  $sSFR$ (middle-bottom panel) and the halo mass (bottom panel) as a function of the cooling time (\tcool). Vertical shows strong cool cores (left side of red line) weak cool core (between red and blue line) and non cool core (right side of blue line). Colour bars indicate the spectroscopic redshift.
}
\label{fig:tcool}	
\end{figure}

The existence of cool cores in galaxy clusters (e.g., \citealt{Rafferty_2008,Pipino_2011,Runge_2018}) provides an important clue about how the cold gas from the ICM enters the galaxy. 
The ICM loses energy by emitting X-ray photons. Since the gas is densest in the central regions of clusters, a cooling flow is generated where the gas cools down and flows into the central BCG. Nonetheless, the $SFR$ (e.g. \citealt{Webb_2015,Bonaventura_2017}) and the reservoirs of molecular gas \citep{Webb_2017} in BCGs are lower than what would be expected if all the inflowing gas went to feed the  star formation (see e.g. \citealt{Fabian_1977,Fabian_1994_ARA}). This cooling flow problem is solved by taking into account the AGN activity in the BCG. By comparing the energy needed to create X-ray cavities, we can estimate the amount of energy deposited by jets and this compensates the cooling in many cases (e.g.\ \citealt{Hlavacek_2015}; see also \citealt{McCarthy_2004, McCarthy_2008}).

To test the influence of ICM cooling on the star-formation activity of BCGs, we matched our catalogue with the Archive of {\it Chandra} Cluster Entropy Profile Tables (ACCEPT, \citealt{Cavagnolo_2009}). The latter catalogue presents the parameters of the ICM entropy profiles of 239 clusters of galaxies from which one can derive the cooling time profiles, characterised as:
\begin{equation}
  \rm t_{cool}(r) = t_{c0} +t_{100} \left(\frac{r}{100~kpc}\right)^{\alpha} yr~,
\nonumber
\end{equation}
where $r$ is the projected distance (in kpc) from the cluster centroid, $\rm t_{c0}$ is the cooling time in the core of the cluster, t$_{100}$ is a normalization at 100 kpc and $\alpha$ is the exponent of the power law.

\citet{Donahue_2005} showed that the best-fit core entropy ($K_0$) is related to $\rm t_{c0}$; then, assuming that free-free interactions are the main cooling mechanism, \citet{Cavagnolo_2009} obtained that the parameter $\rm t_{c0}$ is defined by the equation:
\begin{equation}
    \rm t_{c0} = 10^8 yr \left(\frac{K_0}{10~KeV~cm^2}\right)^{3/2}\left(\frac{kT_x}{5~KeV}\right)^{-1}~,
    \nonumber
\end{equation}
where $\rm kT_x$ is the average cluster energy.

In Paper I we used a sub-sample of 17 BCGs detected in SDSS and WISE ($W1$, $W2$ and $W3$ bands), and other 27 galaxies with upper limits in the WISE $W3$ filter. Here we perform the match between our upgraded BCG sample and the ACCEPT catalogue, using a maximum matching radius of 25.6$^{\prime\prime}$ (corresponding to a linear projected distance of $25.2$ kpc and $143.1$ kpc at $z = 0.05$ and $z = 0.42$, respectively). We obtain a sub-sample of 52 BCGs that are in common between the two data-sets, which we will hereafter identify as the ACCEPT-matched sub-sample. This sub-sample comprises clusters with masses $\rm 14.09 < \log {(M_{200}/\mbox{\msun})} < 15.28$, $SFR$ in the range $ 0.08 $ \msun/yr $< SFR  <$ $ 275.51 $ \msun/yr and stellar masses in the range $11.40 < \log{\mbox{\mstar/\msun})}  < 12.33$.

\begin{table*}
    \centering
    \renewcommand{\arraystretch}{1.5}
    \begin{tabular}{lcccccccc}
\hline

        & Galaxies & $\log$(\mstar)         & $SFR$               & $\log(sSFR)$              & $\log( M_{200})$          & $W1-W2$                    & $\rm t_{cool}$         & $\rm z_{spec}$ \\ 
        & No       &  [\msun]               & [\msun/yr]          & [yr$^{-1}]$               &  [\msun]                  &                            &  [yr]                  &                     \\   
\hline
\hline
SCC     & 25       & 11.76$^{+0.37}_{-0.16}$& 3 $^{+92}_{-2}$     & -11.0$^{+1.5}_{- 0.8}$    &  14.70$^{+0.41}_{-0.22}$  & 0.22$^{+0.07}_{-0.10}$     & 0.26$^{+0.72}_{-0.17 }$& 0.21$^{+0.09}_{-0.08 }$  \\     
WCC     & 23        & 11.96$^{+0.06}_{-0.21}$& 2.2$^{+6.2}_{-1.6}$ & -11.6$^{+0.21}_{- 0.56}$&  14.88$^{+0.30 }_{-0.19 }$& 0.18$^{+0.06}_{-0.11}$    & 3.2$^{+2.5}_{-1.2}$    & 0.21$^{+0.07}_{-0.12}$ \\
CC      & 48       & 11.86$^{+0.19}_{-0.19}$& 3 $^{+24}_{-2}$     & -11.5$^{+1.1}_{- 0.7}$    &  14.78$^{+0.43 }_{-0.21 }$& 0.194$^{+0.088 }_{-0.097}$ & 1.1$^{+2.8 }_{-0.9}$   & 0.22$^{+0.08}_{-0.11}$ \\
NCC     & 4        & 11.82$^{+0.21}_{-0.04}$& 3.5 $^{+4.9}_{-1.3}$& -11.29$^{+0.17}_{- 0.26}$ &  15.03$^{+0.08}_{-0.10}$  & 0.211$^{+0.079}_{-0.009}$& 12$^{+11}_{-1}$        & 0.300$^{+0.051}_{-0.051}$\\
Total   & 52       & 11.86$^{+0.19}_{-0.19}$& 3 $^{+22}_{-2}$     & -11.45$^{+0.99 }_{- 0.61}$&  14.79$^{+0.42 }_{-0.19}$ & 0.197$^{+0.086 }_{-0.093}$ & 1.3$^{+4.2}_{-1.1}$    &0.22$^{+0.08}_{-0.12}$\\  

\hline
    \end{tabular}
    \caption{ Number of galaxies, median and $1\sigma$ widths of the distributions of $\rm \log{(M_*/M_\odot)}$, $SFR$, $\log{(sSFR/\mbox{yr}^{-1})}$, $\rm \log{(M_{200}/M_\odot})$, $(W1 - W2)$, \tcool\ and $\rm {z_{spec}}$
    for, from left to right, BCGs in SCC, WCC, CC and NCC clusters,  and for the BCGs in all the clusters that are in the ACCEPT-matched sub-sample (i.e.\ CC and NCC).
    }
    \label{tab:CC prop}
\end{table*}

Defining whether a cluster has a cool core (CC) or not (NCC) is a matter of debate, since it is not clear what is the best parameter to use for such subdivision. 
Several works show that the thermal instabilities are well described by the ratio between the cooling time (\tcool) and the free-fall time (e.g. \citealt{McCourt_2012, Voit_2015a}).
However, \citet{Hudson_2010} showed that \tcool\ is  also an effective parameter to separate CC from NCC.
This indicates that although the cooling time by itself is not the only parameter that describes thermal instabilities, it is still a good parameter to identify clusters with cool cores. \citet{Hudson_2010} further classified cool-core clusters into strong cool core (SCC; \tcool $<$ 1.2 Gyr), weak cool core (WCC;  1.2 Gyr $<$\tcool $<$ 9.1 Gyr), and no cool core (NCC; \tcool $>$ 9.1 Gyr). 
Our ACCEPT-matched sub-sample has a median \tcool = 1.09 Gyr and, following the \cite{Hudson_2010} criterion, 48\% of it is composed of SCC, 44\% is composed of WCC and 8\% is composed of NCC.

The comparison between the properties of the BCGs in CC (WCC and SCC) and NCC, reveals that the BCGs in CC clusters have higher $SFR$, higher $sSFR$, and are in clusters with lower halo mass than the galaxies in NCC (see Table \ref{tab:CC prop}).
 However, we note that there are only 2 BCGs in NCC in the ACCEPT-matched sub-sample, so they cannot be used to make robust comparisons.

Table \ref{tab:CC prop} shows that the BCGs in SCC have lower median \mstar, higher median $SFR$ and higher median $sSFR$ than BCGs that are in WCC. Furthermore, SCC are in clusters with lower \mh\ in comparison with WCC. These comparisons suggest the existence of a correlation between the internal properties of the BCGs and the strength of the cool core.
As shown in theoretical works such as \cite{McCarthy_2004}, \cite{McCarthy_2008} and \cite{Gaspari_2017}, the AGN in the centre of the BCG acts in regulating the cooling flow. When the ICM gas cools down, it sinks towards the centre of the galaxy. When this happens, the central black hole of the BCG is fed, and an AGN may be activated. As a result, the outflow generated by the AGN warms up the inflowing cool gas around the BCG, and any ongoing star formation is halted. We stress here that there are only 48 clusters with WCC or SCC, corresponding to 0.1\% of the stellar-mass limited sample, and this implies that a larger sample with cooling time information is needed to draw statistically significant conclusions on this point.

If we use the $(W1 - W2)$ colour to select AGN-host galaxies ($W1-W2 > 0.8$, \citealt{Jarrett_2011,Cluver_2014}) we find no AGN host in the BCGs of SCC and WCC (nor in the entire ACCEPT-matched sub-sample). 
We stress, however, that with this criterion, it is only possible to identify galaxies in which the emission at IR wavelengths is dominated by the AGN: for BCGs in which the IR emission is only partially contributed by the AGN, this criterion is unable to recover AGN hosts (see \citealt{Stern_2012}, \citealt{Hogan_2015a, Hogan_2015b}, \citealt{Green_2016}, and Section 4.3 in Paper I).

\begin{table}
    \centering
    \begin{tabular}{lcc}
\hline 
Parameter   & $\rho$  & $p$   \\ 
\hline
\hline
$\rm \log{(M_*/M_\odot)}$       & 0.51  & 0.01$\%$ \\   
$\log{ (SFR/\rm M_\odot\mbox{ yr}^{-1}) }$    & -0.31 & 3.12$\%$ \\   
$\log{(sSFR/\mbox{yr}^{-1})}$           & -0.42 & 0.29$\%$ \\   
$\rm \log{(M_{200}/M_\odot})$   & 0.32  & 2.31$\%$ \\
\hline
    \end{tabular}
    \caption{Spearman's correlation coefficient $\rho$ and p-value for log \tcool\ vs, from top to bottom, $\rm \log{(M_*/M_\odot)}$, $\log{ (SFR/\rm M_\odot\mbox{ yr}^{-1}) }$, $\log{(sSFR/\mbox{yr}^{-1})}$, and $\rm\log{(M_{200}/M_\odot})$. The values refer to the ACCEPT-matched sub-sample.
    }
    \label{tab:pearson coeff}
\end{table}

In order to assess the effects of cooling flows on BCG and cluster properties, we studied the correlations between \tcool\ and \mstar, $SFR$, $sSFR$ and \mh. We find that \tcool\ is anti-correlated with $SFR$ and $sSFR$ and correlated with \mstar\ and \mh\ with Spearman's p-values $<$ 5\%. 
We note that the strongest correlation is with \mstar, for which we find that the probability for \tcool\ and \mstar\ to be uncorrelated is $p = 0.01$ and the correlation coefficient is $\rho = 0.49$ (see Tab.\ref{tab:pearson coeff} and top panel of Fig. \ref{fig:tcool}). 

The correlation between \mstar\ and \tcool\ is also in agreement with the results of Paper I, where we found a tendency for the galaxies with higher \mstar\ to be in clusters with longer \tcool. 
This result suggests that the stellar mass of BCGs may affect the regulation of the temperature of the ICM in the core of the cluster. As shown in \cite{Best_2005} and \cite{Croft_2007}, the AGN fraction in BCGs increases with stellar mass, indicating that low-mass BCGs are less likely to have active nuclei. On the other hand, AGN outflows can heat up the ICM surrounding the BCG, increasing the value of \tcool\ (\citealt{McCarthy_2004, McCarthy_2008}). Thus, our results support the notion that the anti-correlation that we find between \mstar\ and \tcool\ is a reflection of the fact that AGN are more frequent in more massive BCGs and, as a consequence, the ICM around the most massive BCGs is warmer than that around the least massive ones.

The $SFR$ and the $sSFR$ are anti-correlated with \tcool; we find $p <  4\%$ of obtaining higher values of $\rho$ in the case of the ACCEPT-matched sub-sample.
 
The fact that $SFR$ and $sSFR$ are anti-correlated with \tcool\ is clear by looking at the middle panels in Figure \ref{fig:tcool} and is in agreement with the conclusions of Paper I, in which we found that the BCGs with IR colours $(W2 - W3) \geq 1.5$ (star-forming) are in clusters with $\rm t_{cool} < 1$ Gyr, while BCGs with $(W2 - W3) < 1.5$ (quiescent) reside in clusters with a broad range of cooling times.
From these results we concluded that the galaxies in clusters with short \tcool\ are more star forming. 

Unlike Paper I, we find that \tcool\ is also correlated with \mh. We find indeed $\rho = 0.32$ (0.40 if we only consider BCGs at $z > 0.15$) with  $p=2.31\%$ ($p=1.83\%$ if we only consider BCGs at $z > 0.15$; see Table \ref{tab:pearson coeff} and the bottom panel of Figure \ref{fig:tcool}). 
Thus, our results suggest that clusters with higher halo masses also have longer cooling times.

The correlations of \tcool\ with \mstar\ and \mh\ could be explained if we consider the presence of an AGN in the BCG. 
As the stellar mass of a galaxy (not necessarily a BCG) increases, the mass of its central black hole will also increase (\citealt{Reines_2015}).
This implies that the strength of the nuclear activity increases with \mstar\ (e.g. \citealt{Borys_2005,LaMura_2012}). 
At the same time, it is observed that clusters with larger \mh\ contain BCGs with larger \mstar\ (e.g.\ \citealt{Lidman_2012}, \citealt{Lavoie_2016}, \citealt{Bellstedt_2016}). Thus, the correlation between \tcool\ and \mh\ may just be a reflection of the \mstar - \mh\ relation. 

The analysis of AGN in BCGs is beyond the scope of this paper and will be the object of a forthcoming paper in this series. Here we just stress that our results suggest that AGN feedback is responsible for halting the star formation induced by cooling flows in BCGs. 
Since high-\mstar\ galaxies have stronger AGN activity than low-\mstar\ galaxies, star formation is more likely to be quenched in massive BCGs than in low-mass BCGs.   
  
\subsection{The Evolution of the star-forming and quiescent BCG fractions}
\label{subsec:fraction_evolution}

\begin{table}
 \centering
    \renewcommand{\arraystretch}{1.5}
    \begin{tabular}{|c|c|c|c|}
    \hline
    \multicolumn{1}{|c}{\z} & \multicolumn{1}{c}{\fq} & \multicolumn{1}{c}{\fsf} & \multicolumn{1}{c}{$N_g$}\\
    \hline
    \hline
    $0.05-0.15$ & $0.707_{-0.008}^{+0.008}$ & $0.293_{-0.008}^{+0.008}$ & $3,319$ \\
    $0.15-0.25$ & $0.530_{-0.005}^{+0.005}$ & $0.470_{-0.005}^{+0.005}$ & $11,646$ \\
    $0.25-0.35$ & $0.239_{-0.003}^{+0.003}$ & $0.761_{-0.003}^{+0.003}$ & $15,722$ \\
    $0.35-0.42$ & $0.195_{-0.003}^{+0.004}$ & $0.805_{-0.004}^{+0.003}$ & $12,550$ \\
    \hline
    \end{tabular}
    \caption{Fractions of quiescent (\fq) and star-forming (\fsf) BCGs as a function of spectroscopic redshift. The last column in the table indicates the number of galaxies, $N_g$, in each bin.}
    \label{table_pcerulo_1}
\end{table}

\begin{table}
\centering
    \renewcommand{\arraystretch}{1.5}
    \begin{tabular}{|c|c|c|c|}
    \hline
    \multicolumn{1}{|c}{$\rm \log{(M_{200}/M_\odot)}$} & \multicolumn{1}{c}{\fq} & \multicolumn{1}{c}{\fsf} & \multicolumn{1}{c}{$N_g$}\\
    \hline
    \hline
    $13.78-14.18$ & $0.330_{-0.003}^{+0.003}$ & $0.670_{-0.003}^{+0.003}$ & $33,679$ \\
              & $0.260_{-0.004}^{+0.004}$ & $0.740_{-0.004}^{+0.004}$ & $13,880$ \\
              & $0.378_{-0.003}^{+0.003}$ & $0.622_{-0.003}^{+0.003}$ & $19,799$ \\
    $14.18-14.58$ & $0.368_{-0.005}^{+0.005}$ & $0.632_{-0.005}^{+0.005}$ & $8,411$ \\
              & $0.256_{-0.010}^{+0.010}$ & $0.744_{-0.010}^{+0.010}$ & $1,941$ \\
              & $0.402_{-0.006}^{+0.006}$ & $0.598_{-0.006}^{+0.006}$ & $6,470$ \\
    $14.58-14.98$ & $0.458_{-0.015}^{+0.015}$ & $0.542_{-0.015}^{+0.015}$ & $1,097$ \\
              & $0.35_{-0.04}^{+0.05}$ & $0.65_{-0.05}^{+0.04}$ & $117$ \\
              & $0.470_{-0.016}^{+0.016}$ & $0.530_{-0.016}^{+0.016}$ & $980$ \\
    $14.98-15.38$ & $0.48_{-0.07}^{+0.07}$ & $0.52_{-0.07}^{+0.07}$ & $50$ \\
              & $0.21_{-0.15}^{+0.30}$ & $0.79_{-0.30}^{+0.15}$ & $2$ \\
              & $0.50_{-0.07}^{+0.07}$ & $0.50_{-0.07}^{+0.07}$ & $48$ \\
    \hline
    \end{tabular}
        \caption{Fractions of quiescent (\fq) and star-forming (\fsf) BCGs as a function of cluster halo mass. The last column in the table indicates the number of galaxies, $N_g$, in each bin. The three entries in each bin correspond, from top to bottom, to all BCGs, low-\mstar\ BCGs ($\rm 11.35 \leq \log{(M_*/M_\odot)} < 11.5$), and high-\mstar\ BCGs ($\rm \log{(M_*/M_\odot))} \geq 11.5$).}
    \label{table_pcerulo_2}
\end{table}

\begin{table}
\centering
    \renewcommand{\arraystretch}{1.5}
    \begin{tabular}{|c|c|c|c|}
    \hline
    \multicolumn{1}{|c}{$\rm \log{(M_*/M_\odot)}$} & \multicolumn{1}{c}{\fq} & \multicolumn{1}{c}{\fsf} &  \multicolumn{1}{c}{$N_g$}\\
    \hline
    \hline
    $11.35-11.55$ & $0.275_{-0.003}^{+0.003}$ & $0.725_{-0.003}^{+0.003}$ & $21,615$ \\
              & $0.277_{-0.004}^{+0.004}$ & $0.723_{-0.004}^{+0.004}$ & $11,573$ \\
              & $0.273_{-0.004}^{+0.004}$ & $0.727_{-0.004}^{+0.004}$ & $10,042$ \\
    $11.55-11.75$ & $0.380_{-0.004}^{+0.004}$ & $0.620_{-0.004}^{+0.004}$ & $16,308$ \\
              & $0.392_{-0.006}^{+0.006}$ & $0.608_{-0.006}^{+0.006}$ & $6,545$ \\
              & $0.371_{-0.005}^{+0.005}$ & $0.629_{-0.005}^{+0.005}$ & $9,763$ \\
    $11.75-11.95$ & $0.473_{-0.007}^{+0.007}$ & $0.527_{-0.007}^{+0.007}$ & $4,789$ \\
              & $0.478_{-0.015}^{+0.015}$ & $0.522_{-0.015}^{+0.015}$ & $1,060$ \\
              & $0.471_{-0.008}^{+0.008}$ & $0.529_{-0.008}^{+0.008}$ & $3,729$ \\
    $11.95-12.15$ & $0.60_{-0.02}^{+0.02}$ & $0.40_{-0.02}^{+0.02}$ & $495$ \\
              & $0.48_{-0.09}^{+0.09}$ & $0.52_{-0.09}^{+0.09}$ & $27$ \\
              & $0.60_{-0.02}^{+0.02}$ & $0.40_{-0.02}^{+0.02}$ & $468$ \\
    $12.15-12.35$ & $0.82_{-0.09}^{+0.05}$ & $0.18_{-0.05}^{+0.09}$ & $28$ \\
              &                        &                        &          \\
              & $0.82_{-0.09}^{+0.05}$ & $0.18_{-0.05}^{+0.09}$ & $28$ \\
    \hline
    \end{tabular}
    \caption{Fractions of quiescent (\fq) and star-forming (\fsf) BCGs as a function of the BCG \mstar. The last column in the table indicates the number of galaxies, $N_g$, in each bin. The three entries in each bin correspond, from top to bottom, to all clusters, low-\mh\ clusters, and high-\mh\ clusters. Low- and high-\mh\ clusters are defined as those with $6.0 \times 10^{13} \mbox{ } M_\odot \leq M_{200} < M_{200,\rm median}$ and as those with $M_{200} \geq M_{200,\rm median}$, respectively. $M_{200,\rm median}$ is the median cluster mass of the entire sample. It can be seen that there are no BCGs in low-\mh\ clusters in the highest-\mstar\ bin.} 
    
    \label{table_pcerulo_3}
\end{table}

Paper I showed that the fractions of star-forming and quiescent BCGs depend on photometric redshift, cluster mass and BCG stellar mass. In particular, we showed that \fsf\ (\fq) increases (decreases) with redshift and decreases (increases) with cluster mass and BCG stellar mass. Unlike this work, in Paper I we selected star-forming and quiescent galaxies using the $(W1-W2)$ vs $(W3-W4)$ colour-colour diagram and adopting the boundaries proposed in \citet{Wright_2010}. In Appendix \ref{subsec: wise colours} we show that $(W2-W3) = 1.3$ represents a better limit to divide between star-forming and quiescent galaxies with respect to the $(W2-W3) = 1.5$ limit that was used in Paper I. 

\begin{figure*}
\centering
\includegraphics[width=0.85\textwidth, trim=0.0cm 0.0cm 0.0cm 0.0cm, clip, page=1]{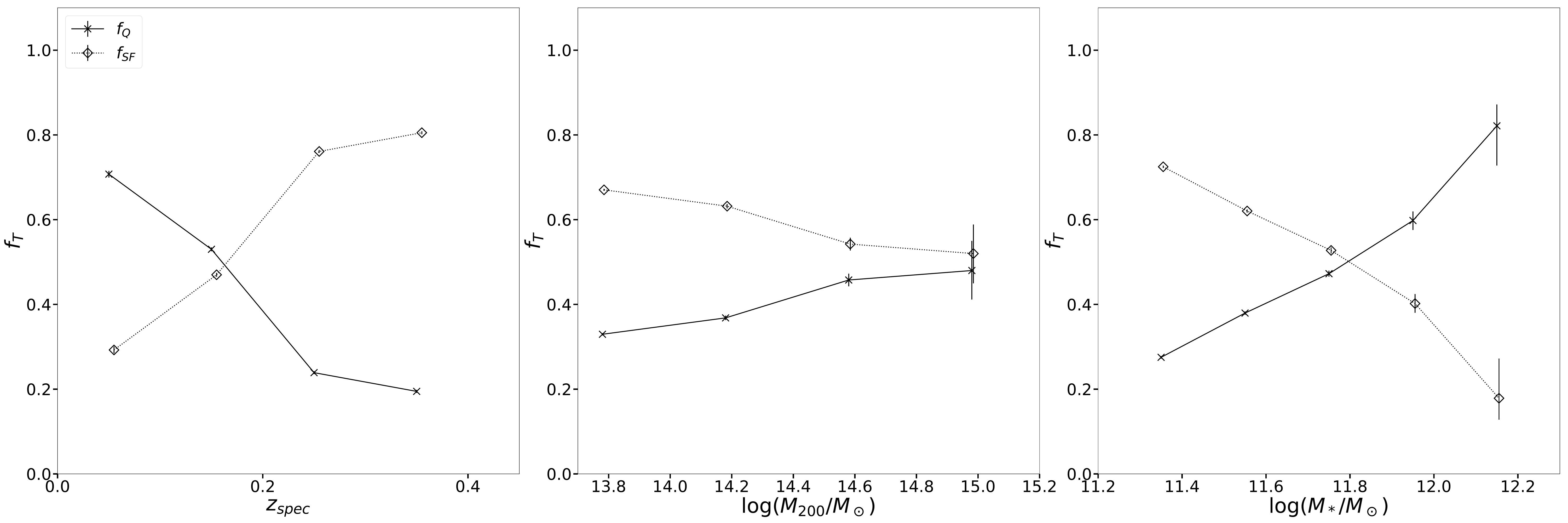}
\caption{Fractions \fq\ and \fsf\ of quiescent and star-forming BCGs as a function of spectroscopic redshift \z, cluster \mh\ and BCG \mstar. In agreement with Paper I we find that \fq\ (\fsf) decreases (increases) with \z\ and decreases (increases) with \mh\ and \mstar. The trends of the fractions with \mh\ are weaker with respect to those reported in Paper I.} 
\label{fig:SED_fractions}	
\end{figure*}

As shown in the same Appendix, a subdivision based on $sSFR$ is more rigorous than one based on colours, and in this section we revisit the results shown in Sections 4.2 and 4.4 of Paper I in the light of the measurements of the stellar population parameters that we obtained from SED fitting. Tables \ref{table_pcerulo_1}, \ref{table_pcerulo_2} and \ref{table_pcerulo_3} show the values of the quiescent and star-forming fractions as a function of \z, \mh, and \mstar. While the values of \fq\ and \fsf\ as a function of each quantity are plotted in Figure \ref{fig:SED_fractions}. The errors on the fractions were determined following \cite{Cameron_2011}.\footnote{The same method was adopted to derive the error quoted in Tables \ref{tab:bins MS} and \ref{tab:bins MS and M200}.}

It can be seen that, while we find again the increasing (decreasing) trend of \fsf\ with redshift (\mstar), the dependence of the fractions on cluster mass is weaker than what we found in Paper I. Thus, while we agree with and strengthen the conclusions of Paper I on the evolution of \fsf\ and \fq\ and on their dependence on stellar mass, we find that cluster mass does not significantly affect the presence of star formation in BCGs. Interestingly, the sample that we use here is complete at \mstar\ $= 10^{11.35} \rm M_\odot$, which is $2.3$ times higher than the stellar mass limit of the sample analysed in Paper I (\mstar\ $= 10^{11.0} \rm M_\odot$). We are therefore sampling a narrower stellar mass range, suggesting that the weak trend of the fractions with cluster mass could be a reflection of the fact that we are sampling the \mstar-\mh\ correlation within a narrower \mstar\ range with respect to Paper I.

In Paper I we noticed that at all redshfits \fsf\ $>$ \fq, which we attributed to the fact that the sample was selected to have reliable photometry in all the $W1$, $W2$ and $W3$ bands. Such a selection was biased towards star-forming galaxies, resulting in high values of $f_{SF}$. Here we see that \fsf\ $<$ \fq\ at $z < 0.2$ and \fsf\ $>$ \fq\ at $z > 0.2$. If we look at the plots of the fractions as a function of \mh, we see that \fsf\ $>$ \fq\ at all values of the cluster mass. This is remarkable, since in this work we are not applying the strict quality criteria that were used to select objects in the AllWISE catalogue and which produced a sample biased towards star-forming BCGs. 

However, we notice that the estimates of $SFR$ and stellar mass are sensitive to the stellar population libraries employed in the SED fitting.
 Running {\ttfamily{CIGALE}} with the \citet{Bruzual_2003} stellar population libraries, we found that although the trends with cluster redshift, cluster halo mass and BCG stellar mass remain unchanged, \fsf\ is always lower than \fq\ at all redshifts, cluster masses and BCG stellar masses.

This result would support the notion that star-forming BCGs are always fewer than quiescent BCGs regardless of cosmic epoch, cluster mass and BCG stellar mass. In particular, the values of \fsf\ at $z < 0.25$ are similar to those reported in \citet{Fraser_McKelvie_2014}, although we remind here that the values of \fq\ and \fsf\ for that sample were obtained using the boundaries in the $(W1-W2)$ vs $(W2-W3)$ colour-colour diagram that we adopted in Paper I.

We notice that the difference between the results obtained using the \citet{Maraston_2005} stellar population library and those obtained with the \citet{Bruzual_2003} library are a consequence of the fact that the adoption of the latter in the SED fitting results in lower estimates of the $SFR$. We find that the stellar masses derived with the two template libraries are similar at all redshifts (similar values for the biweight location and scale; \citealt{Beers_1990}). On the other hand, the values of the biweight location and scale of the $SFR$ distribution derived with the \citet{Bruzual_2003} library range between $SFR_{\rm BW} = 0.2 \pm 0.2$ \msun\ yr$^{-1}$ and $SFR_{\rm BW} = 0.9 \pm 0.8$ \msun\ yr$^{-1}$, while we obtain $SFR_{\rm BW} = 0.7 \pm 0.6$ \msun yr$^{-1}$ and $SFR_{\rm BW} = 3.1 \pm 3.1$ \msun\ yr$^{-1}$ with the \citet{Maraston_2005} libraries. Thus the use of the \citet{Bruzual_2003} libraries results in narrower distributions of $SFR$ with  values of the biweight location that are at most 2 \msun/yr$^{-1}$ lower. It should be stressed that although we find this difference, at any redshift the two values of the biweight location of the $SFR$ are consistent with each other within their respective biweight scales.

It is important to note that although we find that the majority of BCGs at $z> 0.2$ are star-forming, most of them have $SFR$ lower than the field MS at the same redshifts. Furthermore, our definition of star-forming BCGs is based on an arbitrary yet conservative threshold on $sSFR$. If we used the limit $sSFR = 10^{-11}$ yr$^{-1}$ to select star-forming galaxies, we would end up with \fsf $<$ \fq\ at all redshifts.

\begin{figure}
\centering
\includegraphics[width=0.45\textwidth, trim=0.0cm 25.0cm 0.0cm 0.0cm, clip, page=1]{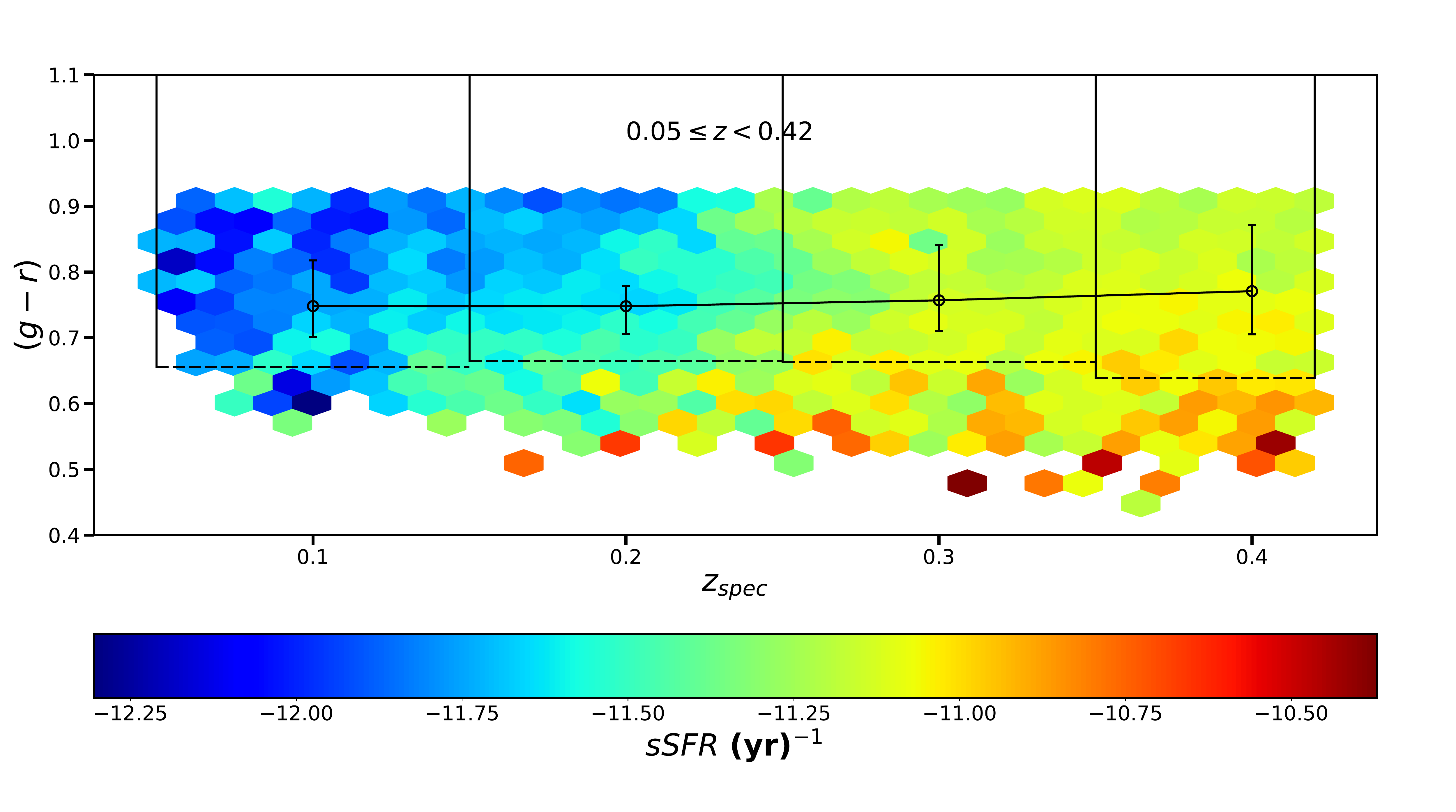}
\caption{Rest-frame $(g-r)$ colour as a function of BCG redshift. The plot is colour coded according to the mean value of $sSFR$ in each cell of the 2d histogram. The vertical solid lines represent the boundaries of the redshift bins used in this work, while the horizontal dashed lines correspond to the boundaries between blue and red BCGs in each redshift bin. The circles with the error bars connected by the solid black line correspond to the median and $1\sigma$ widths of the colour distributions in each redshift bin. Each cell in the histogram contains at least two BCGs. Regions in the planes with only one BCG are not condidered.} 
\label{fig:gr_color_vs_z}	
\end{figure}

Our results on the trends of \fsf\ and \fq\ with redshift agree with \citet{Wen_2020}, who find an increase in the fraction of star-forming BCGs in a sample of clusters detected in the Subaru Hyper-Suprime Cam. They find an increasing fraction of star-forming BCGs at $0.2 < z < 1.5$, although \fsf\ $<0.16$ at the redshifts of our sample.  Our estimates of \fsf\ in the lowest-redshift bin are also consistent with the value derived from the sample of \citet{Fraser_McKelvie_2014}. \citet{Radovich_2020}, on the other hand, find that the fraction of blue BCGs increases from $\sim 20$\% to $\sim 40$\% at $z < 0.4$ in the Kilo Degree Survey (KiDS, \citealt{de_Jong_2013_KIDS}). \cite{Pipino_2011} also report an increase with redshfit of the fraction of blue BCGs, although milder (5 to 10\% at $0.1 < z < 0.3$) than \cite{Radovich_2020}. We split our sample into blue and red BCGs, adopting the same criterion used in Paper I that blue BCGs are those with rest-frame $(g-r)$ colours bluer than $2\sigma$ below the median of the rest-frame $(g-r)$ distribution and all the other BCGs are red and studied \fsf\ and \fq. We find, in agreement with Paper I, that the majority (i.e. $0.777 \pm 0.008$) of the blue BCGs in our sample are star-forming and that the \fsf\ of blue BCGs increases with redshift from $0.45_{-0.04}^{+0.04}$ to $0.871_{-0.012}^{+0.010}$. Thus, our results suggest that an increase with redshift of the fraction of blue BCGs as reported in \cite{Radovich_2020} and \cite{Pipino_2011} is driven by the increase in \fsf. In Figure \ref{fig:gr_color_vs_z} we plotted the rest-frame $(g-r)$ colour as a function of redshift. It can be seen that the boundary between blue and red BCGs is redshift-dependent, and we note, at least qualitatively, that the BCGs with the highest $sSFR$ tend to be blue.

When we consider red BCGs, we find that \fsf\ $= 0.652 \pm 0.002$. Interestingly, the star-forming fraction increases with redshift from $f_{SF} = 0.283 \pm 0.008$ at $z \sim 0.1$ to \fsf\ $= 0.800 \pm 0.004$ at $z \sim 0.4$, showing a trend similar to that observed in the blue and total (i.e.\ blue+red) BCG samples. These results suggest, in agreement with Paper I, that a significant fraction of star-forming BCGs may be dust-rich. Our works are not the only studies that report the presence of dust in BCGs; for example, \cite{O_Dea_2010} showed that the FUV SFR in BCGs at $z < 0.31$ is lower than that estimated from IR luminosity, which they attributed to dust extinction. \cite{Fogarty_2019} reported the presence of dust and molecular gas in the BCG of the cluster MACS 1931.8-2635, at $z=0.35$, while \cite{Edge_2002} presented measurements of the molecular gas-to-dust ratio in BCGs at $z < 0.5$

The trend of \fsf\ with \mstar\ that we find in this work agrees with the results of \citet{Oliva_Altamirano_2014}, although those authors also detected a strong, increasing trend of the quiescent BCG fraction with cluster mass. We argue that the difference with our results may be due to the different definitions of star-forming and quiescent galaxies in \citet{Oliva_Altamirano_2014}, who used the H$\alpha$ line to derive the $SFR$, and to the fact that those authors also considered AGN-hosts in their sample. \citet{Wen_2020} also report no significant trend of the star-forming BCG fraction with cluster halo mass, although these authors use $\rm M_{500}$ as a proxy for cluster mass instead of \mh.

The trends of \fq\ and \fsf\ with \mh\ and \mstar\ hold at all redshifts. However, since \fq\ $\geq$ \fsf\ at $z < 0.25$, star-forming BCGs are proportionally fewer than quiescent BCGs at the low-redsfhit end of our sample. Table \ref{table_pcerulo_2} shows that if we split our sample into low-mass and high-mass BCGs, using the limit \mstar $= 10^{11.5}$ \msun, we find that there is a slight decrease in \fsf\ with \mh\ for high mass BCGs, whereas \fsf\ remains approximately constant with \mh\ for low-mass BCGs. On the other hand,  Table \ref{table_pcerulo_3} shows that when dividing the sample into low- and high-mass clusters, using  the median of \mh\ as separation value, in low-mass clusters \fsf\  decreases with \mstar\ up to \mstar\ $\sim 10^{11.7}$, while \fq\ $\sim$ \fsf\ at higher \mstar. In high-mass clusters the trends of \fq\ and \fsf\ with \mstar\ are similar to those observed for the entire sample. The fact that \fsf\ increases with \mstar\ agrees with other works in the recent literature such as \cite{Oliva_Altamirano_2014}, \cite{Gozaliasl_2016} and \cite{De_Lucia_2019}, and supports the notion that BCGs with low stellar masses quench star-formation at later times with respect to BCGs with high stellar masses.

We can conclude that, using a sample of BCGs larger than the one of Paper I, with deeper IR photometry and a separation between star-forming and quiescent BCGs based on the  $sSFR$ instead of the WISE colour-colour diagram, we find again the trends of \fq\ and \fsf\ seen in Paper I. Therefore, the results of this paper support a scenario in which BCGs quenched their star formation with a time-scale that decreases with \mstar\ (see e.g. \citealt{Hahn_2017}). \cite{Mcdonald_2016} showed that the star formation in BCGs at $z<0.6$ is a result of cooling flows, and several theoretical models such as those of \cite{McCarthy_2004, McCarthy_2008} and \cite{Gaspari_2017} have shown that AGN feedback is the main quenching driver of cooling-flow-induced star formation. In this paper we find a weak but significant correlation between \mstar\ and \tcool, according to which low-\mstar\ BCGs tend to reside in clusters with shorter \tcool. In Section \ref{subsec: cooling flows as driver of SF} we argued that, since the strength of the AGN activity increases with \mstar, the ICM around the most massive BCGs has already been warmed up by the AGN outflows. We do not have estimates of \tcool\ for the entire sample; however, the fact that more massive BCGs tend to be less star-forming than their low-mass counterparts, together with the fact that the AGN fraction increases with stellar mass (\citealt{Croft_2007}), suggest that AGN feedback may be the main driver of quenching in these galaxies.

\section{Summary and conclusions}
We presented a study of the star-formation activity in a sample of 56,399 BCGs at $0.05 < z < 0.42$ drawn from the WHL15 galaxy cluster catalogue. We selected galaxies with spectroscopic redshifts and high S/N 
photometric data
from SDSS (S/N $>$10) and WISE (S/N $>$ 5). We used the {\ttfamily{CIGALE}} code to fit models of SFH, SSP, attenuation law and dust emission, to estimate the \mstar\ and the $SFR$ in the entire sample. We limited our analysis to galaxies with stellar masses larger than the completeness limit at $z = 0.42$, that is  \mstar$> 10^{11.35}$ \msun . The main conclusions of this work are the following:
\begin{itemize}
    \item adopting the criterion $sSFR > 10^{-11.5}$ yr$^{-1}$ to select star-forming galaxies, we find that star formation occurs in $\sim 66$\% of our sample (Figure \ref{fig:SED_fractions} and Table \ref{table_pcerulo_1});
    
    \item both the $SFR$ and the $sSFR$ of BCGs increase with redshift. Although this trend agrees with most published studies, the selection criteria used to define star-forming galaxies affect its shape. 
    In particular, star-forming galaxies selected according to their brightness at $\lambda = 24\mu m$ show the steepest increase in the $SFR$ vs $z$ and $sSFR$ vs $z$ planes (Figures \ref{fig:evolution of SFR} and \ref{fig:evolution of sSFR});
    
    \item our values for the $SFR$ agree with the predictions of the models of  \citet{Mcdonald_2016} at $z < 0.15$. However, at higher redshifts, the median values of our estimates of the $SFR$ are higher than the predictions of the models. This suggests that the transition from merger-induced to ICM-cooling-induced star formation happens below $z=0.6$ (Figure \ref{fig:evolution of sSFR});
    
    \item we find that $SFR$ and $sSFR$ are anti-correlated with the ICM cooling time \tcool\ in the sub-sample of clusters for which there are X-ray data available from the ACCEPT catalogue (Figure \ref{fig:tcool}, central panels, and Table \ref{tab:pearson coeff})
    We also find that \mstar\ and \mh\ are correlated with \tcool (Figure \ref{fig:tcool}, top and bottom panels, and Table \ref{tab:pearson coeff}). 
    
    \item the fraction \fsf\ of star-forming BCGs is not significantly dependent on the cluster halo mass \mh. In particular, for low-mass BCGs the trend of \fsf\ with \mh\ is flat, while for high-mass BCGs \fsf\ shows a slight decrease with \mh (Figure \ref{fig:SED_fractions}, central panel, and Table \ref{table_pcerulo_2}); 
    
    \item \fsf\ and the fraction \fq\ of quiescent BCGs are strongly dependent on BCG stellar mass: \fsf\ decreases with \mstar, while \fq\ increases with \mstar. When splitting the sample into low- and high-mass clusters, we see that in the first case \fsf (\fq) decreases (increases) with \mstar\ up to \mstar $\sim 10^{11.7}$ \msun. At higher stellar masses \fq $\sim$ \fsf. In high-mass clusters the trends of \fsf\ and \fq\ with \mstar\ do not differ from those observed in the entire sample (Figure \ref{fig:SED_fractions}, right-hand panel, and Table \ref{table_pcerulo_3}); 
    
    \item $\sim 80$\% of blue BCGs are also star-forming, and the fraction of star-forming blue BCGs increases by $\sim 40$\% in the redshift range considered in this paper. 
    Our results also suggest that star formation in BCGs is obscured by dust, as indicated by the high fraction  ($\sim 65$\%) of red star-forming BCGs (See Section \ref{subsec:fraction_evolution}).
    
    These results support the hypothesis  
    that the cooling of the ICM induces star formation in BCGs. 
    However, the correlations of \mstar\ and \mh\ with \tcool\ suggest that AGN, which are more frequent in high-mass galaxies, are heating the intra-cluster gas around the BCG, thus quenching or even preventing the formation of stars. 
    With the ACCEPT-matched sample being small, we are not able to draw robust conclusions on this, and we leave the study of the interplay between AGN and cluster and BCG properties to a forthcoming paper.
\end{itemize}

The picture that this work draws is that the occurrence of star-forming BCGs is not rare. However, most BCGs have a low $SFR$ and are located below the MS of star-forming field galaxies. Our study leaves several open questions that will be addressed in forthcoming analyses, particularly the relationships between star formation and molecular gas in the ISM of BCGs, the role of AGN in quenching star formation and the importance of star formation in the stellar mass growth of BCGs.

\section*{Acknowledgements}
We thank the anonymous referee for the helpful and constructive feedback. We thank Dustin Lang (Perimeter Institute for Theoretical Physics) for providing unWISE/SDSS matched photometry.
PC acknowledges the support given by the ALMA-CONICYT grant 31180051 during part of this project.
CC is supported by the National Key R\&D Program of China grant 2017YFA0402704, and the National Natural Science Foundation of China, No. 11803044, 11933003, 12173045.
This work is sponsored (in part) by the Chinese Academy of Sciences (CAS), through a grant to the CAS South America Center for Astronomy (CASSACA). We acknowledge the science research grants from the China Manned Space Project with NO. CMS-CSST-2021-A05.
RD gratefully acknowledges support from the Chilean Centro de Excelencia en Astrof\'isica y Tecnolog\'ias Afines (CATA) BASAL grant AFB-170002.

\section*{Data availability}
Data available on request.




\bibliographystyle{mnras}
\bibliography{gorellana_BCG_SDSS} 

\begin{thebibliography}{}
\makeatletter
\relax
\def\mn@urlcharsother{\let\do\@makeother \do\$\do\&\do\#\do\^\do\_\do\%\do\~}
\def\mn@doi{\begingroup\mn@urlcharsother \@ifnextchar [ {\mn@doi@}
  {\mn@doi@[]}}
\def\mn@doi@[#1]#2{\def\@tempa{#1}\ifx\@tempa\@empty \href
  {http://dx.doi.org/#2} {doi:#2}\else \href {http://dx.doi.org/#2} {#1}\fi
  \endgroup}
\def\mn@eprint#1#2{\mn@eprint@#1:#2::\@nil}
\def\mn@eprint@arXiv#1{\href {http://arxiv.org/abs/#1} {{\tt arXiv:#1}}}
\def\mn@eprint@dblp#1{\href {http://dblp.uni-trier.de/rec/bibtex/#1.xml}
  {dblp:#1}}
\def\mn@eprint@#1:#2:#3:#4\@nil{\def\@tempa {#1}\def\@tempb {#2}\def\@tempc
  {#3}\ifx \@tempc \@empty \let \@tempc \@tempb \let \@tempb \@tempa \fi \ifx
  \@tempb \@empty \def\@tempb {arXiv}\fi \@ifundefined
  {mn@eprint@\@tempb}{\@tempb:\@tempc}{\expandafter \expandafter \csname
  mn@eprint@\@tempb\endcsname \expandafter{\@tempc}}}

\bibitem[\protect\citeauthoryear{{Aihara} et~al.,}{{Aihara}
  et~al.}{2011}]{Aihara_2011}
{Aihara} H.,  et~al., 2011, \mn@doi [\apjs] {10.1088/0067-0049/193/2/29}, \href
  {http://cdsads.u-strasbg.fr/abs/2011ApJS..193...29A} {193, 29}

\bibitem[\protect\citeauthoryear{{Alam} et~al.,}{{Alam}
  et~al.}{2015}]{Alam_2015}
{Alam} S.,  et~al., 2015, \mn@doi [\apjs] {10.1088/0067-0049/219/1/12}, \href
  {http://cdsads.u-strasbg.fr/abs/2015ApJS..219...12A} {219, 12}

\bibitem[\protect\citeauthoryear{{Alberts} et~al.,}{{Alberts}
  et~al.}{2014}]{Alberts_2014}
{Alberts} S.,  et~al., 2014, \mn@doi [\mnras] {10.1093/mnras/stt1897}, \href
  {https://ui.adsabs.harvard.edu/abs/2014MNRAS.437..437A} {437, 437}

\bibitem[\protect\citeauthoryear{{Ascaso}, {Lemaux}, {Lubin}, {Gal},
  {Kocevski}, {Rumbaugh}  \& {Squires}}{{Ascaso} et~al.}{2014}]{Ascaso_2014}
{Ascaso} B.,  {Lemaux} B.~C.,  {Lubin} L.~M.,  {Gal} R.~R.,  {Kocevski} D.~D.,
  {Rumbaugh} N.,   {Squires} G.,  2014, \mn@doi [\mnras]
  {10.1093/mnras/stu877}, \href
  {http://cdsads.u-strasbg.fr/abs/2014MNRAS.442..589A} {442, 589}

\bibitem[\protect\citeauthoryear{{Baldwin}, {Phillips}  \&
  {Terlevich}}{{Baldwin} et~al.}{1981}]{Baldwin_1981}
{Baldwin} J.~A.,  {Phillips} M.~M.,   {Terlevich} R.,  1981, \mn@doi [\pasp]
  {10.1086/130766}, \href
  {https://ui.adsabs.harvard.edu/abs/1981PASP...93....5B} {93, 5}

\bibitem[\protect\citeauthoryear{{Beers}, {Flynn}  \& {Gebhardt}}{{Beers}
  et~al.}{1990}]{Beers_1990}
{Beers} T.~C.,  {Flynn} K.,   {Gebhardt} K.,  1990, \mn@doi [\aj]
  {10.1086/115487}, \href {http://cdsads.u-strasbg.fr/abs/1990AJ....100...32B}
  {100, 32}

\bibitem[\protect\citeauthoryear{{Bellstedt} et~al.,}{{Bellstedt}
  et~al.}{2016}]{Bellstedt_2016}
{Bellstedt} S.,  et~al., 2016, \mn@doi [\mnras] {10.1093/mnras/stw1184}, \href
  {http://cdsads.u-strasbg.fr/abs/2016MNRAS.460.2862B} {460, 2862}

\bibitem[\protect\citeauthoryear{{Best}, {Kauffmann}, {Heckman}  \&
  {Ivezi{\'c}}}{{Best} et~al.}{2005}]{Best_2005}
{Best} P.~N.,  {Kauffmann} G.,  {Heckman} T.~M.,   {Ivezi{\'c}} {\v Z}.,  2005,
  \mn@doi [\mnras] {10.1111/j.1365-2966.2005.09283.x}, \href
  {http://cdsads.u-strasbg.fr/abs/2005MNRAS.362....9B} {362, 9}

\bibitem[\protect\citeauthoryear{{Best}, {von der Linden}, {Kauffmann},
  {Heckman}  \& {Kaiser}}{{Best} et~al.}{2007}]{best_2007}
{Best} P.~N.,  {von der Linden} A.,  {Kauffmann} G.,  {Heckman} T.~M.,
  {Kaiser} C.~R.,  2007, \mn@doi [\mnras] {10.1111/j.1365-2966.2007.11937.x},
  \href {http://cdsads.u-strasbg.fr/abs/2007MNRAS.379..894B} {379, 894}

\bibitem[\protect\citeauthoryear{{Bildfell}, {Hoekstra}, {Babul}  \&
  {Mahdavi}}{{Bildfell} et~al.}{2008}]{Bildfell_2008}
{Bildfell} C.,  {Hoekstra} H.,  {Babul} A.,   {Mahdavi} A.,  2008, \mn@doi
  [\mnras] {10.1111/j.1365-2966.2008.13699.x}, \href
  {http://cdsads.u-strasbg.fr/abs/2008MNRAS.389.1637B} {389, 1637}

\bibitem[\protect\citeauthoryear{{Bonaventura} et~al.,}{{Bonaventura}
  et~al.}{2017}]{Bonaventura_2017}
{Bonaventura} N.~R.,  et~al., 2017, \mn@doi [\mnras] {10.1093/mnras/stx722},
  \href {http://cdsads.u-strasbg.fr/abs/2017MNRAS.469.1259B} {469, 1259}

\bibitem[\protect\citeauthoryear{{Boquien}, {Burgarella}, {Roehlly}, {Buat},
  {Ciesla}, {Corre}, {Inoue}  \& {Salas}}{{Boquien}
  et~al.}{2019}]{Boquien_2019}
{Boquien} M.,  {Burgarella} D.,  {Roehlly} Y.,  {Buat} V.,  {Ciesla} L.,
  {Corre} D.,  {Inoue} A.~K.,   {Salas} H.,  2019, \mn@doi [\aap]
  {10.1051/0004-6361/201834156}, \href
  {https://ui.adsabs.harvard.edu/abs/2019A&A...622A.103B} {622, A103}

\bibitem[\protect\citeauthoryear{{Borys}, {Smail}, {Chapman}, {Blain},
  {Alexander}  \& {Ivison}}{{Borys} et~al.}{2005}]{Borys_2005}
{Borys} C.,  {Smail} I.,  {Chapman} S.~C.,  {Blain} A.~W.,  {Alexander} D.~M.,
   {Ivison} R.~J.,  2005, \mn@doi [\apj] {10.1086/491617}, \href
  {https://ui.adsabs.harvard.edu/abs/2005ApJ...635..853B} {635, 853}

\bibitem[\protect\citeauthoryear{{Bouwens} et~al.,}{{Bouwens}
  et~al.}{2012}]{Bouwens_2012}
{Bouwens} R.~J.,  et~al., 2012, \mn@doi [\apj] {10.1088/0004-637X/754/2/83},
  \href {https://ui.adsabs.harvard.edu/abs/2012ApJ...754...83B} {754, 83}

\bibitem[\protect\citeauthoryear{{Brinchmann}, {Charlot}, {White}, {Tremonti},
  {Kauffmann}, {Heckman}  \& {Brinkmann}}{{Brinchmann}
  et~al.}{2004}]{Brinchmann_2004}
{Brinchmann} J.,  {Charlot} S.,  {White} S.~D.~M.,  {Tremonti} C.,  {Kauffmann}
  G.,  {Heckman} T.,   {Brinkmann} J.,  2004, \mn@doi [\mnras]
  {10.1111/j.1365-2966.2004.07881.x}, \href
  {https://ui.adsabs.harvard.edu/abs/2004MNRAS.351.1151B} {351, 1151}

\bibitem[\protect\citeauthoryear{{Brough}, {Proctor}, {Forbes}, {Couch},
  {Collins}, {Burke}  \& {Mann}}{{Brough} et~al.}{2007}]{Brough_2007}
{Brough} S.,  {Proctor} R.,  {Forbes} D.~A.,  {Couch} W.~J.,  {Collins} C.~A.,
  {Burke} D.~J.,   {Mann} R.~G.,  2007, \mn@doi [\mnras]
  {10.1111/j.1365-2966.2007.11900.x}, \href
  {http://cdsads.u-strasbg.fr/abs/2007MNRAS.378.1507B} {378, 1507}

\bibitem[\protect\citeauthoryear{{Brough}, {Couch}, {Collins}, {Jarrett},
  {Burke}  \& {Mann}}{{Brough} et~al.}{2008}]{Brough_2008}
{Brough} S.,  {Couch} W.~J.,  {Collins} C.~A.,  {Jarrett} T.,  {Burke} D.~J.,
  {Mann} R.~G.,  2008, \mn@doi [\mnras] {10.1111/j.1745-3933.2008.00442.x},
  \href {http://cdsads.u-strasbg.fr/abs/2008MNRAS.385L.103B} {385, L103}

\bibitem[\protect\citeauthoryear{{Brown}, {Jarrett}  \& {Cluver}}{{Brown}
  et~al.}{2014}]{Brown_2014}
{Brown} M.~J.~I.,  {Jarrett} T.~H.,   {Cluver} M.~E.,  2014, \mn@doi [\pasa]
  {10.1017/pasa.2014.44}, \href
  {http://cdsads.u-strasbg.fr/abs/2014PASA...31...49B} {31, HASH}

\bibitem[\protect\citeauthoryear{{Bruzual} \& {Charlot}}{{Bruzual} \&
  {Charlot}}{2003}]{Bruzual_2003}
{Bruzual} G.,  {Charlot} S.,  2003, \mn@doi [\mnras]
  {10.1046/j.1365-8711.2003.06897.x}, \href
  {http://adsabs.harvard.edu/abs/2003MNRAS.344.1000B} {344, 1000}

\bibitem[\protect\citeauthoryear{{Calzetti}, {Armus}, {Bohlin}, {Kinney},
  {Koornneef}  \& {Storchi-Bergmann}}{{Calzetti} et~al.}{2000}]{Calzetti_2000}
{Calzetti} D.,  {Armus} L.,  {Bohlin} R.~C.,  {Kinney} A.~L.,  {Koornneef} J.,
   {Storchi-Bergmann} T.,  2000, \mn@doi [\apj] {10.1086/308692}, \href
  {https://ui.adsabs.harvard.edu/abs/2000ApJ...533..682C} {533, 682}

\bibitem[\protect\citeauthoryear{{Cameron}}{{Cameron}}{2011}]{Cameron_2011}
{Cameron} E.,  2011, \mn@doi [\pasa] {10.1071/AS10046}, \href
  {http://adsabs.harvard.edu/abs/2011PASA...28..128C} {28, 128}

\bibitem[\protect\citeauthoryear{{Cavagnolo}, {Donahue}, {Voit}  \&
  {Sun}}{{Cavagnolo} et~al.}{2008}]{Cavagnolo_2008}
{Cavagnolo} K.~W.,  {Donahue} M.,  {Voit} G.~M.,   {Sun} M.,  2008, \mn@doi
  [\apjl] {10.1086/591665}, \href
  {http://cdsads.u-strasbg.fr/abs/2008ApJ...683L.107C} {683, L107}

\bibitem[\protect\citeauthoryear{{Cavagnolo}, {Donahue}, {Voit}  \&
  {Sun}}{{Cavagnolo} et~al.}{2009}]{Cavagnolo_2009}
{Cavagnolo} K.~W.,  {Donahue} M.,  {Voit} G.~M.,   {Sun} M.,  2009, \mn@doi
  [\apjs] {10.1088/0067-0049/182/1/12}, \href
  {http://cdsads.u-strasbg.fr/abs/2009ApJS..182...12C} {182, 12}

\bibitem[\protect\citeauthoryear{{Cerulo} et~al.,}{{Cerulo}
  et~al.}{2017}]{Cerulo_2017}
{Cerulo} P.,  et~al., 2017, \mn@doi [\mnras] {10.1093/mnras/stx1687}, \href
  {https://ui.adsabs.harvard.edu/abs/2017MNRAS.472..254C} {472, 254}

\bibitem[\protect\citeauthoryear{{Cerulo}, {Orellana}  \& {Covone}}{{Cerulo}
  et~al.}{2019}]{Cerulo_2019}
{Cerulo} P.,  {Orellana} G.~A.,   {Covone} G.,  2019, \mn@doi [\mnras]
  {10.1093/mnras/stz1495}, \href
  {https://ui.adsabs.harvard.edu/abs/2019MNRAS.487.3759C} {487, 3759}

\bibitem[\protect\citeauthoryear{{Chabrier}}{{Chabrier}}{2003}]{Chabrier_2003}
{Chabrier} G.,  2003, \mn@doi [\apjl] {10.1086/374879}, \href
  {http://adsabs.harvard.edu/abs/2003ApJ...586L.133C} {586, L133}

\bibitem[\protect\citeauthoryear{{Cluver} et~al.,}{{Cluver}
  et~al.}{2014}]{Cluver_2014}
{Cluver} M.~E.,  et~al., 2014, \mn@doi [\apj] {10.1088/0004-637X/782/2/90},
  \href {http://cdsads.u-strasbg.fr/abs/2014ApJ...782...90C} {782, 90}

\bibitem[\protect\citeauthoryear{{Contini}, {Yi}  \& {Kang}}{{Contini}
  et~al.}{2018}]{Contini_2018}
{Contini} E.,  {Yi} S.~K.,   {Kang} X.,  2018, \mn@doi [\mnras]
  {10.1093/mnras/sty1518}, \href
  {http://cdsads.u-strasbg.fr/abs/2018MNRAS.tmp.1448C} {}

\bibitem[\protect\citeauthoryear{{Contini}, {Yi}  \& {Kang}}{{Contini}
  et~al.}{2019}]{Contini_2019}
{Contini} E.,  {Yi} S.~K.,   {Kang} X.,  2019, \mn@doi [\apj]
  {10.3847/1538-4357/aaf41f}, \href
  {http://cdsads.u-strasbg.fr/abs/2019ApJ...871...24C} {871, 24}

\bibitem[\protect\citeauthoryear{{Cooke}, {Kartaltepe}, {Tyler}, {Darvish},
  {Casey}, {Le F{\`e}vre}, {Salvato}  \& {Scoville}}{{Cooke}
  et~al.}{2019}]{Cooke_2019}
{Cooke} K.~C.,  {Kartaltepe} J.~S.,  {Tyler} K.~D.,  {Darvish} B.,  {Casey}
  C.~M.,  {Le F{\`e}vre} O.,  {Salvato} M.,   {Scoville} N.,  2019, \mn@doi
  [\apj] {10.3847/1538-4357/ab30c9}, \href
  {https://ui.adsabs.harvard.edu/abs/2019ApJ...881..150C} {881, 150}

\bibitem[\protect\citeauthoryear{{Correa}, {Wyithe}, {Schaye}  \&
  {Duffy}}{{Correa} et~al.}{2015a}]{Correa_2015a}
{Correa} C.~A.,  {Wyithe} J.~S.~B.,  {Schaye} J.,   {Duffy} A.~R.,  2015a,
  \mn@doi [\mnras] {10.1093/mnras/stv689}, \href
  {http://cdsads.u-strasbg.fr/abs/2015MNRAS.450.1514C} {450, 1514}

\bibitem[\protect\citeauthoryear{{Correa}, {Wyithe}, {Schaye}  \&
  {Duffy}}{{Correa} et~al.}{2015b}]{Correa_2015b}
{Correa} C.~A.,  {Wyithe} J.~S.~B.,  {Schaye} J.,   {Duffy} A.~R.,  2015b,
  \mn@doi [\mnras] {10.1093/mnras/stv697}, \href
  {http://cdsads.u-strasbg.fr/abs/2015MNRAS.450.1521C} {450, 1521}

\bibitem[\protect\citeauthoryear{{Correa}, {Wyithe}, {Schaye}  \&
  {Duffy}}{{Correa} et~al.}{2015c}]{Correa_2015c}
{Correa} C.~A.,  {Wyithe} J.~S.~B.,  {Schaye} J.,   {Duffy} A.~R.,  2015c,
  \mn@doi [\mnras] {10.1093/mnras/stv1363}, \href
  {http://cdsads.u-strasbg.fr/abs/2015MNRAS.452.1217C} {452, 1217}

\bibitem[\protect\citeauthoryear{{Covone}, {Sereno}, {Kilbinger}  \&
  {Cardone}}{{Covone} et~al.}{2014}]{Covone_2014}
{Covone} G.,  {Sereno} M.,  {Kilbinger} M.,   {Cardone} V.~F.,  2014, \mn@doi
  [\apjl] {10.1088/2041-8205/784/2/L25}, \href
  {http://adsabs.harvard.edu/abs/2014ApJ...784L..25C} {784, L25}

\bibitem[\protect\citeauthoryear{{Croft}, {de Vries}  \& {Becker}}{{Croft}
  et~al.}{2007}]{Croft_2007}
{Croft} S.,  {de Vries} W.,   {Becker} R.~H.,  2007, \mn@doi [\apjl]
  {10.1086/522086}, \href {http://cdsads.u-strasbg.fr/abs/2007ApJ...667L..13C}
  {667, L13}

\bibitem[\protect\citeauthoryear{{Daddi} et~al.,}{{Daddi}
  et~al.}{2007}]{Daddi_2007}
{Daddi} E.,  et~al., 2007, \mn@doi [\apj] {10.1086/521818}, \href
  {https://ui.adsabs.harvard.edu/abs/2007ApJ...670..156D} {670, 156}

\bibitem[\protect\citeauthoryear{{De Lucia}, {Springel}, {White}, {Croton}  \&
  {Kauffmann}}{{De Lucia} et~al.}{2006}]{De_Lucia_2006}
{De Lucia} G.,  {Springel} V.,  {White} S. D.~M.,  {Croton} D.,   {Kauffmann}
  G.,  2006, \mn@doi [\mnras] {10.1111/j.1365-2966.2005.09879.x}, \href
  {https://ui.adsabs.harvard.edu/abs/2006MNRAS.366..499D} {366, 499}

\bibitem[\protect\citeauthoryear{{De Lucia} et~al.,}{{De Lucia}
  et~al.}{2007}]{De_Lucia_and_Blaizot_2007}
{De Lucia} G.,  et~al., 2007, \mn@doi [\mnras]
  {10.1111/j.1365-2966.2006.11199.x}, \href
  {http://cdsads.u-strasbg.fr/abs/2007MNRAS.374..809D} {374, 809}

\bibitem[\protect\citeauthoryear{{De Lucia}, {Hirschmann}  \& {Fontanot}}{{De
  Lucia} et~al.}{2019}]{De_Lucia_2019}
{De Lucia} G.,  {Hirschmann} M.,   {Fontanot} F.,  2019, \mn@doi [\mnras]
  {10.1093/mnras/sty3059}, \href
  {https://ui.adsabs.harvard.edu/abs/2019MNRAS.482.5041D} {482, 5041}

\bibitem[\protect\citeauthoryear{{Donahue}, {Voit}, {O'Dea}, {Baum}  \&
  {Sparks}}{{Donahue} et~al.}{2005}]{Donahue_2005}
{Donahue} M.,  {Voit} G.~M.,  {O'Dea} C.~P.,  {Baum} S.~A.,   {Sparks} W.~B.,
  2005, \mn@doi [\apjl] {10.1086/462416}, \href
  {https://ui.adsabs.harvard.edu/abs/2005ApJ...630L..13D} {630, L13}

\bibitem[\protect\citeauthoryear{{Donzelli}, {Muriel}  \& {Madrid}}{{Donzelli}
  et~al.}{2011}]{Donzelli_2011}
{Donzelli} C.~J.,  {Muriel} H.,   {Madrid} J.~P.,  2011, \mn@doi [\apjs]
  {10.1088/0067-0049/195/2/15}, \href
  {http://cdsads.u-strasbg.fr/abs/2011ApJS..195...15D} {195, 15}

\bibitem[\protect\citeauthoryear{{Draine} \& {Li}}{{Draine} \&
  {Li}}{2007}]{Draine_2007}
{Draine} B.~T.,  {Li} A.,  2007, \mn@doi [\apj] {10.1086/511055}, \href
  {https://ui.adsabs.harvard.edu/abs/2007ApJ...657..810D} {657, 810}

\bibitem[\protect\citeauthoryear{{Dunne} et~al.,}{{Dunne}
  et~al.}{2009}]{Dunne_2009}
{Dunne} L.,  et~al., 2009, \mn@doi [\mnras] {10.1111/j.1365-2966.2008.13900.x},
  \href {https://ui.adsabs.harvard.edu/abs/2009MNRAS.394....3D} {394, 3}

\bibitem[\protect\citeauthoryear{{Edge}}{{Edge}}{2001}]{Edge_2001}
{Edge} A.~C.,  2001, \mn@doi [\mnras] {10.1046/j.1365-8711.2001.04802.x}, \href
  {https://ui.adsabs.harvard.edu/abs/2001MNRAS.328..762E} {328, 762}

\bibitem[\protect\citeauthoryear{{Edge}, {Wilman}, {Johnstone}, {Crawford},
  {Fabian}  \& {Allen}}{{Edge} et~al.}{2002}]{Edge_2002}
{Edge} A.~C.,  {Wilman} R.~J.,  {Johnstone} R.~M.,  {Crawford} C.~S.,  {Fabian}
  A.~C.,   {Allen} S.~W.,  2002, \mn@doi [\mnras]
  {10.1046/j.1365-8711.2002.05790.x}, \href
  {https://ui.adsabs.harvard.edu/abs/2002MNRAS.337...49E} {337, 49}

\bibitem[\protect\citeauthoryear{{Elbaz} et~al.,}{{Elbaz}
  et~al.}{2007}]{Elbaz_2007}
{Elbaz} D.,  et~al., 2007, \mn@doi [\aap] {10.1051/0004-6361:20077525}, \href
  {http://cdsads.u-strasbg.fr/abs/2007A%26A...468...33E} {468, 33}

\bibitem[\protect\citeauthoryear{{Elbaz} et~al.,}{{Elbaz}
  et~al.}{2011}]{Elbaz_2011}
{Elbaz} D.,  et~al., 2011, \mn@doi [\aap] {10.1051/0004-6361/201117239}, \href
  {https://ui.adsabs.harvard.edu/abs/2011A&A...533A.119E} {533, A119}

\bibitem[\protect\citeauthoryear{{Fabian}}{{Fabian}}{1994}]{Fabian_1994_ARA}
{Fabian} A.~C.,  1994, \mn@doi [\araa] {10.1146/annurev.aa.32.090194.001425},
  \href {https://ui.adsabs.harvard.edu/abs/1994ARA&A..32..277F} {32, 277}

\bibitem[\protect\citeauthoryear{{Fabian} \& {Nulsen}}{{Fabian} \&
  {Nulsen}}{1977}]{Fabian_1977}
{Fabian} A.~C.,  {Nulsen} P.~E.~J.,  1977, \mn@doi [\mnras]
  {10.1093/mnras/180.3.479}, \href
  {https://ui.adsabs.harvard.edu/abs/1977MNRAS.180..479F} {180, 479}

\bibitem[\protect\citeauthoryear{{Fogarty}, {Postman}, {Larson}, {Donahue}  \&
  {Moustakas}}{{Fogarty} et~al.}{2017}]{Fogarty_2017}
{Fogarty} K.,  {Postman} M.,  {Larson} R.,  {Donahue} M.,   {Moustakas} J.,
  2017, \mn@doi [\apj] {10.3847/1538-4357/aa82b9}, \href
  {https://ui.adsabs.harvard.edu/abs/2017ApJ...846..103F} {846, 103}

\bibitem[\protect\citeauthoryear{{Fogarty} et~al.,}{{Fogarty}
  et~al.}{2019}]{Fogarty_2019}
{Fogarty} K.,  et~al., 2019, \mn@doi [\apj] {10.3847/1538-4357/ab22a4}, \href
  {https://ui.adsabs.harvard.edu/abs/2019ApJ...879..103F} {879, 103}

\bibitem[\protect\citeauthoryear{{Fraser-McKelvie}, {Brown}  \&
  {Pimbblet}}{{Fraser-McKelvie} et~al.}{2014}]{Fraser_McKelvie_2014}
{Fraser-McKelvie} A.,  {Brown} M.~J.~I.,   {Pimbblet} K.~A.,  2014, \mn@doi
  [\mnras] {10.1093/mnrasl/slu117}, \href
  {http://cdsads.u-strasbg.fr/abs/2014MNRAS.444L..63F} {444, L63}

\bibitem[\protect\citeauthoryear{{Gaspari} \& {S{\k{a}}dowski}}{{Gaspari} \&
  {S{\k{a}}dowski}}{2017}]{Gaspari_2017}
{Gaspari} M.,  {S{\k{a}}dowski} A.,  2017, \mn@doi [\apj]
  {10.3847/1538-4357/aa61a3}, \href
  {https://ui.adsabs.harvard.edu/abs/2017ApJ...837..149G} {837, 149}

\bibitem[\protect\citeauthoryear{{Gozaliasl}, {Finoguenov}, {Khosroshahi},
  {Mirkazemi}, {Erfanianfar}  \& {Tanaka}}{{Gozaliasl}
  et~al.}{2016}]{Gozaliasl_2016}
{Gozaliasl} G.,  {Finoguenov} A.,  {Khosroshahi} H.~G.,  {Mirkazemi} M.,
  {Erfanianfar} G.,   {Tanaka} M.,  2016, \mn@doi [\mnras]
  {10.1093/mnras/stw448}, \href
  {http://cdsads.u-strasbg.fr/abs/2016MNRAS.458.2762G} {458, 2762}

\bibitem[\protect\citeauthoryear{{Graham}, {Lauer}, {Colless}  \&
  {Postman}}{{Graham} et~al.}{1996}]{Graham_1996}
{Graham} A.,  {Lauer} T.~R.,  {Colless} M.,   {Postman} M.,  1996, \mn@doi
  [\apj] {10.1086/177440}, \href
  {http://adsabs.harvard.edu/abs/1996ApJ...465..534G} {465, 534}

\bibitem[\protect\citeauthoryear{{Green} et~al.,}{{Green}
  et~al.}{2016}]{Green_2016}
{Green} T.~S.,  et~al., 2016, \mn@doi [\mnras] {10.1093/mnras/stw1338}, \href
  {https://ui.adsabs.harvard.edu/abs/2016MNRAS.461..560G} {461, 560}

\bibitem[\protect\citeauthoryear{{Haarsma} et~al.,}{{Haarsma}
  et~al.}{2010}]{Haarsma_2010}
{Haarsma} D.~B.,  et~al., 2010, \mn@doi [\apj] {10.1088/0004-637X/713/2/1037},
  \href {https://ui.adsabs.harvard.edu/abs/2010ApJ...713.1037H} {713, 1037}

\bibitem[\protect\citeauthoryear{{Hahn}, {Tinker}  \& {Wetzel}}{{Hahn}
  et~al.}{2017}]{Hahn_2017}
{Hahn} C.,  {Tinker} J.~L.,   {Wetzel} A.,  2017, \mn@doi [\apj]
  {10.3847/1538-4357/aa6d6b}, \href
  {https://ui.adsabs.harvard.edu/abs/2017ApJ...841....6H} {841, 6}

\bibitem[\protect\citeauthoryear{{Hinshaw} et~al.,}{{Hinshaw}
  et~al.}{2009}]{Hinshaw_2009}
{Hinshaw} G.,  et~al., 2009, \mn@doi [\apjs] {10.1088/0067-0049/180/2/225},
  \href {http://cdsads.u-strasbg.fr/abs/2009ApJS..180..225H} {180, 225}

\bibitem[\protect\citeauthoryear{{Hlavacek-Larrondo}
  et~al.,}{{Hlavacek-Larrondo} et~al.}{2015}]{Hlavacek_2015}
{Hlavacek-Larrondo} J.,  et~al., 2015, \mn@doi [\apj]
  {10.1088/0004-637X/805/1/35}, \href
  {https://ui.adsabs.harvard.edu/abs/2015ApJ...805...35H} {805, 35}

\bibitem[\protect\citeauthoryear{{Hlavacek-Larrondo}
  et~al.,}{{Hlavacek-Larrondo} et~al.}{2020}]{Hlavacek_2020}
{Hlavacek-Larrondo} J.,  et~al., 2020, \mn@doi [\apjl]
  {10.3847/2041-8213/ab9ca5}, \href
  {https://ui.adsabs.harvard.edu/abs/2020ApJ...898L..50H} {898, L50}

\bibitem[\protect\citeauthoryear{{Hogan} et~al.,}{{Hogan}
  et~al.}{2015a}]{Hogan_2015b}
{Hogan} M.~T.,  et~al., 2015a, \mn@doi [\mnras] {10.1093/mnras/stv1517}, \href
  {https://ui.adsabs.harvard.edu/abs/2015MNRAS.453.1201H} {453, 1201}

\bibitem[\protect\citeauthoryear{{Hogan} et~al.,}{{Hogan}
  et~al.}{2015b}]{Hogan_2015a}
{Hogan} M.~T.,  et~al., 2015b, \mn@doi [\mnras] {10.1093/mnras/stv1518}, \href
  {https://ui.adsabs.harvard.edu/abs/2015MNRAS.453.1223H} {453, 1223}

\bibitem[\protect\citeauthoryear{{Hu}, {Cowie}  \& {Wang}}{{Hu}
  et~al.}{1985}]{Hu_1985}
{Hu} E.~M.,  {Cowie} L.~L.,   {Wang} Z.,  1985, \mn@doi [\apjs]
  {10.1086/191081}, \href {http://cdsads.u-strasbg.fr/abs/1985ApJS...59..447H}
  {59, 447}

\bibitem[\protect\citeauthoryear{{Hudson}, {Mittal}, {Reiprich}, {Nulsen},
  {Andernach}  \& {Sarazin}}{{Hudson} et~al.}{2010}]{Hudson_2010}
{Hudson} D.~S.,  {Mittal} R.,  {Reiprich} T.~H.,  {Nulsen} P.~E.~J.,
  {Andernach} H.,   {Sarazin} C.~L.,  2010, \mn@doi [\aap]
  {10.1051/0004-6361/200912377}, \href
  {https://ui.adsabs.harvard.edu/abs/2010A&A...513A..37H} {513, A37}

\bibitem[\protect\citeauthoryear{{Hunter}, {Elmegreen}  \& {Martin}}{{Hunter}
  et~al.}{2006}]{Hunter_2006}
{Hunter} D.~A.,  {Elmegreen} B.~G.,   {Martin} E.,  2006, \mn@doi [\aj]
  {10.1086/505202}, \href
  {https://ui.adsabs.harvard.edu/abs/2006AJ....132..801H} {132, 801}

\bibitem[\protect\citeauthoryear{{Jarrett} et~al.,}{{Jarrett}
  et~al.}{2011}]{Jarrett_2011}
{Jarrett} T.~H.,  et~al., 2011, \mn@doi [\apj] {10.1088/0004-637X/735/2/112},
  \href {http://adsabs.harvard.edu/abs/2011ApJ...735..112J} {735, 112}

\bibitem[\protect\citeauthoryear{{Jarrett} et~al.,}{{Jarrett}
  et~al.}{2017}]{Jarrett_2017}
{Jarrett} T.~H.,  et~al., 2017, \mn@doi [\apj] {10.3847/1538-4357/836/2/182},
  \href {http://adsabs.harvard.edu/abs/2017ApJ...836..182J} {836, 182}

\bibitem[\protect\citeauthoryear{{Karim} et~al.,}{{Karim}
  et~al.}{2011}]{Karim_2011}
{Karim} A.,  et~al., 2011, \mn@doi [\apj] {10.1088/0004-637X/730/2/61}, \href
  {https://ui.adsabs.harvard.edu/abs/2011ApJ...730...61K} {730, 61}

\bibitem[\protect\citeauthoryear{{Kewley}, {Heisler}, {Dopita}  \&
  {Lumsden}}{{Kewley} et~al.}{2001}]{Kewley_2001}
{Kewley} L.~J.,  {Heisler} C.~A.,  {Dopita} M.~A.,   {Lumsden} S.,  2001,
  \mn@doi [\apjs] {10.1086/318944}, \href
  {https://ui.adsabs.harvard.edu/abs/2001ApJS..132...37K} {132, 37}

\bibitem[\protect\citeauthoryear{{La Mura}, {Bindoni}, {Ciroi}, {Cracco},
  {D'Abrusco}, {Rafanelli}  \& {Vaona}}{{La Mura} et~al.}{2012}]{LaMura_2012}
{La Mura} G.,  {Bindoni} D.,  {Ciroi} S.,  {Cracco} V.,  {D'Abrusco} R.,
  {Rafanelli} P.,   {Vaona} L.,  2012, \mn@doi [\mnras]
  {10.1111/j.1365-2966.2012.21840.x}, \href
  {https://ui.adsabs.harvard.edu/abs/2012MNRAS.426.1893L} {426, 1893}

\bibitem[\protect\citeauthoryear{{Lang}}{{Lang}}{2014}]{Lang_2014_unWISE}
{Lang} D.,  2014, \mn@doi [\aj] {10.1088/0004-6256/147/5/108}, \href
  {http://cdsads.u-strasbg.fr/abs/2014AJ....147..108L} {147, 108}

\bibitem[\protect\citeauthoryear{{Lang}, {Hogg}  \& {Schlegel}}{{Lang}
  et~al.}{2016}]{Lang_2016_unWISE}
{Lang} D.,  {Hogg} D.~W.,   {Schlegel} D.~J.,  2016, \mn@doi [\aj]
  {10.3847/0004-6256/151/2/36}, \href
  {https://ui.adsabs.harvard.edu/abs/2016AJ....151...36L} {151, 36}

\bibitem[\protect\citeauthoryear{{Lavoie} et~al.,}{{Lavoie}
  et~al.}{2016}]{Lavoie_2016}
{Lavoie} S.,  et~al., 2016, \mn@doi [\mnras] {10.1093/mnras/stw1906}, \href
  {http://cdsads.u-strasbg.fr/abs/2016MNRAS.462.4141L} {462, 4141}

\bibitem[\protect\citeauthoryear{{Lee} et~al.,}{{Lee} et~al.}{2015}]{Lee_2015}
{Lee} N.,  et~al., 2015, \mn@doi [\apj] {10.1088/0004-637X/801/2/80}, \href
  {https://ui.adsabs.harvard.edu/abs/2015ApJ...801...80L} {801, 80}

\bibitem[\protect\citeauthoryear{{Lidman} et~al.,}{{Lidman}
  et~al.}{2012}]{Lidman_2012}
{Lidman} C.,  et~al., 2012, \mn@doi [\mnras]
  {10.1111/j.1365-2966.2012.21984.x}, \href
  {http://adsabs.harvard.edu/abs/2012MNRAS.427..550L} {427, 550}

\bibitem[\protect\citeauthoryear{{Lidman} et~al.,}{{Lidman}
  et~al.}{2013}]{Lidman_2013}
{Lidman} C.,  et~al., 2013, \mn@doi [\mnras] {10.1093/mnras/stt777}, \href
  {http://adsabs.harvard.edu/abs/2013MNRAS.433..825L} {433, 825}

\bibitem[\protect\citeauthoryear{{Lin}, {Brodwin}, {Gonzalez}, {Bode},
  {Eisenhardt}, {Stanford}  \& {Vikhlinin}}{{Lin} et~al.}{2013}]{Lin_2013}
{Lin} Y.-T.,  {Brodwin} M.,  {Gonzalez} A.~H.,  {Bode} P.,  {Eisenhardt}
  P.~R.~M.,  {Stanford} S.~A.,   {Vikhlinin} A.,  2013, \mn@doi [\apj]
  {10.1088/0004-637X/771/1/61}, \href
  {http://cdsads.u-strasbg.fr/abs/2013ApJ...771...61L} {771, 61}

\bibitem[\protect\citeauthoryear{{Loubser}, {S{\'a}nchez-Bl{\'a}zquez},
  {Sansom}  \& {Soechting}}{{Loubser} et~al.}{2009}]{Loubser_2009}
{Loubser} S.~I.,  {S{\'a}nchez-Bl{\'a}zquez} P.,  {Sansom} A.~E.,   {Soechting}
  I.~K.,  2009, \mn@doi [\mnras] {10.1111/j.1365-2966.2009.15171.x}, \href
  {http://cdsads.u-strasbg.fr/abs/2009MNRAS.398..133L} {398, 133}

\bibitem[\protect\citeauthoryear{{Mancone} \& {Gonzalez}}{{Mancone} \&
  {Gonzalez}}{2012}]{Mancone_2012}
{Mancone} C.~L.,  {Gonzalez} A.~H.,  2012, \mn@doi [\pasp] {10.1086/666502},
  \href {http://cdsads.u-strasbg.fr/abs/2012PASP..124..606M} {124, 606}

\bibitem[\protect\citeauthoryear{{Maraston}}{{Maraston}}{2005}]{Maraston_2005}
{Maraston} C.,  2005, \mn@doi [\mnras] {10.1111/j.1365-2966.2005.09270.x},
  \href {https://ui.adsabs.harvard.edu/abs/2005MNRAS.362..799M} {362, 799}

\bibitem[\protect\citeauthoryear{{McCarthy}, {Balogh}, {Babul}, {Poole}  \&
  {Horner}}{{McCarthy} et~al.}{2004}]{McCarthy_2004}
{McCarthy} I.~G.,  {Balogh} M.~L.,  {Babul} A.,  {Poole} G.~B.,   {Horner}
  D.~J.,  2004, \mn@doi [\apj] {10.1086/423267}, \href
  {http://cdsads.u-strasbg.fr/abs/2004ApJ...613..811M} {613, 811}

\bibitem[\protect\citeauthoryear{{McCarthy}, {Babul}, {Bower}  \&
  {Balogh}}{{McCarthy} et~al.}{2008}]{McCarthy_2008}
{McCarthy} I.~G.,  {Babul} A.,  {Bower} R.~G.,   {Balogh} M.~L.,  2008, \mn@doi
  [\mnras] {10.1111/j.1365-2966.2008.13141.x}, \href
  {http://cdsads.u-strasbg.fr/abs/2008MNRAS.386.1309M} {386, 1309}

\bibitem[\protect\citeauthoryear{{McCourt}, {Sharma}, {Quataert}  \&
  {Parrish}}{{McCourt} et~al.}{2012}]{McCourt_2012}
{McCourt} M.,  {Sharma} P.,  {Quataert} E.,   {Parrish} I.~J.,  2012, \mn@doi
  [\mnras] {10.1111/j.1365-2966.2011.19972.x}, \href
  {https://ui.adsabs.harvard.edu/abs/2012MNRAS.419.3319M} {419, 3319}

\bibitem[\protect\citeauthoryear{{McDonald} et~al.,}{{McDonald}
  et~al.}{2016}]{Mcdonald_2016}
{McDonald} M.,  et~al., 2016, \mn@doi [\apj] {10.3847/0004-637X/817/2/86},
  \href {https://ui.adsabs.harvard.edu/abs/2016ApJ...817...86M} {817, 86}

\bibitem[\protect\citeauthoryear{{McDonald}, {Gaspari}, {McNamara}  \&
  {Tremblay}}{{McDonald} et~al.}{2018}]{Mcdonald_2018}
{McDonald} M.,  {Gaspari} M.,  {McNamara} B.~R.,   {Tremblay} G.~R.,  2018,
  \mn@doi [\apj] {10.3847/1538-4357/aabace}, \href
  {https://ui.adsabs.harvard.edu/abs/2018ApJ...858...45M} {858, 45}

\bibitem[\protect\citeauthoryear{{McNamara} \& {O'Connell}}{{McNamara} \&
  {O'Connell}}{1989}]{McNamara_1989}
{McNamara} B.~R.,  {O'Connell} R.~W.,  1989, \mn@doi [\aj] {10.1086/115275},
  \href {https://ui.adsabs.harvard.edu/abs/1989AJ.....98.2018M} {98, 2018}

\bibitem[\protect\citeauthoryear{{Meidt} et~al.,}{{Meidt}
  et~al.}{2012}]{Meidt_2012}
{Meidt} S.~E.,  et~al., 2012, \mn@doi [\apj] {10.1088/0004-637X/744/1/17},
  \href {https://ui.adsabs.harvard.edu/abs/2012ApJ...744...17M} {744, 17}

\bibitem[\protect\citeauthoryear{{Nantais} et~al.,}{{Nantais}
  et~al.}{2020}]{Nantais_2020}
{Nantais} J.,  et~al., 2020, \mn@doi [\mnras] {10.1093/mnras/staa2872}, \href
  {https://ui.adsabs.harvard.edu/abs/2020MNRAS.499.3061N} {499, 3061}

\bibitem[\protect\citeauthoryear{{Noeske} et~al.,}{{Noeske}
  et~al.}{2007}]{Noeske_2007}
{Noeske} K.~G.,  et~al., 2007, \mn@doi [\apjl] {10.1086/517927}, \href
  {https://ui.adsabs.harvard.edu/abs/2007ApJ...660L..47N} {660, L47}

\bibitem[\protect\citeauthoryear{{O'Dea} et~al.,}{{O'Dea}
  et~al.}{2010}]{O_Dea_2010}
{O'Dea} K.~P.,  et~al., 2010, \mn@doi [\apj] {10.1088/0004-637X/719/2/1619},
  \href {https://ui.adsabs.harvard.edu/abs/2010ApJ...719.1619O} {719, 1619}

\bibitem[\protect\citeauthoryear{{Old} et~al.,}{{Old} et~al.}{2020}]{Old_2020}
{Old} L.~J.,  et~al., 2020, arXiv e-prints, \href
  {https://ui.adsabs.harvard.edu/abs/2020arXiv200211735O} {p. arXiv:2002.11735}

\bibitem[\protect\citeauthoryear{{Oliva-Altamirano} et~al.,}{{Oliva-Altamirano}
  et~al.}{2014}]{Oliva_Altamirano_2014}
{Oliva-Altamirano} P.,  et~al., 2014, \mn@doi [\mnras] {10.1093/mnras/stu277},
  \href {http://cdsads.u-strasbg.fr/abs/2014MNRAS.440..762O} {440, 762}

\bibitem[\protect\citeauthoryear{{Olivares} et~al.,}{{Olivares}
  et~al.}{2019}]{Olivares_2019}
{Olivares} V.,  et~al., 2019, \mn@doi [\aap] {10.1051/0004-6361/201935350},
  \href {https://ui.adsabs.harvard.edu/abs/2019A&A...631A..22O} {631, A22}

\bibitem[\protect\citeauthoryear{{Oliver} et~al.,}{{Oliver}
  et~al.}{2010}]{Oliver_2010}
{Oliver} S.,  et~al., 2010, \mn@doi [\mnras]
  {10.1111/j.1365-2966.2010.16643.x}, \href
  {https://ui.adsabs.harvard.edu/abs/2010MNRAS.405.2279O} {405, 2279}

\bibitem[\protect\citeauthoryear{{Orellana} et~al.,}{{Orellana}
  et~al.}{2017}]{Orellana_2017}
{Orellana} G.,  et~al., 2017, \mn@doi [\aap] {10.1051/0004-6361/201629009},
  \href {https://ui.adsabs.harvard.edu/abs/2017A&A...602A..68O} {602, A68}

\bibitem[\protect\citeauthoryear{{Pahre}, {Ashby}, {Fazio}  \&
  {Willner}}{{Pahre} et~al.}{2004}]{Pahre_2004}
{Pahre} M.~A.,  {Ashby} M.~L.~N.,  {Fazio} G.~G.,   {Willner} S.~P.,  2004,
  \mn@doi [\apjs] {10.1086/423320}, \href
  {https://ui.adsabs.harvard.edu/abs/2004ApJS..154..229P} {154, 229}

\bibitem[\protect\citeauthoryear{{Pipino}, {Kaviraj}, {Bildfell}, {Babul},
  {Hoekstra}  \& {Silk}}{{Pipino} et~al.}{2009}]{Pipino_2009}
{Pipino} A.,  {Kaviraj} S.,  {Bildfell} C.,  {Babul} A.,  {Hoekstra} H.,
  {Silk} J.,  2009, \mn@doi [\mnras] {10.1111/j.1365-2966.2009.14534.x}, \href
  {http://cdsads.u-strasbg.fr/abs/2009MNRAS.395..462P} {395, 462}

\bibitem[\protect\citeauthoryear{{Pipino}, {Szabo}, {Pierpaoli}, {MacKenzie}
  \& {Dong}}{{Pipino} et~al.}{2011}]{Pipino_2011}
{Pipino} A.,  {Szabo} T.,  {Pierpaoli} E.,  {MacKenzie} S.~M.,   {Dong} F.,
  2011, \mn@doi [\mnras] {10.1111/j.1365-2966.2011.19444.x}, \href
  {http://cdsads.u-strasbg.fr/abs/2011MNRAS.417.2817P} {417, 2817}

\bibitem[\protect\citeauthoryear{{Postman} et~al.,}{{Postman}
  et~al.}{2012}]{Postman_2012}
{Postman} M.,  et~al., 2012, \mn@doi [\apjs] {10.1088/0067-0049/199/2/25},
  \href {http://cdsads.u-strasbg.fr/abs/2012ApJS..199...25P} {199, 25}

\bibitem[\protect\citeauthoryear{{Pozzetti} et~al.,}{{Pozzetti}
  et~al.}{2010}]{Pozzetti_2010}
{Pozzetti} L.,  et~al., 2010, \mn@doi [\aap] {10.1051/0004-6361/200913020},
  \href {http://cdsads.u-strasbg.fr/abs/2010A%26A...523A..13P} {523, A13}

\bibitem[\protect\citeauthoryear{{Radovich} et~al.,}{{Radovich}
  et~al.}{2020}]{Radovich_2020}
{Radovich} M.,  et~al., 2020, \mn@doi [\mnras] {10.1093/mnras/staa2705}, \href
  {https://ui.adsabs.harvard.edu/abs/2020MNRAS.498.4303R} {498, 4303}

\bibitem[\protect\citeauthoryear{{Rafferty}, {McNamara}  \&
  {Nulsen}}{{Rafferty} et~al.}{2008}]{Rafferty_2008}
{Rafferty} D.~A.,  {McNamara} B.~R.,   {Nulsen} P.~E.~J.,  2008, \mn@doi [\apj]
  {10.1086/591240}, \href {http://adsabs.harvard.edu/abs/2008ApJ...687..899R}
  {687, 899}

\bibitem[\protect\citeauthoryear{{Reines} \& {Volonteri}}{{Reines} \&
  {Volonteri}}{2015}]{Reines_2015}
{Reines} A.~E.,  {Volonteri} M.,  2015, \mn@doi [\apj]
  {10.1088/0004-637X/813/2/82}, \href
  {https://ui.adsabs.harvard.edu/abs/2015ApJ...813...82R} {813, 82}

\bibitem[\protect\citeauthoryear{{Rodighiero} et~al.,}{{Rodighiero}
  et~al.}{2011}]{Rodighiero_2011}
{Rodighiero} G.,  et~al., 2011, \mn@doi [\apjl] {10.1088/2041-8205/739/2/L40},
  \href {https://ui.adsabs.harvard.edu/abs/2011ApJ...739L..40R} {739, L40}

\bibitem[\protect\citeauthoryear{{Runge} \& {Yan}}{{Runge} \&
  {Yan}}{2018}]{Runge_2018}
{Runge} J.,  {Yan} H.,  2018, \mn@doi [\apj] {10.3847/1538-4357/aaa020}, \href
  {http://cdsads.u-strasbg.fr/abs/2018ApJ...853...47R} {853, 47}

\bibitem[\protect\citeauthoryear{{Salpeter}}{{Salpeter}}{1955}]{Salpeter_1955}
{Salpeter} E.~E.,  1955, \mn@doi [\apj] {10.1086/145971}, \href
  {http://cdsads.u-strasbg.fr/abs/1955ApJ...121..161S} {121, 161}

\bibitem[\protect\citeauthoryear{{Santini} et~al.,}{{Santini}
  et~al.}{2009}]{Santini_2009}
{Santini} P.,  et~al., 2009, \mn@doi [\aap] {10.1051/0004-6361/200811434},
  \href {https://ui.adsabs.harvard.edu/abs/2009A&A...504..751S} {504, 751}

\bibitem[\protect\citeauthoryear{{Schlegel}, {Finkbeiner}  \&
  {Davis}}{{Schlegel} et~al.}{1998}]{Schlegel_1998}
{Schlegel} D.~J.,  {Finkbeiner} D.~P.,   {Davis} M.,  1998, \mn@doi [\apj]
  {10.1086/305772}, \href {http://adsabs.harvard.edu/abs/1998ApJ...500..525S}
  {500, 525}

\bibitem[\protect\citeauthoryear{{Scoville} et~al.,}{{Scoville}
  et~al.}{2007}]{Scoville_2007_COSMOS}
{Scoville} N.,  et~al., 2007, \mn@doi [\apjs] {10.1086/516585}, \href
  {http://cdsads.u-strasbg.fr/abs/2007ApJS..172....1S} {172, 1}

\bibitem[\protect\citeauthoryear{{Sobral}, {Best}, {Smail}, {Mobasher}, {Stott}
   \& {Nisbet}}{{Sobral} et~al.}{2014}]{Sobral_2014}
{Sobral} D.,  {Best} P.~N.,  {Smail} I.,  {Mobasher} B.,  {Stott} J.,
  {Nisbet} D.,  2014, \mn@doi [\mnras] {10.1093/mnras/stt2159}, \href
  {https://ui.adsabs.harvard.edu/abs/2014MNRAS.437.3516S} {437, 3516}

\bibitem[\protect\citeauthoryear{{Speagle}, {Steinhardt}, {Capak}  \&
  {Silverman}}{{Speagle} et~al.}{2014}]{Speagle_2014}
{Speagle} J.~S.,  {Steinhardt} C.~L.,  {Capak} P.~L.,   {Silverman} J.~D.,
  2014, \mn@doi [\apjs] {10.1088/0067-0049/214/2/15}, \href
  {http://adsabs.harvard.edu/abs/2014ApJS..214...15S} {214, 15}

\bibitem[\protect\citeauthoryear{{Steinhardt} et~al.,}{{Steinhardt}
  et~al.}{2014}]{Steinhardt_2014}
{Steinhardt} C.~L.,  et~al., 2014, \mn@doi [\apjl]
  {10.1088/2041-8205/791/2/L25}, \href
  {https://ui.adsabs.harvard.edu/abs/2014ApJ...791L..25S} {791, L25}

\bibitem[\protect\citeauthoryear{{Stern} et~al.,}{{Stern}
  et~al.}{2012}]{Stern_2012}
{Stern} D.,  et~al., 2012, \mn@doi [\apj] {10.1088/0004-637X/753/1/30}, \href
  {http://cdsads.u-strasbg.fr/abs/2012ApJ...753...30S} {753, 30}

\bibitem[\protect\citeauthoryear{{Tonini}, {Bernyk}, {Croton}, {Maraston}  \&
  {Thomas}}{{Tonini} et~al.}{2012}]{Tonini_2012}
{Tonini} C.,  {Bernyk} M.,  {Croton} D.,  {Maraston} C.,   {Thomas} D.,  2012,
  \mn@doi [\apj] {10.1088/0004-637X/759/1/43}, \href
  {http://cdsads.u-strasbg.fr/abs/2012ApJ...759...43T} {759, 43}

\bibitem[\protect\citeauthoryear{{Tonry}}{{Tonry}}{1987}]{Tonry_1987}
{Tonry} J.~L.,  1987, in {de Zeeuw} P.~T.,  ed.,  IAU Symposium Vol. 127,
  Structure and Dynamics of Elliptical Galaxies. pp 89--96

\bibitem[\protect\citeauthoryear{{Voit} \& {Donahue}}{{Voit} \&
  {Donahue}}{2015}]{Voit_2015a}
{Voit} G.~M.,  {Donahue} M.,  2015, \mn@doi [\apjl]
  {10.1088/2041-8205/799/1/L1}, \href
  {https://ui.adsabs.harvard.edu/abs/2015ApJ...799L...1V} {799, L1}

\bibitem[\protect\citeauthoryear{{Voit}, {Meece}, {Li}, {O'Shea}, {Bryan}  \&
  {Donahue}}{{Voit} et~al.}{2017}]{Voit_2017}
{Voit} G.~M.,  {Meece} G.,  {Li} Y.,  {O'Shea} B.~W.,  {Bryan} G.~L.,
  {Donahue} M.,  2017, \mn@doi [\apj] {10.3847/1538-4357/aa7d04}, \href
  {https://ui.adsabs.harvard.edu/abs/2017ApJ...845...80V} {845, 80}

\bibitem[\protect\citeauthoryear{{Webb} et~al.,}{{Webb}
  et~al.}{2015}]{Webb_2015}
{Webb} T.~M.~A.,  et~al., 2015, \mn@doi [\apj] {10.1088/0004-637X/814/2/96},
  \href {http://cdsads.u-strasbg.fr/abs/2015ApJ...814...96W} {814, 96}

\bibitem[\protect\citeauthoryear{{Webb} et~al.,}{{Webb}
  et~al.}{2017}]{Webb_2017}
{Webb} T.~M.~A.,  et~al., 2017, \mn@doi [\apjl] {10.3847/2041-8213/aa7749},
  \href {http://cdsads.u-strasbg.fr/abs/2017ApJ...844L..17W} {844, L17}

\bibitem[\protect\citeauthoryear{{Wen} \& {Han}}{{Wen} \&
  {Han}}{2015}]{Wen_2015b}
{Wen} Z.~L.,  {Han} J.~L.,  2015, \mn@doi [\mnras] {10.1093/mnras/stu2722},
  \href {http://cdsads.u-strasbg.fr/abs/2015MNRAS.448....2W} {448, 2}

\bibitem[\protect\citeauthoryear{{Wen} \& {Han}}{{Wen} \&
  {Han}}{2020}]{Wen_2020}
{Wen} Z.~L.,  {Han} J.~L.,  2020, \mn@doi [\mnras] {10.1093/mnras/staa3308},
  \href {https://ui.adsabs.harvard.edu/abs/2020MNRAS.tmp.3120W} {}

\bibitem[\protect\citeauthoryear{{Wen}, {Han}  \& {Liu}}{{Wen}
  et~al.}{2012}]{Wen_2012}
{Wen} Z.~L.,  {Han} J.~L.,   {Liu} F.~S.,  2012, \mn@doi [\apjs]
  {10.1088/0067-0049/199/2/34}, \href
  {http://cdsads.u-strasbg.fr/abs/2012ApJS..199...34W} {199, 34}

\bibitem[\protect\citeauthoryear{{Wetzel}, {Tinker}  \& {Conroy}}{{Wetzel}
  et~al.}{2012}]{Wetzel_2012}
{Wetzel} A.~R.,  {Tinker} J.~L.,   {Conroy} C.,  2012, \mn@doi [\mnras]
  {10.1111/j.1365-2966.2012.21188.x}, \href
  {https://ui.adsabs.harvard.edu/abs/2012MNRAS.424..232W} {424, 232}

\bibitem[\protect\citeauthoryear{{Whiley} et~al.,}{{Whiley}
  et~al.}{2008}]{Whiley_2008}
{Whiley} I.~M.,  et~al., 2008, \mn@doi [\mnras]
  {10.1111/j.1365-2966.2008.13324.x}, \href
  {http://cdsads.u-strasbg.fr/abs/2008MNRAS.387.1253W} {387, 1253}

\bibitem[\protect\citeauthoryear{{Wright} et~al.,}{{Wright}
  et~al.}{2010}]{Wright_2010}
{Wright} E.~L.,  et~al., 2010, \mn@doi [\aj] {10.1088/0004-6256/140/6/1868},
  \href {http://cdsads.u-strasbg.fr/abs/2010AJ....140.1868W} {140, 1868}

\bibitem[\protect\citeauthoryear{{York} et~al.,}{{York}
  et~al.}{2000}]{York_2000}
{York} D.~G.,  et~al., 2000, \mn@doi [\aj] {10.1086/301513}, \href
  {http://cdsads.u-strasbg.fr/abs/2000AJ....120.1579Y} {120, 1579}

\bibitem[\protect\citeauthoryear{{Zahid}, {Dima}, {Kewley}, {Erb}  \&
  {Dav{\'e}}}{{Zahid} et~al.}{2012}]{zahid_2012}
{Zahid} H.~J.,  {Dima} G.~I.,  {Kewley} L.~J.,  {Erb} D.~K.,   {Dav{\'e}} R.,
  2012, \mn@doi [\apj] {10.1088/0004-637X/757/1/54}, \href
  {https://ui.adsabs.harvard.edu/abs/2012ApJ...757...54Z} {757, 54}

\bibitem[\protect\citeauthoryear{{Zhao}, {Arag{\'o}n-Salamanca}  \&
  {Conselice}}{{Zhao} et~al.}{2015}]{Zhao_2015b}
{Zhao} D.,  {Arag{\'o}n-Salamanca} A.,   {Conselice} C.~J.,  2015, \mn@doi
  [\mnras] {10.1093/mnras/stv1940}, \href
  {http://cdsads.u-strasbg.fr/abs/2015MNRAS.453.4444Z} {453, 4444}

\bibitem[\protect\citeauthoryear{{de Jong} et~al.,}{{de Jong}
  et~al.}{2013}]{de_Jong_2013_KIDS}
{de Jong} J.~T.~A.,  et~al., 2013, The Messenger, \href
  {https://ui.adsabs.harvard.edu/abs/2013Msngr.154...44D} {154, 44}

\bibitem[\protect\citeauthoryear{{von der Linden}, {Best}, {Kauffmann}  \&
  {White}}{{von der Linden} et~al.}{2007}]{von_der_Linden_2007}
{von der Linden} A.,  {Best} P.~N.,  {Kauffmann} G.,   {White} S.~D.~M.,  2007,
  \mn@doi [\mnras] {10.1111/j.1365-2966.2007.11940.x}, \href
  {http://cdsads.u-strasbg.fr/abs/2007MNRAS.379..867V} {379, 867}

\makeatother
\end{thebibliography}

\appendix

\section{Near Infrared colours as tracers of star formation}
\label{subsec: wise colours}

\begin{figure*}
\centering
\includegraphics[width=0.3\textwidth]{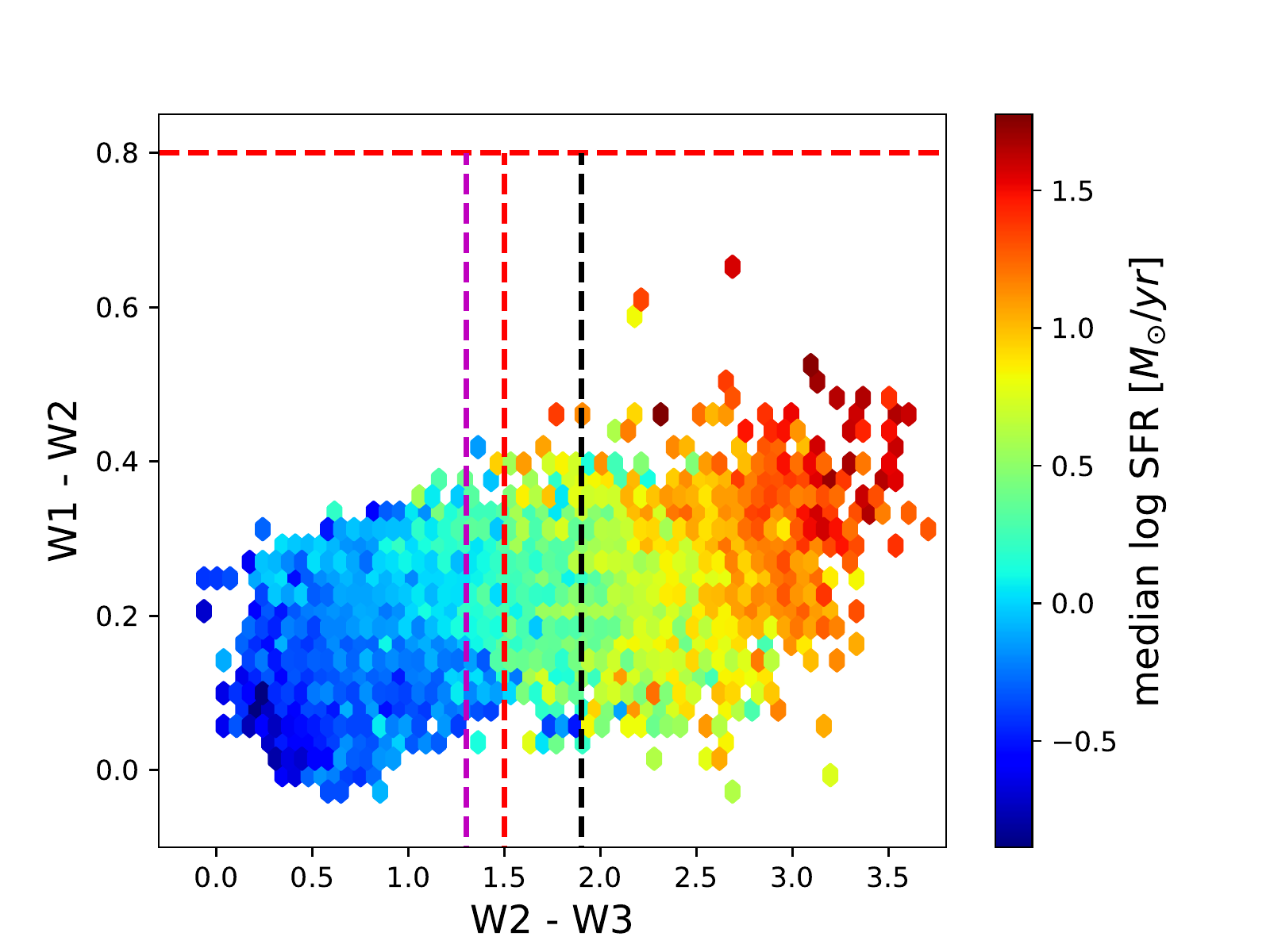}
\includegraphics[width=0.3\textwidth]{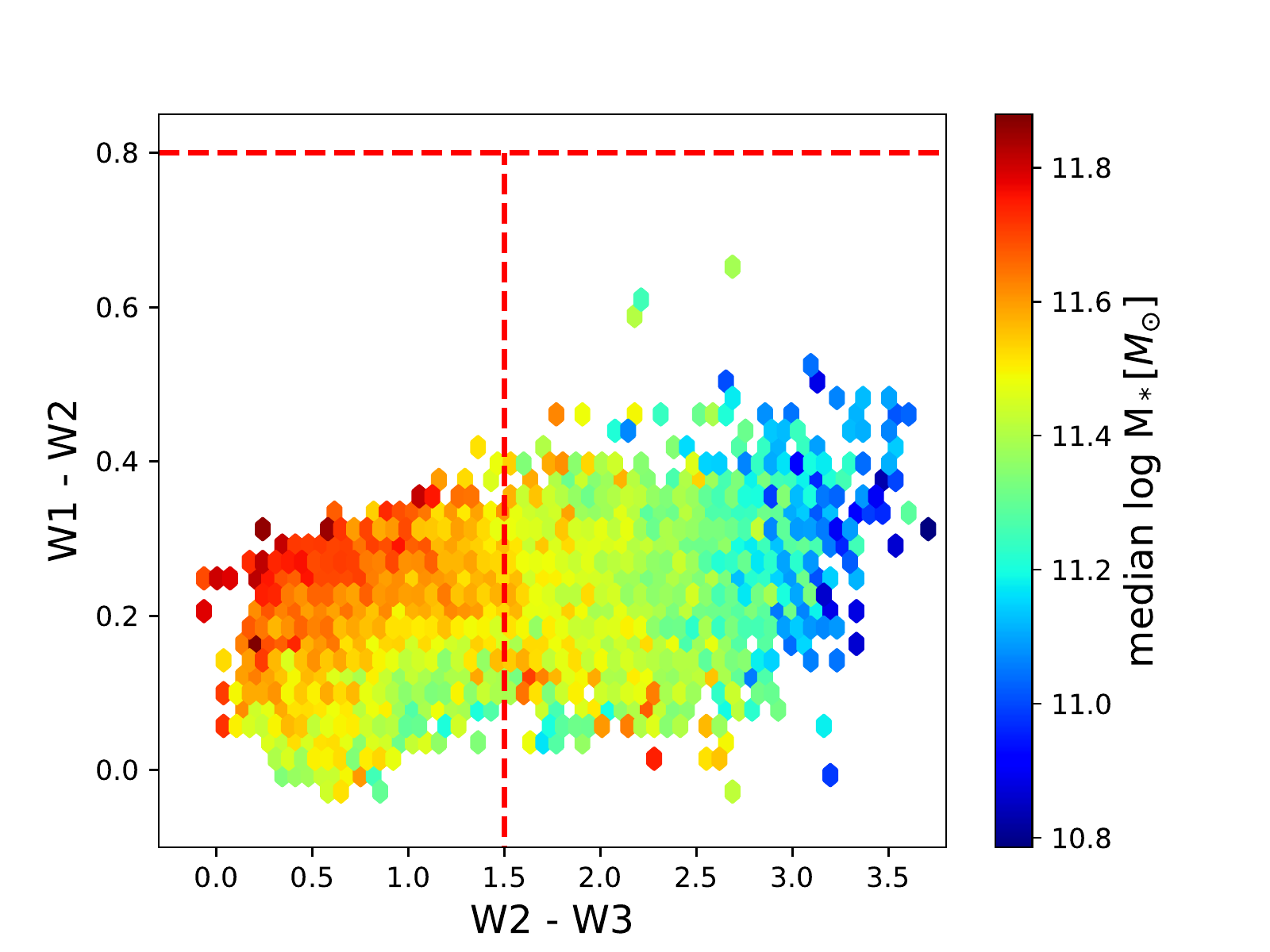}
\includegraphics[width=0.3\textwidth]{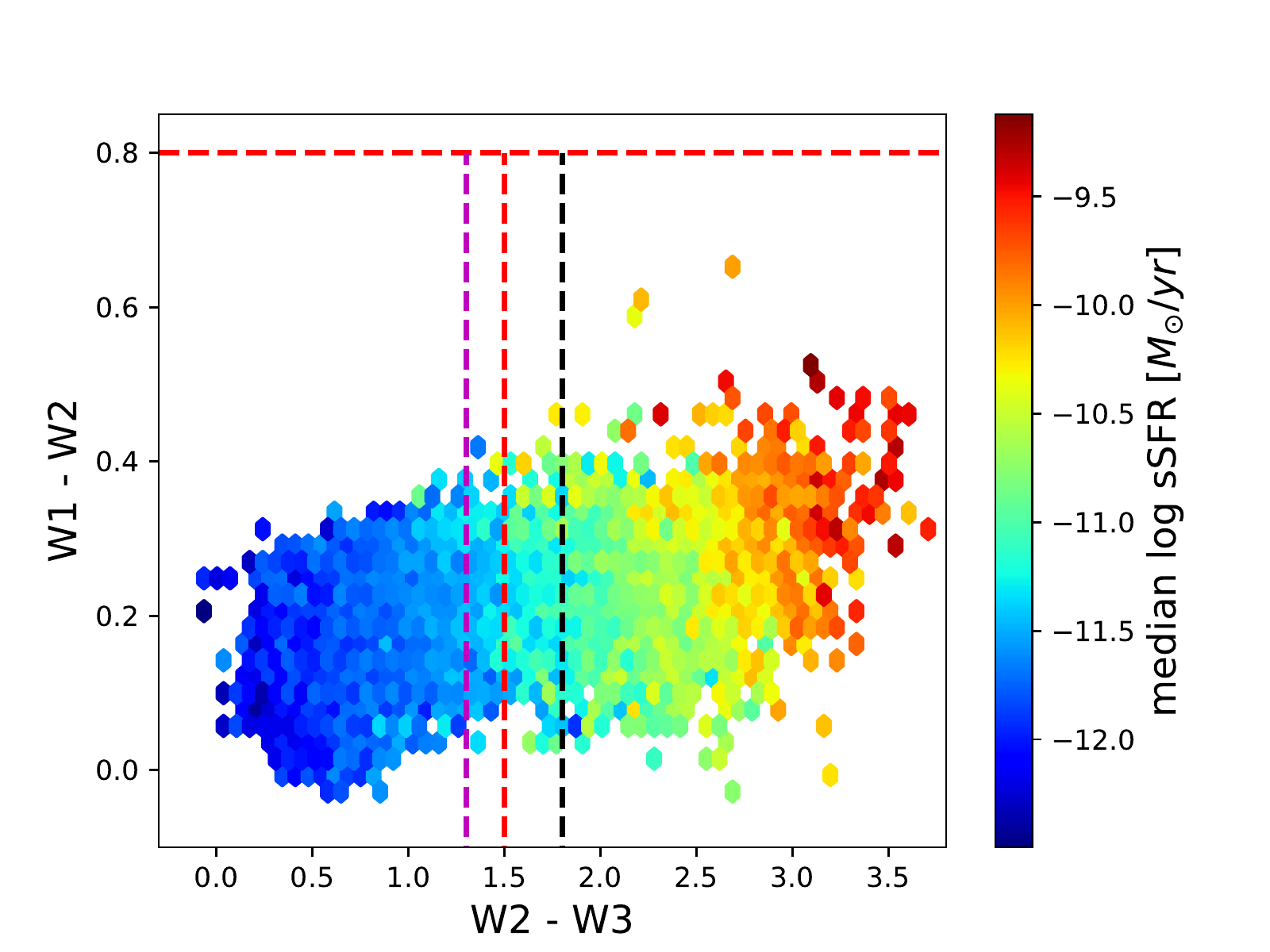}
\includegraphics[width=0.3\textwidth]{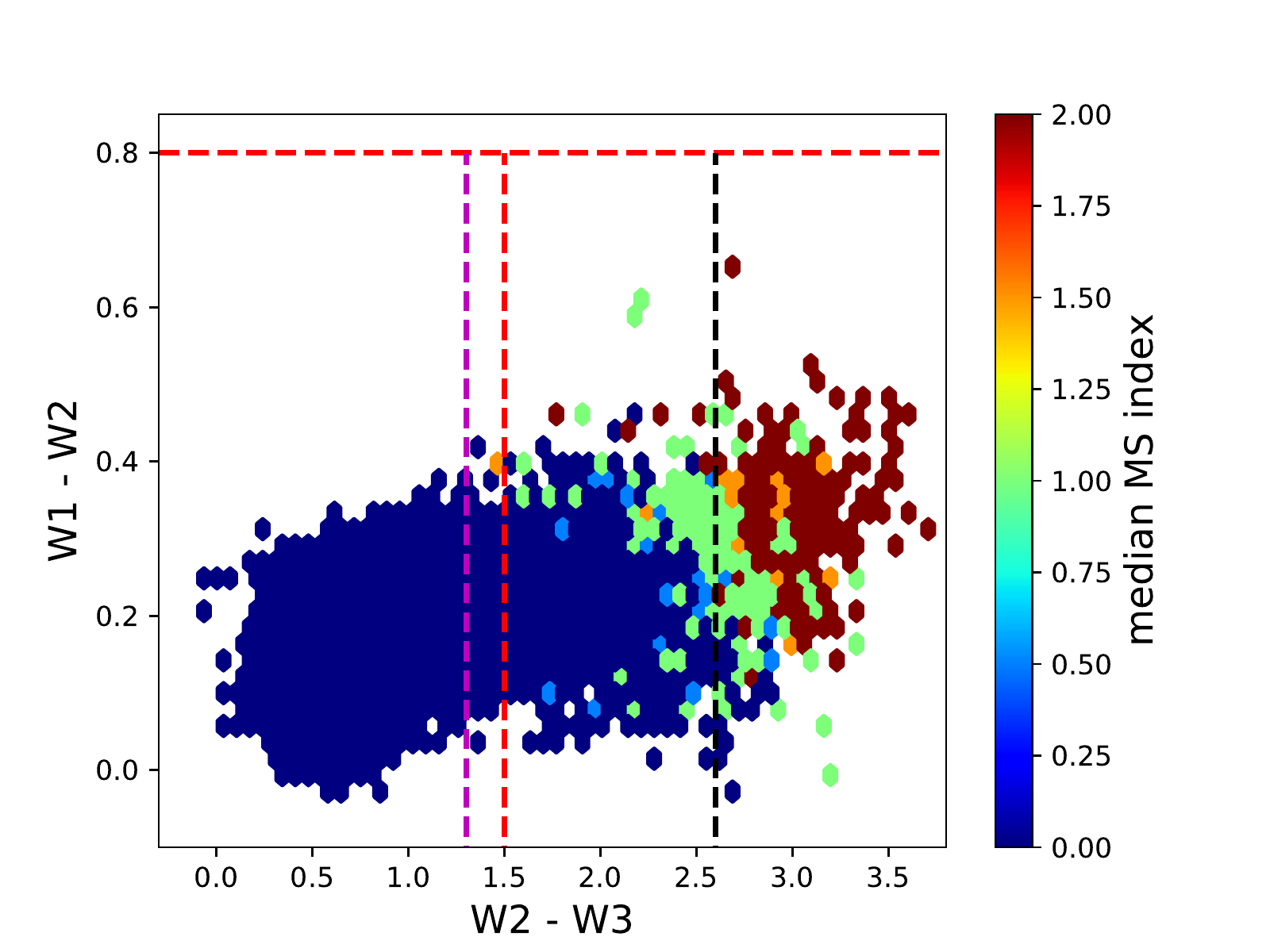}
\includegraphics[width=0.3\textwidth]{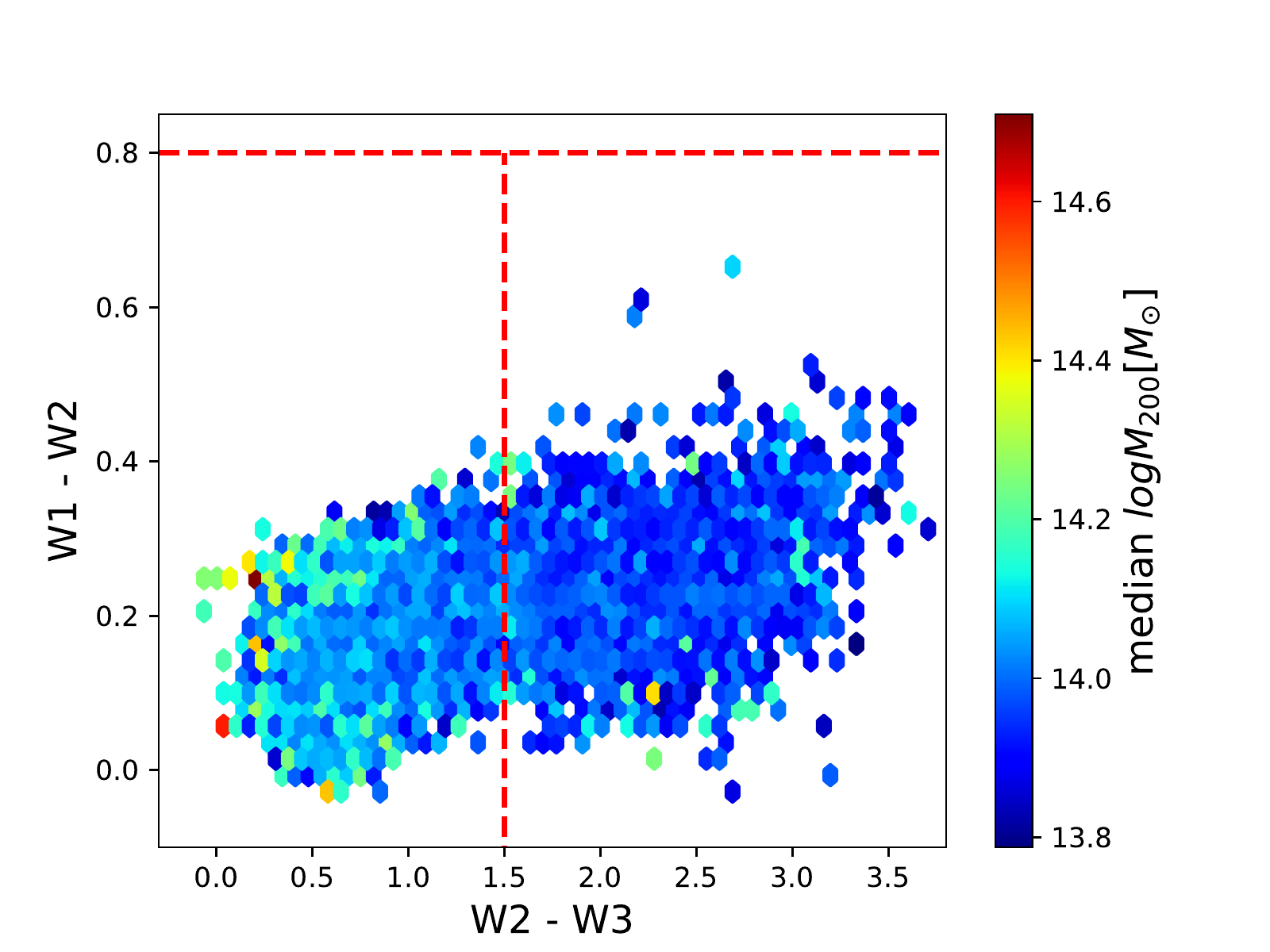}
\includegraphics[width=0.3\textwidth]{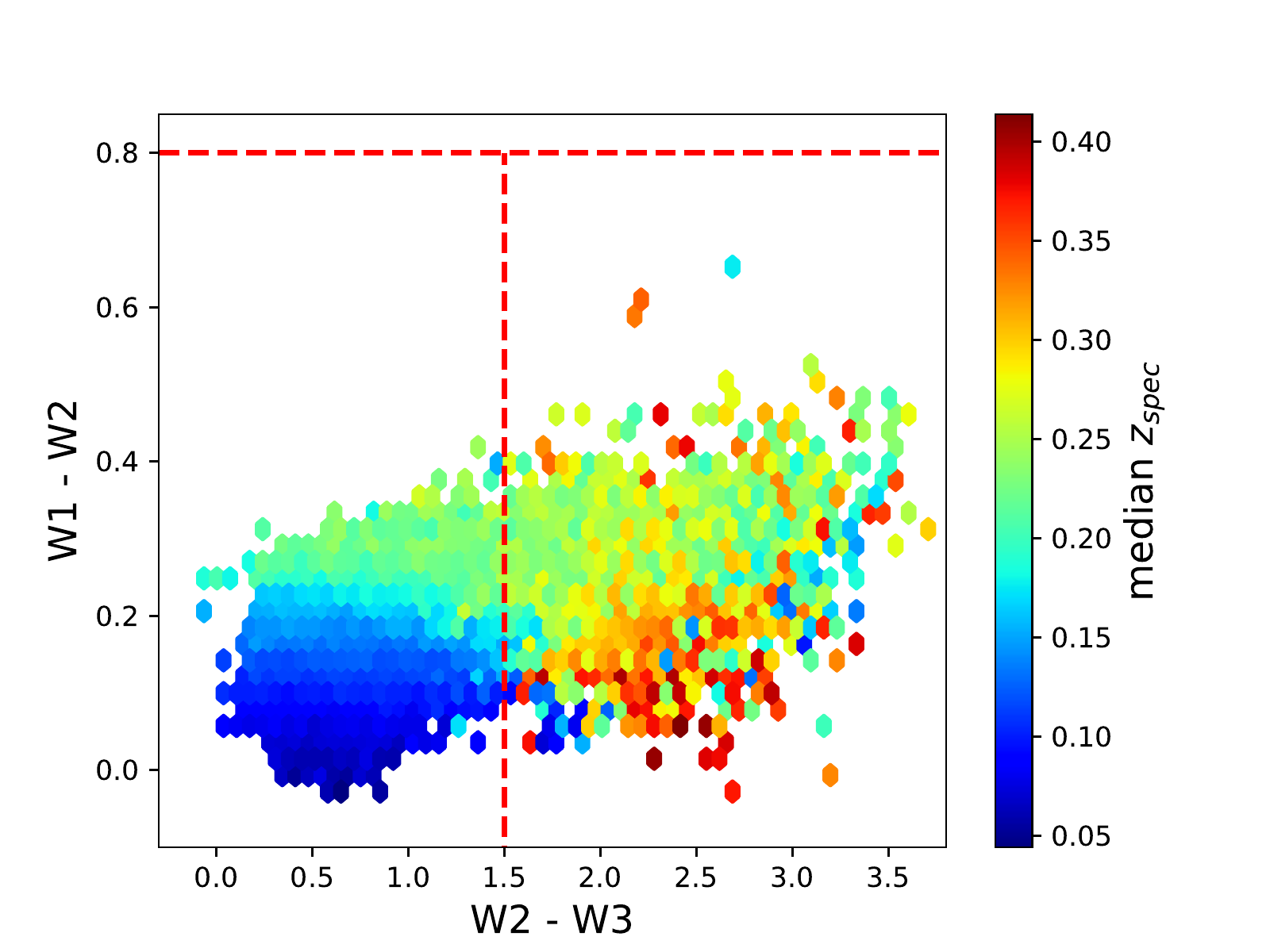}

\caption{WISE colour-colour diagram represented as a 2D density plot in which each cell contains at least two BCGs.
The colour bars indicate the median value of $\log{(SFR)}$ (top left), log \mstar\ (top middle), $\log{(sSFR)}$ (top right), the $\rm MS_{index}$ (bottom left), \mh\ (bottom middle) and spectroscopic redshift (bottom right), in each cell.
The horizontal red line shows the limit for AGN-host galaxies according to \citet{Wright_2010}, while the vertical red line shows the separation between galaxies with and without star formation, used in Paper I.
The vertical black line shows the separation where non passive galaxies dominate over the passive ones, following the MS criterion. The magenta vertical line shows the separation between passive and star-forming galaxies, according to their $sSFR$ (see Section \ref{sec:results}).
}
\label{fig:wise colours}	
\end{figure*}

In Paper I we used the WISE IR colours to identify BCGs with ongoing star formation. In this appendix we relate that definition of star-forming galaxies with those based on the MS and the $sSFR$ adopted in this work.

\citet{Jarrett_2011} showed that the WISE $W1$ and $W2$ bands are reliable tracers of the continuum emission from low-mass, evolved stars, meaning that the fluxes in these filters can be used to trace the \mstar\ of galaxies \citep{Pahre_2004, Meidt_2012, Hunter_2006, Cluver_2014}. 
Furthermore, they showed that the flux measurements in these filters are only marginally affected by dust attenuation.

On the other hand, $W3$ traces the emission from PAH and the $\rm[Ne II]$ line, indicating that the flux at these wavelengths is sensitive to the $SFR$, the total IR luminosity (TIR) and the hot dust heated by AGN emission (e.g., \citealt{Meidt_2012, Hunter_2006, Elbaz_2011,Cluver_2014, Orellana_2017} ). Finally, the $W4$ filter traces the continuum emission related with AGN-heated dust, the $SFR$ and the TIR (e.g., \citealt{Elbaz_2011,Cluver_2014,Orellana_2017}).
 
In Paper I we used the $(W1-W2)$ vs $(W2-W3)$ colour-colour diagram to identify quiescent, star-forming and AGN-host BCGs and investigated the trends with redshift, cluster mass and BCG \mstar\ for each galaxy type. We now perform again the identification of passive, star-forming and AGN-host galaxies in the sub-sample of BCGs with measurements in $W1$, $W2$ and $W3$, and compare this classification with the subdivision into quiescent and star-forming BCGs based on $sSFR$. Although we find that only 12,625 galaxies (22.4\% of the sample) meet this requirement, we note that this sub-sample is 83.1\% larger than the one used in Paper I, highlighting the higher quality of the unWISE photometry with respect to AllWISE.

Fig. \ref{fig:wise colours} shows different versions of the $(W1-W2)$ vs $(W2-W3)$ colour-colour plot. The diagram is represented as a 2D density plot in which only cells with at least two galaxies are plotted. The colour bars indicate the median $SFR$ (top-left panel), median \mstar\ (top-middle panel), median $\log{(sSFR)}$ (top-right panel), median value of the $\rm MS_{index}$ (bottom-left panel), median \mh\ (bottom-middle panel), and  median $z_{spec}$ (bottom right panel) in each cell.

We find that only $0.33\%$ of the BCGs in the sub-sample fall in the AGN-host locus ($(W1 - W2) > 0.8$). If we consider the BCGs with $W1$ and $W2$, but not necessarily $W3$, this fraction drops to $0.07\%$. We note that AGN hosts are extremely sparse in the colour-colour plot, and in no case there are regions of the AGN-host locus in which we can make cells with at least two objects. For this reason these galaxies do not appear in Fig. \ref{fig:wise colours} (above the horizontal red line). 

As it can be seen in the top-left panel in Figure \ref{fig:wise colours}, the $(W1 - W2)$ colour  is not a good tracer of the $SFR$ in galaxies. Only at $(W1 - W2) > 0.4$, it is possible to see that the galaxies have  higher $SFR$. As already shown in \citet{Jarrett_2011}, the $(W2- W3)$ colour is a more sensitive tracer of $SFR$: as this colour increases, the increase in $SFR$ in the top-left panel of Figure \ref{fig:wise colours} is also visible. In Paper I, we used $(W2 - W3)$ to separate star forming ($(W2 - W3) \geq 1.5$) from quiescent ($(W2 - W3) < 1.5$) galaxies, following the boundaries in \citet{Jarrett_2011, Jarrett_2017} and \citet{Fraser_McKelvie_2014}. Using the $(W2 - W3) = 1.5$ boundary as in Paper I, we find that the median $SFR$ for galaxies with $(W2 - W3) < 1.5$ is 0.71 \msun/yr. Meanwhile, for galaxies with $(W2 - W3) \geq 1.5$, the median SFR is 5.69 \msun/yr, showing a clear separation between star-forming and passive galaxies.

To test this colour boundary for the identification of star-forming BCGs, we divide the  $(W2 - W3)$ range in bins with width of $\Delta (W2 - W3) = 0.5$, and we measure the median $SFR$ and the fraction of passive and star-forming galaxies following our $sSFR$ criterion in each colour bin.
Then, we try to define a boundary between star-forming and quiescent galaxies identifying the bin where the fraction of star-forming BCGs is larger than the fraction of quenched BCGs.
Following this procedure, we find that the separation between star-forming and quenched BCGs is at $(W2 - W3) = 1.3$ where the median $SFR$ for galaxies with $(W2 - W3) < 1.3$ is 0.64 \msun/yr while for galaxies with $(W2 - W3) \geq 1.3$ is 4.61 \msun/yr. 

Using a similar procedure to separate the galaxies according to the MS classification, we find that the boundary between passive and non-passive (starburst plus MS) galaxies is at $(W2 - W3) = 2.5$. Below this colour we find that the  median $SFR$ is 1.25 \msun/yr and above it it is 14.63 \msun/yr. These $SFR$ values are higher than those obtained from the use of the $sSFR$ criterion to divide into star-forming and quiescent BCGs.

Thus, regardless of the criterion used to split between star-forming and quiescent galaxies, we see that in agreement with \cite{Jarrett_2011} and \cite{Jarrett_2017} the $(W2-W3)$ colour can be used as a powerful yet simple tool to separate between galaxies with and without star formation. However, we present here updated boundaries in $(W2-W3)$ to separate between star-forming and quiescent BCGs, applicable with both definitions of star-forming galaxies, based on $sSFR$ or the position with respect to the MS.

We can see from the central panel in the top row of Figure \ref{fig:wise colours} that we can define zones in the colour-colour plot populated by galaxies with higher and lower \mstar. High-\mstar\ galaxies are located at $(W2-W3) < 1.5$, where we find that the median stellar mass is \mstar\ $= 10^{11.56}$ \msun. Intermediate-\mstar\ BCGs are located at $1.5 < (W2-W3) < 2.8$, where the median stellar mass is \mstar $= 10^{11.41}$ \msun. Finally, low-\mstar\ BCGs are located at $(W2-W3) > 2.8$, where the median stellar mass is \mstar $= 10^{11.14}$ \msun.  We note that, as seen in Section \ref{subsec:fraction_evolution}, the fraction \fsf\ of star-forming BCGs decreases with increasing \mstar. Since the $(W2-W3)$ colour is sensitive to the $SFR$, the fact that lower-mass BCGs have redder $(W2-W3)$ colours is just a reflection of the fact that star-forming galaxies tend to be concentrated at red $(W2-W3)$. 

The WISE colour-colour diagram can also be used to split the sample into regions with different values of $sSFR$. 
Using the same colour limits used to split BCGs into high-, intermediate- and low-\mstar, we find that at $(W2-W3) < 1.5$ the median $sSFR$ is $\log{(sSFR/\mbox{yr}^{-1})} = -11.69$, at $1.5 < (W2-W3) < 2.8$, the median $sSFR$ is $\log{(sSFR/\mbox{yr}^{-1})} = -10.81$ and, finally, at $W2-W3 > 2.8$ the median $sSFR$ is $\log{(sSFR/\mbox{yr}^{-1})} = -9.84$ (see Figure \ref{fig:wise colours}, right-hand panel in the top row). The cluster halo mass shows no apparent relation with the IR colours, neither is it possible to distinguish  different regions in the colour-colour plot that comprise clusters with certain ranges of \mh. 

Finally, we find that the WISE colours depend on the BCG spectroscopic redshift. As it can be seen from the right-hand panel in the bottom row of Figure \ref{fig:wise colours}, the highest- and lowest-redshift BCGs appear confined at $(W1-W2) < 0.2$. However, the lowest-redshift BCGs appear to have $(W2-W3)$ colours below 1.5, while the highest-redshift BCGs have $2.0 < (W2 - W3) < 2.9$. 
Galaxies with redshifts $0.15 < z < 0.33$ are spread across the entire colour-colour range, although with a preference for galaxies with $0.2 < z < 0.28$ to have $(W2-W3) > 1.5$.

\bsp	
\label{lastpage}
\end{document}